\documentclass[onecolumn, draftcls, peerreview, 12pt, romanappendices]{IEEEtran}

\newcommand{\sss}{\scriptscriptstyle}
\newcommand{\mbs}[1]{\boldsymbol{#1}}
\renewcommand{\mbs}[1]{\mathbf{#1}}
\renewcommand{\mbs}[1]{\pmb{#1}}

\newcommand{\vect}[1]{{\lowercase{\mbs{#1}}}}
\newcommand{\mat}[1]{{\uppercase{\mbs{#1}}}}

\newcommand{\asympteq}{\doteq} %

\usepackage{footnote} 
\makesavenoteenv{tabular}
\usepackage{color}
\usepackage{url}
\usepackage{cite}
\usepackage{graphicx}
\usepackage{graphics}
\usepackage{epsfig}
\usepackage{subfigure}
\usepackage{amsmath,amssymb,amsfonts,mathrsfs,amscd}
\usepackage{pst-node,pst-text,pst-3d}

\newtheorem{proposition}{Proposition}[section]
\newtheorem{remark}{Remark}[section]
\newtheorem{definition}{Definition}[section]
\newtheorem{theorem}{Theorem}[section]
\newtheorem{example}{Example}[section]

\newtheorem{corollary}{Corollary}[section]

\newtheorem{lemma}{Lemma}[section]

\newcommand{\bb}[1]{\mathbb{#1}}

\newcommand{\Fig}[1]{Fig.~\!\ref{#1}}

\newcommand{\Eq}[1]{(\ref{#1})}

\newcommand{\ie}{\emph{i.e.}}
\newcommand{\eg}{\emph{e.g.}}

\newcommand{\D}{\displaystyle}

\newcommand{\eqncaseslabel}[6]{
  \begin{equation}
    \setlength{\nulldelimiterspace}{0pt}
    #1=\left\{
      \begin{IEEEeqnarraybox}[\relax][c]{l's}
        #2, &for #3\\
        #4, &for #5%
      \end{IEEEeqnarraybox}\right.\label{#6}
  \end{equation}
}

\newcommand{\eqncasesasympt}[5]{
  \begin{equation}
    \setlength{\nulldelimiterspace}{0pt}
    #1\asympteq\left\{
      \begin{IEEEeqnarraybox}[\relax][c]{l's}
        #2, &#3\\
        #4, &#5%
      \end{IEEEeqnarraybox}\right.
  \end{equation}
}

\newcommand{\eqncasesasymptlabel}[6]{
  \begin{equation}
    \setlength{\nulldelimiterspace}{0pt}
    #1\asympteq\left\{
      \begin{IEEEeqnarraybox}[\relax][c]{l's}
        #2, &#3\\
        #4, &#5%
      \end{IEEEeqnarraybox}\right.\label{#6}
  \end{equation}
}

\renewcommand{\matrix}[1]{\begin{bmatrix}#1\end{bmatrix}}

\newcommand\transcsymbol{\scriptscriptstyle \dag \!}
\newcommand\transsymbol{\scriptscriptstyle \mathsf{T} \!}

\newcommand\Abs[1]{\left|#1\right|}
\newcommand\floor[1]{\left\lfloor{#1}\right\rfloor}
\newcommand\ceil[1]{\left\lceil{#1}\right\rceil}
\newcommand\Abssqr[1]{\left|#1\right|^2}

\newcommand{\inv}[1]{{#1}^{\scriptscriptstyle -1 \!}}
\newcommand{\transc}[1]{{#1}^{\transcsymbol}}
\newcommand{\trans}[1]{{#1}^{\transsymbol}}
\newcommand{\pstv}[1]{{#1}^{\sss +}}

\newcommand\diag{\mathrm{diag}}

\newcommand\Det{\mathrm{Det}}
\newcommand\Norm[1]{\left\|{#1}\right\|}
\newcommand\Frob[1]{\Norm{#1}^2_{\textrm{F}}}

\newcommand\defeq{\triangleq}

\newcommand\Prob[1]{\textrm{Prob}\left\{#1\right\}}
\newcommand\prob{\textrm{Prob}}

\newcommand{\Id}{\mathbf{I}}
\newcommand{\CN}[1][\Id]{\Ccal\Ncal\!\left(0,#1\right)}

\newcommand{\SNR}{{\mathsf{SNR}} }

\newcommand\Pe{P_{\textrm{e}}}
\newcommand\Pout{P_{\textrm{out}}}

\renewcommand\d{\mathrm{d}}

\newcommand\CC{\bb{C}}
\newcommand\EE{\bb{E}}
\newcommand\FF{\bb{F}}
\newcommand\KK{\bb{K}}
\newcommand\LL{\bb{L}}
\newcommand\QQ{\bb{Q}}
\newcommand\RR{\bb{R}}
\newcommand\ZZ{\bb{Z}}

\newcommand\Acal{\mathcal{A}}
\newcommand\Bcal{\mathcal{B}}
\newcommand\Ccal{\mathcal{C}}
\newcommand\Dcal{\mathcal{D}}

\newcommand\Gcal{\mathcal{G}}

\newcommand\Ical{\mathcal{I}}

\newcommand\Ncal{\mathcal{N}}
\newcommand\Ocal{\mathcal{O}}
\newcommand\Pcal{\mathcal{P}}

\newcommand\Rcal{\mathcal{R}}
\newcommand\Scal{\mathcal{S}}
\newcommand\Tcal{\mathcal{T}}

\newcommand\Wcal{\mathcal{W}}
\newcommand\Xcal{\mathcal{X}}

\newcommand{\mD}{\mat{D}}

\newcommand{\mF}{\mat{F}}
\newcommand{\mG}{\mat{G}}
\newcommand{\mH}{\mat{H}}

\newcommand{\mJ}{\mat{J}}

\newcommand{\mM}{\mat{M}}

\newcommand{\mP}{\mat{P}}
\newcommand{\mQ}{\mat{Q}}
\newcommand{\mR}{\mat{R}}
\newcommand{\mS}{\mat{S}}
\newcommand{\mT}{\mat{T}}
\newcommand{\mU}{\mat{U}}
\newcommand{\mV}{\mat{V}}
\newcommand{\mW}{\mat{W}}
\newcommand{\mX}{\mat{X}}

\newcommand{\mc}{\vect{c}}

\newcommand{\mh}{\vect{h}}

\newcommand{\mm}{\vect{m}}
\newcommand{\mn}{\vect{n}}

\newcommand{\mx}{\vect{x}}
\newcommand{\my}{\vect{y}}
\newcommand{\mz}{\vect{z}}

\newcommand{\mXi}{\mbs{\Xi}}

\newcommand{\mPi}{\mat{\Pi}}

\newcommand{\mOmega}{\mbs{\Omega}}

\newcommand{\mSigma}{\mbs{\Sigma}}

\newcommand{\mPhi}{\mbs{\Phi}}

\newcommand{\mlambda}{\boldsymbol{\lambda}}
\newcommand{\malpha}{\boldsymbol{\alpha}}

\newcommand{\mmu}{\boldsymbol{\mu}}
\newcommand{\mbeta}{\boldsymbol{\beta}}

\usepackage{enumerate}

\usepackage{cases}

\begin{document}

 \title{Diversity of MIMO Multihop Relay Channels}
  \author{%
    \IEEEauthorblockN{Sheng Yang and Jean-Claude Belfiore\thanks{Manuscript
        submitted to the IEEE Transactions on Information Theory.  The
        authors are with the Department of Communications and
        Electronics, \'{E}cole Nationale Sup\'{e}rieure des
        T\'{e}l\'{e}communications, 46, rue Barrault,
     75013 Paris, France~(e-mail: syang@enst.fr; belfiore@enst.fr).}}}

\maketitle

\renewcommand{\IEEEQED}{\IEEEQEDopen}
\newcommand{\nT}{n_\text{T}}
\newcommand{\nR}{n_\text{R}}
\newcommand{\nS}{n_\text{S}}
\newcommand{\PhiT}{\mPhi_\text{T}}
\newcommand{\PhiR}{\mPhi_\text{R}}
\newcommand{\PhiS}{\mPhi_\text{S}}
\newcommand{\asymptleq}{\ \dot{\leq}\,}
\newcommand{\asymptgeq}{\ \dot{\geq}\,}
\newcommand{\mnt}{\mbs{\tilde{n}}}
\newcommand{\E}{E}
\newcommand{\Ec}{\hat{E}}
\newcommand{\ilb}{\underline{i}}
\newcommand{\jlb}{\underline{j}}
\newcommand{\wrt}{\emph{w.r.t.}}

\newcommand{\nminp}{n'_{\min}}
\newcommand{\nmin}{n_{\min}}
\newcommand{\DMT}{\textsf{DMT}}
\renewcommand{\div}{\textsf{div}}

\newcommand{\dAF}{d^{\text{AF}}}
\newcommand{\dDF}{d^{\text{DF}}}
\newcommand{\dPF}{d^{\text{PF}}}
\newcommand{\dRP}{d^{\text{RP}}}

\newcommand{\Ndc}{N_{\text{d/c}}}
\newcommand{\Ndcs}{N^*_{\text{d/c}}}

\newcommand{\dmax}{d_{\max}}
\newcommand{\rmax}{r_{\max}}
\newcommand{\RP}{\mPi}
\newcommand{\RPC}{{RP channel}}
\newcommand{\F}[2]{\mF_{{#1,#2}}}
\newcommand{\J}[2]{\mJ_{{#1,#2}}}
\renewcommand{\mod}[2]{#1\ \text{mod}\ #2}

\newcommand{\dFF}{d^{\text{FF}}}
\renewcommand{\dAF}{d^{\text{AF}}}

\newcommand{\umH}{\underline{\mH}}
\newcommand{\umD}{\underline{\mD}}
\newcommand{\umQ}{\underline{\mQ}}
\newcommand{\umG}{\underline{\mG}}
\newcommand{\umS}{\underline{\mS}}
\newcommand{\RPPF}{\RP_{\text{PF}}}
\newcommand{\RPPFtran}{\transc{\RP}_{\text{PF}}}
\newcommand{\RPpPF}{\RP'_{\text{PF}}}
\newcommand{\RPpPFtran}{\transc{\left(\RP'_{\text{PF}}\right)}}

\newcommand{\llra}{\Longleftrightarrow} 
\newcommand{\layer}[1]{layer~$#1$}
\newcommand{\R}{R}
\newcommand{\dUB}{d_{\text{UB}}} \newcommand{\dLB}{d_{\text{LB}}}

\renewcommand{\S}{\mathcal{S}} 
\renewcommand{\P}{\mathcal{P}} 
\newcommand{\Psize}{\Abs{\P}} 
\renewcommand{\H}{\mPi} 
\newcommand{\Path}{\Pcal}
\newcommand{\G}{\Gcal}

\newcommand{\sr}[1]{\tilde{#1}}
\newcommand{\prl}[1]{\hat{#1}}
\renewcommand{\prl}[1]{\H}
\newcommand{\Tmin}{T_{\min}}

\newcommand{\Gal}{\text{Gal}}
\renewcommand{\ie}{i.e.}
\renewcommand{\eg}{e.g.}

\newcommand{\nt}{n_{\text{t}}}
\newcommand{\nr}{n_{\text{r}}}

\newcommand{\ntk}{n_{\text{t},k}}
\newcommand{\nrk}{n_{\text{r},k}}
\newcommand{\nti}{n_{\text{t},i}}
\newcommand{\ntj}{n_{\text{t},j}}

\newcommand{\dbar}{\bar{d}}
\renewcommand{\th}[1]{$#1$\,{th}}
\newcommand{\ntilde}{\tilde{n}}
\newcommand{\mntilde}{\mbs{\tilde{n}}}
\newcommand{\nav}{n_{\text{av}}}
\renewcommand{\r}{\textsf{r}}
\newcommand{\mul}{\,}
\renewcommand\Prob[1]{\textsf{P}\left\{#1\right\}}
\renewcommand\prob{\textsf{P}}
\newcommand{\XcalF}{\Xcal_{\text{full}}}
\newcommand{\XF}{\mX_{\text{full}}}
\newcommand{\nsum}{n_{\text{sum}}}

\begin{abstract} 
  
  We consider \emph{slow} fading relay channels with a single
  multi-antenna source-destination terminal pair. The source signal
  arrives at the destination via $N$ hops through $N-1$ layers of
  relays. We analyze the diversity of such channels with \emph{fixed}
  network size at \emph{high SNR}. In the clustered case where the
  relays within the same layer can have full cooperation, the
  cooperative decode-and-forward~(DF) scheme is shown to be optimal in
  terms of the diversity-multiplexing tradeoff~(DMT). The upper bound
  on the DMT, the cut-set bound, is attained. In the non-clustered
  case, we show that the naive amplify-and-forward~(AF) scheme has the
  maximum multiplexing gain of the channel but is suboptimal in
  diversity, as compared to the cut-set bound. To improve the
  diversity, space-time relay processing is introduced through the
  parallel partition of the multihop channel. The idea is to let the
  source signal go through $K$ different ``AF paths'' in the multihop
  channel. This \emph{parallel AF scheme} creates a parallel channel
  in the time domain and has the maximum diversity if the partition is
  properly designed.  Since this scheme does not achieve the maximum
  multiplexing gain in general, we propose a
  \emph{flip-and-forward}~(FF) scheme that is built from the parallel
  AF scheme. It is shown that the FF scheme achieves both the maximum
  diversity and multiplexing gains in a distributed multihop channel
  of arbitrary size.  In order to realize the DMT promised by the
  relaying strategies, approximately universal coding schemes are also
  proposed.
\end{abstract}


\IEEEpeerreviewmaketitle

\section{Introduction}
\label{sec:intro}

Recent years have seen a surge of interest in wireless networks.
Unlike the traditional point-to-point communication, elementary modes
of cooperation such as relaying are needed to improve both the
throughput and reliability in a wireless network. Although capacity of
a relay channel~\cite{vanderMeulen, Cover_relay} is still unknown in
general, considerable progress has been made on several aspects,
including some achievable capacity
results~\cite{Kramer_relay,Wang_relay} and capacity scaling laws of
large
networks~\cite{Gupta_Kumar,Xie_Kumar,Gastpar_Vetterli,Morgen_Bolcskei,Ayfer_Adhoc}.
In parallel, research on the cooperative
diversity~\cite{Sendonaris1,Sendonaris2}, where the relays help the
source exploit the spatial diversity of a slow fading channel in a
distributed fashion, has attracted significant
attention~\cite{LTW1,LTW2,Hunter2,Nabar,ElGamal_coop,SY_JCB_SAF,Elza_coop}.


In small relay networks where the source signal can reach the
destination terminal via a direct link, many results have been known
in both the channel capacity~\cite{Cover_relay,Kramer_relay} and the
cooperative diversity. The capacity results are mostly based on the
decode-and-forward~(DF) and the compress-and-forward~(CF) strategies.
The amplify-and-forward~(AF) scheme, however, is rarely considered in
this scenario due to the noise accumulation at the relays.  On the
other hand, the AF scheme is widely used for cooperative diversity. It
has been shown in \cite{LTW2,Nabar} that the AF scheme is as good as
the DF scheme at high SNR as far as the diversity is concerned.
Furthermore, it is pointed out in \cite{SY_JCB_SAF} that not needing
to decode the source signal makes the relays more capable of
protecting the source signal in some cases. The CF scheme, which works
with perfect global channel state information~(CSI), is usually
excluded in the cooperative diversity scenario for practical
considerations. In larger relay networks, where direct
source-destination links are generally absent, substantial results on
the capacity scaling laws have been obtained in the large network size
regime~\cite{Gupta_Kumar,Xie_Kumar,Gastpar_Vetterli,Ayfer_Adhoc} .
However, much less is known about the cooperative diversity than in
the case of small networks.

This paper analyzes the cooperative diversity in relay networks with a
single multi-antenna source-destination terminal pair.  The source
signal arrives at the destination via a sequence of $N$ hops through
$N-1$ layers of relays. Similar channel setting with a single layer
has been studied in \cite{Bolcskei_relay,Jing,Jing2} in different
contexts. Using large random matrix theory, the ergodic capacity
results of some particular relaying schemes have been established for
large networks~\cite{Bolcskei_relay}. Recently, the study has been
extended to the case with multiple layers of relays~\cite{Yeh_Leveque}
and the case with multiple source-destination
pairs~\cite{Morgen_Bolcskei}. Cooperative diversity in this setting
was first studied in \cite{Jing} for the single-antenna case then in
\cite{Jing2} for the multi-antenna case, with distributed space-time
coding. All the mentioned works assume linear processing at the relays
and the DF scheme is not considered.  Actually, one can figure out
immediately that the DF scheme is not suitable for the multi-antenna
setting due to the suboptimality in terms of degrees of freedom.
Requiring the relays to decode the source signal restricts the
achievable degrees of freedom. This is one of the fundamental
differences between the large networks and small networks~: the
degrees of freedom of the latter are determined by the
source-destination link and not by the relaying strategy.

In this work, we suppose that the network size is arbitrary~(but
fixed) and the signal-to-noise ratio~(SNR) is large. The multihop
channel is investigated in terms of the diversity-multiplexing
tradeoff~(DMT). The DMT was introduced in~\cite{Zheng_Tse} for the
point-to-point multi-antenna~(MIMO) channels to capture the
fundamental tradeoff between the throughput and reliability in a slow
fading channel at high SNR. It was then extensively used in multiuser
channels such as the multiple access channels~\cite{Tse_DMT_MAC} and
the relay channels~\cite{LTW1,LTW2,ElGamal_coop,SY_JCB_SAF,Elza_coop}
as performance measure and design criterion of different schemes.
Our main contributions are summarized in the following paragraphs.

First, we use the information theoretic cut-set bound~\cite{Cover3} to
derive an upper bound on the DMT of any relaying strategy. In the
clustered case where the relays in the same layer can fully cooperate,
this bound is shown to be tight. An optimal scheme is the cooperative
DF scheme, where the clustered relays perform joint decoding and joint
re-encoding.

While the clustered channel is equivalent to a series-channel and does
not feature the distributed nature of wireless networks, the
non-clustered case is studied as the main focus of the paper. Since no
within-layer cooperation is considered, linear processing at the
relays is assumed. We start by the AF strategy, which seems to be the
natural first choice as a linear relaying scheme. We show that the AF
scheme is, in the DMT sense, equivalent to the {Rayleigh product}~(RP)
channel, a point-to-point channel whose channel matrix is defined by a
product of $N$ Gaussian matrices. That being said, we examine the RP
channel in great detail. It turns out that the DMT of a RP channel has
a nice recursive structure and lends some intuitive insights into the
typical outage events in such channels. The study of the RP channel
leads directly to an exact DMT characterization for the AF scheme in
multihop channels of arbitrary size. The closed-form DMT provides
simple guidelines on how to efficiently use the available relays with
the AF scheme. One such example is how to reduce the number of relays
while keeping the same diversity. While the maximum multiplexing gain
is achieved, the achievable diversity gain of the AF scheme can be far
from maximum diversity gain suggested by the cut-set bound.
Specifically, the DMT of the AF scheme is limited by a virtual
``bottleneck'' channel.

The following question is then raised~: is the DMT cut-set bound tight
in the non-clustered case? The question is partially answered in this
work~: there exists a scheme that achieves both extremes of the
cut-set bound, that is, the maximum diversity extreme and maximum
multiplexing extreme. In order to achieve the maximum diversity gain,
the key is space-time relay processing. Noting that the AF scheme is
space-only, we incorporate the temporal processing into the AF scheme.
The first scheme that we propose is the \emph{parallel AF} scheme. By
partitioning the multihop channel into $K$ ``AF paths'', we create a
set of $K$ parallel sub-channels in the time domain. A packet that
goes through the parallel channel attains an improved diversity if the
partition is properly designed. It is shown that there is at least one
partition such that the maximum diversity is achieved.  However, the
parallel AF scheme does not have the maximum multiplexing gain in
general, since the achievable degrees of freedom by the scheme are
restricted by those of the individual AF paths. In most cases, the AF
paths are not as ``wide'' as the original channel in terms of the
degrees of freedom. In order to overcome the loss of degrees of
freedom, we linearly transform the set of parallel AF channels into
another set in which each sub-channel has the same degrees of freedom
as the multihop channel.  In the new parallel channel, each relay only
need to flip the received signal in a pre-assigned mode, hence the
name \emph{flip-and-forward}~(FF). It is shown that the FF scheme
achieves both the maximum diversity and multiplexing gains.
Furthermore, the DMT of the FF scheme is lower-bounded by that of the
AF scheme.

Using the results obtained in the non-clustered case, we revisit the
clustered case by pointing out that the cooperative DF operation might
not be needed in all clusters to get the maximum diversity. We also
indicate that cross-antenna linear processing in each cluster helps to
improve the DMT only when both transmitter CSI and receiver CSI are
known to the relays.

Finally, coding schemes are proposed for all the studied relaying
strategies. In the clustered case, a series of Perfect space-time
block codes~(STBCs)~\cite{Oggier_perfect,Elia} with appropriate rates
and dimensions are used at the source and each relay cluster that
performs the cooperative DF operation. In the non-clustered case,
construction of Perfect STBCs for general parallel MIMO channels is
first provided.  The constructed codes can be applied directly to the
parallel AF scheme and the FF scheme. All suggested coding schemes
achieve the DMT despite of the fading statistics and are thus
approximately universal~\cite{Tavildar}.


Regarding the notations, we use boldface lower case letters $\mbs{v}$
to denote vectors, boldface capital letters $\mbs{M}$ to denote
matrices. $\Ccal\Ncal(\mu,\sigma^2)$ represents a complex Gaussian
random variable with mean $\mu$ and variance $\sigma^2$. $\EE[\cdot]$
stands for the expectation operator.
$\trans{[\cdot]},\transc{[\cdot]}$ respectively denote the matrix
transposition and conjugated transposition operations.  $\Norm{\cdot}$
is the vector norm.  $\Norm{\cdot}_\text{F}$ is the Frobenius matrix
norm. We define $\prod_{i=1}^N \mM_i \defeq \mM_N \cdots \mM_1$ for
any matrices $\mM_i$'s. The square root $\sqrt{\mP}$ of a positive
semi-definite matrix $\mP$ is defined as a positive semi-definite
matrix such that $\mP=\sqrt{\mP}\transc{\bigl(\sqrt{\mP}\bigr)}$.
$\lambda_{\max}(\mP)$ and $\lambda_{\min}(\mP)$ denote respectively
the maximum and minimum eigenvalues of a semi-definite matrix $\mP$.
$\pstv{(x)}$ means $\max(0,x)$.  $\left\lceil a
\right\rceil$~(respectively, $\left\lfloor a \right\rfloor$) is the
closest integer that is not smaller~(respectively, not larger) than
$a$. $(a)_b$ means $\mod{a}{b}$. $\log(\cdot)$ stands for the base-$2$
logarithm. For any quantity $q$,
\begin{equation*}
  q \asympteq \SNR^{a}\quad \textrm{means}\quad \lim_{\SNR\to\infty}\frac{\log q}{\log \SNR} = a
\end{equation*}
and similarly for $\asymptleq$ and $\asymptgeq$. The tilde notation
$\mntilde$ is used to denote the (increasingly)~ordered version of
$\mn$. Let $\mm$ and $\mn$ be two vectors of respective length
$L_{\mm}$ and $L_{\mn}$, then $\mm\preceq\mn$ means
$\tilde{m}_i\leq\tilde{n}_i$,
$\forall\,i=1,\ldots,\min\{L_{\mm},L_{\mn}\}-1$. $\mm\subseteq \mn$
means that $\mm$ is a sub-vector of some permutated version of $\mn$.

The rest of this paper is organized as follows.
Section~\ref{sec:system-model} describes the system model and some
basic assumptions in our work. The DMT cut-set bound and the clustered
case with the DF scheme are presented. In Section~\ref{sec:AF}, we
study the non-clustered case with the AF scheme. The parallel AF and
the FF schemes are proposed in Section~\ref{sec:para-partition}. In
section~\ref{sec:clustered}, the clustered case is revisited. The
approximately universal coding schemes are proposed in
Section~\ref{sec:stc}. Section~\ref{sec:nr} provides some selected
numerical examples. Finally, a brief conclusion is drawn in
Section~\ref{sec:conclusion}.  Most detailed proofs are deferred to
the appendices.

\section{System Model and Basic Assumptions}
\label{sec:system-model}

\subsection{Channel Model}
\label{sec:channel-model}

The considered $N$-hop relay channel model is composed of one
source~(\layer{0}), one destination~(\layer{N}), and $N-1$ layers of
relays~(\layer{1} to \layer{N-1}). Each terminal is equipped with
multiple antennas. The total number of antennas in \layer{i} is
denoted by $n_i$. For convenience, we define $\nt\defeq n_0$,
$\nr\defeq n_N$, and $\nmin=\min_{i=0,\ldots,N} n_i$. We assume that
the source signal arrives at the destination via a sequence of $N$
hops through the $N-1$ layers and that terminals in \layer{i} can only
receive the signal from \layer{i-1}. The fading sub-channel between
\layer{i-1} and \layer{i} is denoted by the matrix $\mH_i$.
Sub-channels are assumed to be mutually independent, flat
Rayleigh-fading and quasi-static. That is, the channel coefficients
are independent and identically distributed~(i.i.d.) complex circular
symmetric Gaussian with unit variance. And they remain constant during
a coherence interval of length $L$ and change independently from one
coherence interval to another. Furthermore, the transmission is
supposed to be perfectly synchronized. Under these assumptions, the
signal model within a coherence interval can be written as
\begin{equation*}
  \my_{i}[l] = \mH_{i}\mul \mx_{i-1}[l] + \mz_{i}[l],\quad l=1,\ldots L,
\end{equation*}%
where $\mx_i[l],\my_i[l]\in\CC^{n_i\times 1}$ denote the transmitted
and received signal at \layer{i}; $\mz_i[l]\in\CC^{n_i\times 1}$ is
the additive white Gaussian noise~(AWGN) at \layer{i} with i.i.d.
$\CN[1]$ entries. Since we consider the non-ergodic case where the
coherence time interval $L$ is large enough, we drop the time index
$l$ hereafter. It is assumed that all relays work in
full-duplex\footnote{This assumption is merely for simplicity of
  notation. Since we assume that no cross-talk exists between
  different channels, the half-duplex constraint is directly
  translated to a reduction of degrees of freedom by a factor of two
  and does not impact the relaying strategy. This is achieved by
  letting all even-numbered~(respectively, odd-numbered) nodes
  transmit~(respective, receive) in even-numbered time slot and
  received~(respective, transmit) in odd-numbered time slots.} mode
and the transmission is subject to the short-term power constraint
\begin{equation}
  \label{eq:power-constraint}
  \EE\left\{\Frob{\mx_i}\right\} \leq \SNR,\quad\forall\,i
\end{equation}%
with $\SNR$ being the average transmitted SNR per layer.  All
terminals are supposed to have perfect channel state information at
the receiver\footnote{As we will see, assuming no CSI at all at the
  relays will not change the results of our work.}  and no CSI at the
transmitter. From now on, we denote the channel as a
$(n_0,n_1,\ldots,n_N)$ multihop channel.

\subsection{Diversity-Multiplexing Tradeoff}

Slow fading channels are outage-limited, \ie, there is an \emph{outage
  probability} $\Pout(\SNR,R)$ that the channel cannot support a
target data rate of $R$ bits per channel use at signal-to-noise ratio
$\SNR$. In the high SNR regime, this fundamental interplay between
throughput and reliability is characterized by the
diversity-multiplexing tradeoff~\cite{Zheng_Tse}.
\begin{definition}
  The \emph{multiplexing gain} $r$ and \emph{diversity gain} $d$ of a
  fading channel are defined by
\begin{equation*}
  r \defeq \lim_{{\sss\textsf{SNR}}\to\infty} \frac{R(\SNR)}{\log\SNR} \quad
\textrm{and}\quad 
  d \defeq - \lim_{{\sss\textsf{SNR}}\to\infty} \frac{\log\Pout(\SNR,R)}{\log\SNR}.
\end{equation*}
A more compact form is
\begin{equation}
  \label{eq:dmt-compact}
  \Pout(\SNR,r\log\SNR) \asympteq \SNR^{-d(r)}.
\end{equation}%
\end{definition}
Note that in the definition we use the outage probability instead of
the error probability, since it is shown in \cite{Zheng_Tse} that the
error probability of any particular coding scheme with maximum
likelihood~(ML) decoding is dominated by the outage probability at
high SNR and that the thus defined DMT is the best that one can
achieve with any coding scheme. In the Rayleigh MIMO channel, the DMT
has the following closed form.
\begin{lemma}[\cite{Zheng_Tse}]\label{lemma:dmt-Rayleigh}
  The DMT of a $\nt \times \nr$ Rayleigh channel is a
  piecewise-linear function connecting the points~$(k,d(k))$,
  $k=0,1,\ldots,\min{(\nt,\nr)}$, where
  \begin{equation}\label{eq:dmt-MIMO}
    d(k) = (\nt-k)(\nr-k).
  \end{equation}%
\end{lemma}
In the following, we will use the DMT as our performance measure. For
convenience of presentation, we provide the following definition.
\begin{definition}
  Two channels are said to be \emph{DMT-equivalent} or
  \emph{equivalent} if they have the same DMT.
\end{definition}

\subsection{Upper Bound on the DMT}
\label{sec:ub}

Before studying any specific relaying strategy, we establish an upper
bound on the DMT of the multihop system as a benchmark. 
\begin{proposition}[Cut-set bound]\label{prop:dmt-ub}
  For any relaying strategy $\Tcal$, we have
  \begin{equation*}
    d^{\Tcal}(r) \leq \dbar(r) 
  \end{equation*}
  with
  \begin{equation}
    \label{eq:dmt-ub}
    \dbar(r) \defeq \min_{i=1,\ldots,N} d_i(r),
  \end{equation}%
  where $d_i(r)$ is the DMT of the point-to-point channel between
  \layer{i-1} and \layer{i}. In particular, by defining the maximum
  diversity gain and multiplexing gain as $\dmax\defeq \dbar(0)$ and
  $\rmax\defeq \sup \{\dbar(r)>0\}$, respectively, we have
  \begin{align}
    \dmax &= \min_{i=1,\ldots,N} n_{i-1} n_i,\ \text{and}  \label{eq:dmax}\\
    \rmax &= \min_{i=0,\ldots,N} n_i \label{eq:rmax}.
  \end{align}%
\end{proposition}
\begin{IEEEproof}
  From the information theoretic cut-set bound~\cite{Cover3}, the
  mutual information between the source and the destination satisfies
  \begin{equation*}
    I_\Tcal(\mx_0;\my_N | \mH_1,\ldots,\mH_N) \leq I(\mx_{i-1};\my_i | \mH_i),\quad \forall\,i,
  \end{equation*}
  for any relaying strategy $\Tcal$. Thus, the outage probability
  using a relaying scheme $\Tcal$ is
  \begin{align}
    \Pout^{\Tcal}(R) &\defeq \prob_{\{\mH_i\}_i} \left\{I_\Tcal(\mx_0;\my_N  | \mH_1,\ldots,\mH_N )<R \right\} \nonumber\\
    &\geq \max_i \prob_{\mH_i} \left\{I(\mx_{i-1};\my_i | \mH_i )<R \right\} \nonumber \\
    &= \max_i P_{\text{out},i}(R), \label{eq:tmp999}
  \end{align}%
  where $P_{\text{out},i}(R)$ is the outage probability of the \th{i}
  sub-channel. From \Eq{eq:dmt-compact} and \Eq{eq:tmp999}, we prove
  \Eq{eq:dmt-ub}. Finally, \Eq{eq:dmax} and \Eq{eq:rmax} are from the
  direct application of Lemma~\ref{lemma:dmt-Rayleigh}.
\end{IEEEproof}

\subsection{The Clustered Case and Decode-and-Forward}
\label{sec:CCnDF}

If we assume that the relays within the same layer are clustered, \ie,
they can perform joint decoding and joint re-coding operations, then
each layer can act as a virtual multi-antenna terminal. This could
happen either when the relays are controlled by a central unit via
wired links or when they are close enough to each other to exchange
information perfectly. In this case, the relay channel model is
equivalent to a serial concatenation of $N$ independent MIMO channels.
Let us consider the following cooperative decode-and-forward scheme.
Each layer tries to cooperatively decode the received signal. When a
successful decoding is assumed, the embedded message is re-encoded and
then forwarded to the next layer. We can show that this simple scheme
is DMT optimal.
\begin{proposition} \label{prop:DF}
  When the relays are clustered, the cooperative DF scheme achieves
  the DMT cut-set bound $\dbar(r)$ defined in \Eq{eq:dmt-ub}.
\end{proposition}
\begin{IEEEproof}
  To show the achievability, note that the cooperative DF scheme being
  in outage implies the outage of at least one of the sub-channels. By
  the union bound,
  \begin{equation*} 
    \Pout^{\text{DF}}(R) \leq  \sum_{i=1}^N P_{\text{out},i}(R). 
  \end{equation*}%
  At high SNR, the probability is dominated by the largest term in the
  sum of the right-hand side~(RHS). From \Eq{eq:dmt-compact}, we get
  \begin{equation*}
    \dDF(r) \geq \min_{i=1,\ldots,N} d_i(r) = \dbar(r).
  \end{equation*}%
\end{IEEEproof}
In the high SNR regime, the union bound defined by the sum operation
coincides in the SNR exponent with the cut-set bound defined by the
minimum operation. Hence, the DMT cut-set bound is tight in the
clustered case. However, relays in wireless networks are not clustered
in general. In fact, one of the important and interesting attributes
of wireless networks is the distributed nature. In the following two
sections, we will concentrate on the non-clustered case and analyze
the achievable DMT.

\section{Amplify-and-Forward}
\label{sec:AF}

In this section, we consider the non-clustered case, where the relays
work in a distributed manner and no within-layer communication is
allowed. In this case, applying the DF scheme at each individual relay
might incur loss of degrees of freedom. To see this, take the
single-layer channel as an example. In the best case where all the
relays succeed in decoding, they transmit the message using a
pre-assigned codebook.  This scheme transforms the relays-destination
channel into a $n_1\times n_2$ virtual MIMO channel. Before this could
possibly happen, however, the success decoding at the relays must be
guaranteed with high probability. This constraint imposes that the
degrees of freedom in this scheme must not be larger than
$\min_k\{n_{1,k}\}$ with $n_{1,k}$ being the number of antennas at the
\th{k} relay. While this scheme achieves the maximum multiplexing gain
in the single-antenna case, it could fail in the multi-antenna case.

Since we do not know how to cooperate efficiently in this case, we
start by the most obvious and naivest relaying scheme~: the
amplify-and-forward scheme. This scheme in the considered setting has
been studied in \cite{Bolcskei_relay, Yeh_Leveque} for the capacity
scaling laws, and in \cite{Borade_Zheng_Gallager} for the DMT. It is
worth noting\footnote{The authors found \cite{Borade_Zheng_Gallager}
  at the very end of the preparation for this manuscript.} that, in
\cite{Borade_Zheng_Gallager}, a lower bound on the DMT of the AF
scheme in a symmetric network~($n_i=n$, $\forall\,i$) was obtained,
while our work derives the exact DMT for a network of arbitrary
dimension with a different approach.

\subsection{Signal Model}

In the considered AF scheme, each antenna node normalizes the received
signal to the same power level and then retransmits it. This linear
operation can be expressed as
\begin{align*}
  \mx_{i} &= \mD_{i}\mul \my_{i}, \quad i=1,\ldots,N-1,
\end{align*}%
where, by the power constraint \Eq{eq:power-constraint},
$$\EE\left(\Abs{\mx_i(j)}^2\right) \leq \frac{\SNR}{n_i},\quad
j=1,\ldots,n_i;$$
the scaling matrix $\mD_i\in\CC^{n_i\times n_i}$ is
diagonal due to the antenna-wise nature of the relaying scheme, with
the normalization factors\footnote{In the case where long-term power
  constraint is imposed, we simply replace the channel coefficients
  $\Abs{\mH_i(j,k)}$ in \Eq{eq:mD} by $1$'s.}
\begin{equation}
  \label{eq:mD}
\mD_i(j,j) =
\sqrt{\frac{1}{\frac{\SNR}{n_{i-1}}\left(\sum_{k=1}^{n_{i-1}}\Abs{\mH_i(j,k)}^2\right)+1}}\cdot
\sqrt{\frac{\SNR}{n_{i}}}.  
\end{equation}%
Thus, the signal model of the end-to-end channel is
\begin{equation}\label{eq:multihop_AF1}
  \my_N = \left(\prod_{i=1}^N \mD_i \mH_i\right) \mx_0 + \sum_{j=1}^N \left( \prod_{i=j}^N \mH_{i+1}\mD_i \right) \mz_j,
\end{equation}%
where, for the sake of simplicity, we defined $\mH_{N+1}\defeq\Id$ and
$\mD_N\defeq \Id$. The whitened form of this channel is
\begin{equation*}
  \my = \sqrt{\mR} \left(\prod_{i=1}^N \mD_i \mH_i\right) \mx_0 + \mz,
\end{equation*}%
where $\mz$ is the whitened noise and $\sqrt{\mR}$ is the whitening
matrix with $\inv{\mR}$ being the covariance matrix of the noise in
\Eq{eq:multihop_AF1}. Since it can be shown that
$\lambda_{\max}(\mR)\asympteq \lambda_{\min}(\mR) \asympteq \SNR^0$,
$\mR$ can be neglected in the DMT analysis and the AF
channel\footnote{Here, with a slight abuse of terminology, we call the
  multihop channel with AF scheme an \emph{AF channel}.} is equivalent
to the MIMO channel defined by the following matrix
\begin{equation}
  \label{eq:AFeqchannel}
  \mH_N\mD_{N-1}\cdots\mH_2\mD_1\mH_1.
\end{equation}%
The rest of the section is devoted to the DMT analysis of this
channel.

\subsection{The Rayleigh Product Channel}
\label{sec:rp}

\begin{definition}\label{def:RP}
  Let $\mH_i\in\CC^{n_{i-1}\times n_i}$, $i=1,2,\ldots,N$, be $N$
  independent complex Gaussian matrices with i.i.d. $\CN[1]$ entries.
  A $(n_0,n_1,\ldots,n_N)$ \emph{Rayleigh product}~(RP) channel is a
  $n_N \times n_0$ MIMO channel defined by
  \begin{equation}
    \label{eq:channel-model}
    \my = \sqrt{\frac{\SNR}{n_1\cdots n_{N}}}\mul \RP\mul\mx + \mz,
  \end{equation}%
  where $\RP \defeq \mH_1\mH_2\cdots\mH_N$ is the channel matrix;
  $\mx\in\CC^{n_N\times1}$ is the transmitted signal with normalized
  power, \ie, $\EE\{\Norm{\mx}^2\}=n_N$; and $\my\in\CC^{n_0\times 1}$
  is the received signal; $\mz\in\CC^{n_0\times1}\sim\CN$ is the AWGN;
  $\SNR$ is the SNR per receive antenna. $(n_0,n_1,\ldots,n_N)$ is
  called the \emph{dimension} of the channel and $N$ is called the
  \emph{length} of the channel.
\end{definition}%
While this channel model has been studied in terms of the asymptotic
eigenvalue distribution in the large dimension
regime~\cite{Muller_productmatrix}, we are particularly interested in
the fixed dimension case in the high SNR regime. In this regime, we
can define a more general RP channel as
\begin{equation}
  \label{eq:general-RP}
  \RP_g \defeq \mH_1 \mT_{1,2} \mH_2\cdots \mH_{N-1} \mT_{N-1,N}
\mH_N.  
\end{equation}%
\begin{proposition}\label{prop:pdf-general-RP}
  The general RP channel is equivalent to
  \begin{itemize}
  \item a $(n_0,n_1,\ldots,n_N)$ RP channel, if all the matrices
    $\mT_{i,i+1}$'s are square and their singular values satisfy
    $\sigma_j(\mT_{i,i+1})\asympteq \SNR^0$, $\forall i,j$;
  \item a $(n_0,n'_1,\ldots,n'_{N-1}, n_N)$ RP channel, with $n'_i$
    being the rank of the matrix $\mT_{i,i+1}$, if the matrices
    $\mT_{i,i+1}$'s are constant.
  \end{itemize}
\end{proposition}
\begin{IEEEproof}
  See Appendix~\ref{sec:proof-pdf-general-RP}.
\end{IEEEproof}
Hence, we can consider the RP channel from Definition~\ref{def:RP}
without loss of generality.

\subsubsection{Direct Characterization}

Recall that $\mnt$ is the ordered version of $\mn$ with $\ntilde_N\geq
\ntilde_{N-1}\geq\cdots\geq \ntilde_0$ and $\nmin\defeq \ntilde_0$.
\begin{theorem}\label{thm:dmt-rp}
  The DMT of a RP channel~$(n_0,n_1,\ldots,n_N)$ is a piecewise-linear
  function connecting the points~$(k,\dRP(k))$, $k=0,1,\ldots,\nmin$,
  where
  \begin{equation}
    \dRP(k) = \sum_{i=k+1}^{\nmin} c_i\label{eq:dk}
  \end{equation}%
  with
  \begin{equation}\label{eq:ci}
    c_i \defeq 1-i + \min_{k=1,\ldots, N} \left\lfloor\frac{\sum_{l=0}^{k}\ntilde_l - i}{k} \right\rfloor,\quad i=1,\ldots,{\nmin}.
  \end{equation}%
\end{theorem}
\begin{IEEEproof}
  The DMT depends on the ``near zero'' probability of the singular
  values of channel matrix. While this probability for the given
  product matrix is intractable, we can characterize it by induction
  on the length $N$. The main idea is that, conditioned on a given
  product matrix $\mH_1 \mH_2\cdots\mH_{N-1}$, $\mH_1
  \mH_2\cdots\mH_{N}$ is Gaussian whose singular distribution is
  tractable. See Appendix~\ref{sec:proof-thm-dmt-rp} for details.
\end{IEEEproof}
The following corollaries are given without proofs.
\begin{corollary}[Permutation invariance]\label{coro:per-inv}
  The DMT of a RP channel depends only on the ordered dimension
  $\mnt$.
\end{corollary}
\begin{corollary}[Monotonicity]\label{coro:increase}
The DMT is monotonic in the following senses~:
  \begin{itemize}
  \item if $\mn_1 \succeq \mn_2$, 
    then
    $$\dRP_{\mn_1}(r) \geq \dRP_{\mn_2}(r),\quad \forall\,r;$$
  \item if $\mn_1 \supseteq \mn_2$, then
    $$\dRP_{\mn_1}(r) \leq \dRP_{\mn_2}(r),\quad \forall\,r.$$
  \end{itemize}
\end{corollary}

\begin{corollary}[Symmetric Rayleigh product channels]\label{coro:sym}
  When $n_0=\ldots=n_N=n$, we have
  \begin{equation}\label{eq:dmt_sym}
    \dRP_n(k) = \frac{(n-k)(n+1-k)}{2} + \frac{a(k)}{2}((a(k)-1)N+2b(k)),
  \end{equation}%
  where $a(k) \defeq \left\lfloor \frac{n-k}{N} \right\rfloor$ and
  $b(k) \defeq (n-k)_N$. 
\end{corollary}

\subsubsection{DMT Equivalent Classes}
Corollary~\ref{coro:per-inv} implies that RP channels with the same
ordered dimension belong to the same DMT equivalent class. In the
following, a precise characterization of the DMT class is obtained.
Before that, we need the following definitions.

\begin{definition} A $(m_0,m_1,\ldots,m_k)$
  RP channel is said to be a \emph{reduction} of a
  $(n_0,n_1,\ldots,n_N)$ RP channel if 1)~they are equivalent,
  2)~$k\leq N$, and 3)~$\mm \preceq \mn$. In particular, if $k=N$,
  then it is called a \emph{vertical reduction}.  Similarly, if
  $\tilde{m}_i=\ntilde_i,\ \forall\,i\in[0,k]$, it is a
  \emph{horizontal reduction}.
\end{definition}

\begin{definition}
  $(\ntilde_0,\ntilde_1,\ldots,\ntilde_{N^*})$ is said to be a
  \emph{minimal form} if no reduction other than itself exists.
  Similarly, it is called a \emph{minimal vertical
    form}~(respectively, \emph{minimal horizontal form}) if no
  vertical~(respectively, horizontal) reduction other than itself
  exists. A RP channel is said to have \emph{order} $N^*$ if its
  minimal form is of length $N^*+1$.
\end{definition}%

\begin{theorem}\label{thm:reduction}
  A $(n_0,n_1,\ldots,n_N)$ RP channel can be reduced to a
  $(\ntilde_0,\ntilde_1,\ldots,\ntilde_k)$ channel if and only if
  \begin{equation}
    \label{eq:reduction-cond}
    k(\ntilde_{k+1} + 1) \geq \sum_{l=0}^k \ntilde_l.    
  \end{equation}%
  In particular, it can be reduced to a Rayleigh channel if and only if
  \begin{equation}
    \label{eq:reduction-cond-0}
    \ntilde_{2} + 1 \geq \ntilde_0+\ntilde_1.    
  \end{equation}%
\end{theorem}
\begin{IEEEproof}
  See Appendix~\ref{sec:proof-thm-reduction}.
\end{IEEEproof}
\begin{corollary}\label{coro:reduc}
  The channel order $N^*$ is the minimum integer such that
  \Eq{eq:reduction-cond} is satisfied. The minimal horizontal form is
  the minimal form $(\ntilde_0,\ntilde_1,\ldots,\ntilde_{N^*})$ and
  the minimal vertical form is
  $(\ntilde_0,\ntilde_1,\ldots,\ntilde_{N^*},\bar{n},\ldots,\bar{n})$
  with
\begin{equation}
  \label{eq:minimum_number}
  \bar{n} \defeq \left\lceil \frac{\sum_{l=0}^{N^*} \ntilde_i}{N^*}\right\rceil - 1.  
\end{equation}%
\end{corollary}
For instance, the minimal form of a $(1,n_1,\ldots,n_N)$ RP channel is
$(1,\ntilde_1)$, \ie, a $1 \times \ntilde_1$ or $\ntilde_1 \times 1$
Rayleigh channel.

\begin{theorem}\label{thm:minimal}
  The DMT equivalent class is \emph{uniquely} identified by the
  minimal form, \ie, two RP channels are equivalent \emph{if and only
    if} they have the same minimal form.
\end{theorem}
\begin{IEEEproof}
  See Appendix~\ref{sec:proof-thm-minimal}.
\end{IEEEproof}

\subsubsection{Recursive Characterization}\label{sec:recur}

In order to interpret the closed-form DMT of Theorem~\ref{thm:dmt-rp},
we derive an equivalent recursive form as shown in the following
theorem.
\begin{figure*}[!t]
\begin{center}
  \subfigure[Interpretation of
  $\R_1^{(N)}(k)$]{\includegraphics[angle=0,height=0.3\textwidth]{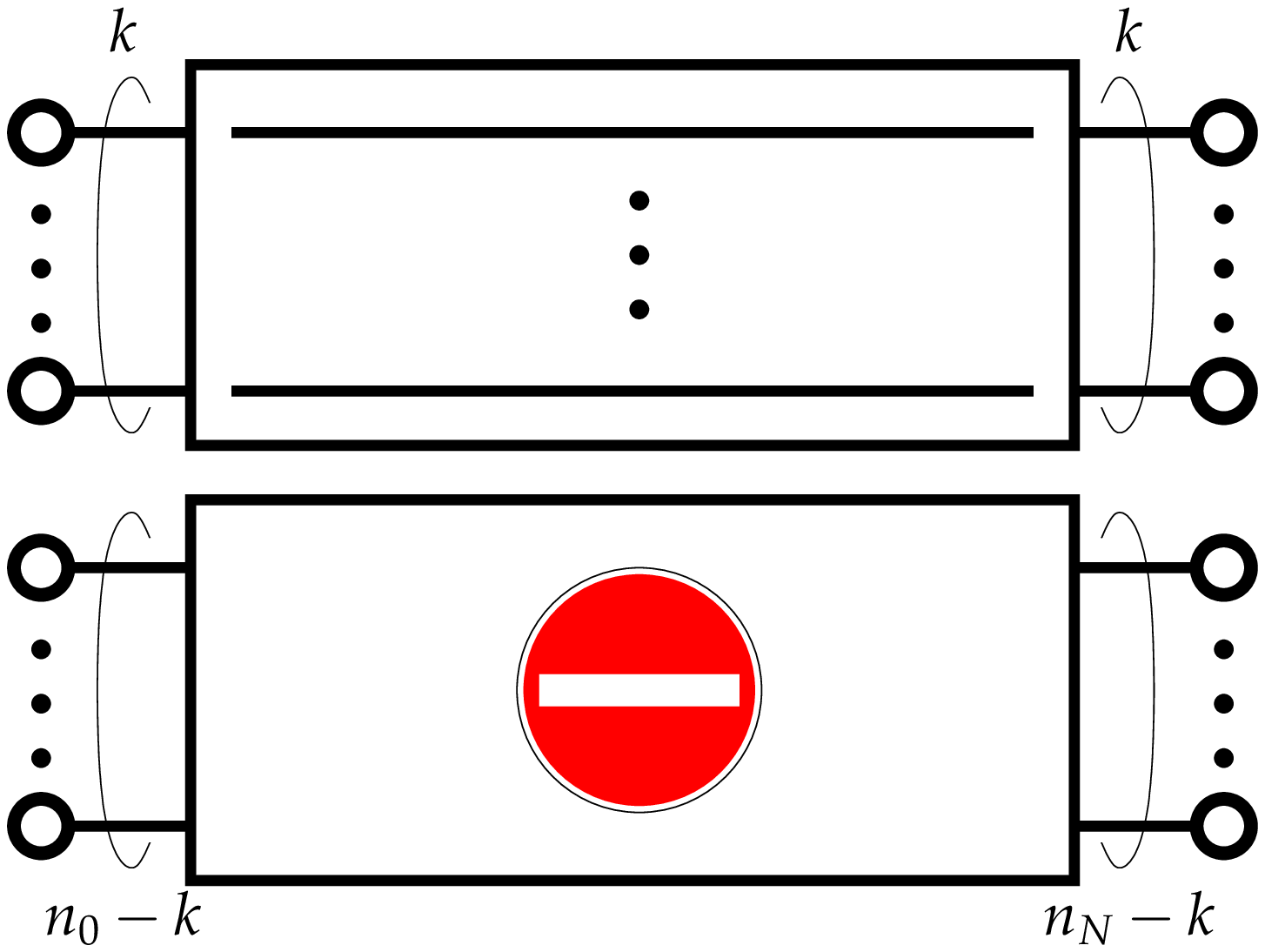}\label{fig:interp1}}
  \hspace{2cm} \subfigure[Interpretation of
  $\R_2^{(N)}(i)$]{\includegraphics[angle=0,height=0.3\textwidth]{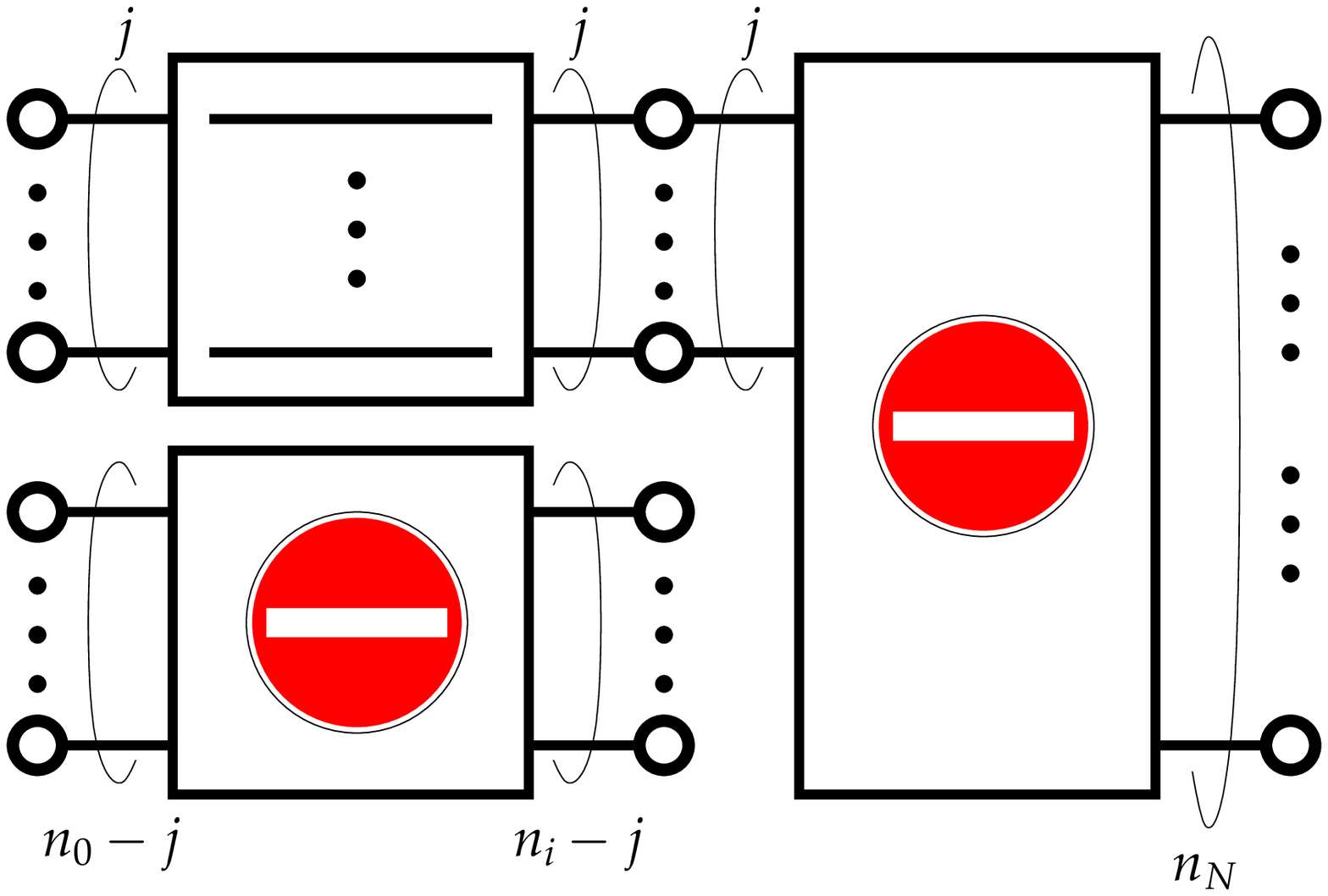}\label{fig:interp2}}
\caption{Interpretations of the DMT of the RP channel.}
\label{fig:interp}  
\end{center}
\end{figure*}

\begin{theorem}\label{thm:recursive}
  The DMT $\dRP(k)$ defined in \Eq{eq:dk} can be alternatively
  characterized by
  \begin{align} 
    \R_1^{(N)}(k)~:\quad \dRP_{(n_0,\ldots,n_N)}(k) &= \dRP_{(n_0-k,\ldots,n_N-k)}(0),\quad\forall k;\label{eq:interp1} \\
    \R_2^{(N)}(i)~:\quad \dRP_{(n_0,\ldots,n_N)}(0) &= \min_{j\geq 0} \left\{
    \dRP_{(n_0,\ldots,n_i)}(j) +
    \dRP_{(j,n_{i+1},\ldots,n_N)}(0)\right\},\quad\forall i;
    \label{eq:interp2} \\
    \R_3^{(N)}(i,k)~:\quad \dRP_{(n_0,\ldots,n_N)}(k) &= \min_{j\geq
      k} \left\{ \dRP_{(n_0,\ldots,n_i)}(j) +
      \dRP_{(j,n_{i+1},\ldots,n_N)}(k) \right\},\quad\forall i,k.
    \label{eq:interp3} 
  \end{align}%
\end{theorem}%
\begin{IEEEproof}
  See Appendix~\ref{sec:proof-thm-recursive}.
\end{IEEEproof}
A new interpretation of the DMT is as follows.  \emph{Let us consider
  a multi-layer network of dimension $(n_0,n_1,\ldots,n_N)$. Then,
  $\dRP(k)$ is the {minimum ``cost''} to limit the ``{network flow}''
  between the source and the destination to $k$~(the flow-$k$ event).
  In particular, the maximum diversity $\dRP(0)$ is the
  {``disconnection cost''}. } Now, we can apply the new interpretation
to the results of Theorem~\ref{thm:recursive}.  First, $\R_1(k)$ says
that the most efficient way to limit the flow to $k$ is to keep a
$(k,k,\ldots,k)$ channel fully connected and to disconnect the
$(n_0-k,n_1-k,\ldots,n_N-k)$ residual channel, as shown in
\Fig{fig:interp1}. Then, $\R_2(i)$ suggests that in order to
disconnect a $(n_0,n_1,\ldots,n_N)$ channel, if we allow for $j$ flows
from the source to some \layer{i}, then the $(j,n_{i+1},\ldots,n_N)$
channel from the $j$ ``ends'' of the flows at \layer{i} to the
destination must be disconnected~(\Fig{fig:interp2}). Obviously, the
most efficient way is such that the total cost is minimized with
respect to $j$.  Finally, the flow-$k$ event takes place when both the
flow-$j$~($j\geq k$) event in the $(n_0,\ldots,n_i)$ channel and the
flow-$k$ event in the $(j,n_{i+1},\ldots,n_k)$ channel happen at the
same time. We can easily verify that $(\R_1(k),\R_3(i,k))$ is
equivalent to $(\R_1(k),\R_2(i))$. Also note that $\R_2(i)$ and
$\R_3(i,k)$ hold for any layer $i$, which guarantees the coherence of
the interpretation.

The recursive characterization sheds lights on the typical outage
event of the RP channel. In the trivial case of $N=1$ (the Rayleigh
channel), the typical and only way for the channel to be in outage at
multiplexing gain $r$ approaching to zero is that all the
$\ntilde_0\times\ntilde_1$ paths are bad, \ie, all channel gains are
close to zero. And the disconnection cost is
$\ntilde_0\times\ntilde_1$. In the non-trivial cases~($N>1$) where
channels are concatenated, there are several types of outage event.
Each type is numbered by the index $j$ in \Eq{eq:interp2} and
\Eq{eq:interp3}. The cost of the type-$j$ event is given by
$\dRP_{(n_0,\ldots,n_i)}(j) + \dRP_{(j,n_{i+1},\ldots,n_N)}(0)$ for a
certain $j$. Hence, the typical outage event is the one with the
minimum cost and it does not necessarily happen when one of the
sub-channels being totally bad~($j=0$ or $j=n_i$). The \emph{mismatch}
of two partially bad sub-channels can also cause outage. This
phenomenon will be detailed later on.

\subsection{DMT of the AF Scheme}

From the equivalent channel matrix \Eq{eq:AFeqchannel} and
Proposition~\ref{prop:pdf-general-RP}, the AF channel is equivalent to
a $(n_N,n_{N-1},\ldots,n_0)$ RP channel.\footnote{Theoretically, this
  is true only when the singular values $\sigma_j(\mD_i)\asympteq
  \SNR^0$, $\forall i,j$. To this end, it is enough to modify the
  matrices as ${\mD}_i(j,j) = \min\left\{\mD_i(j,j), \kappa \right\}$
  where $0<\kappa<\infty$ is a constant independent of $\SNR$. Note
  that the $\kappa$ is only for theoretical proof and is not used in
  practice, since we can always set $\kappa$ a very large constant but
  independent of $\SNR$.}  Therefore, the DMT of the AF channel is
$$\dAF(r) = \dRP(r),\quad \forall\,r.$$

\subsubsection{Implications}
From the results of Section~\ref{sec:rp}, several interesting
implications on the AF scheme with respect to the DMT are summarized
below.
\begin{figure}[!t]
\begin{center}
  \includegraphics[angle=0,width=0.45\textwidth]{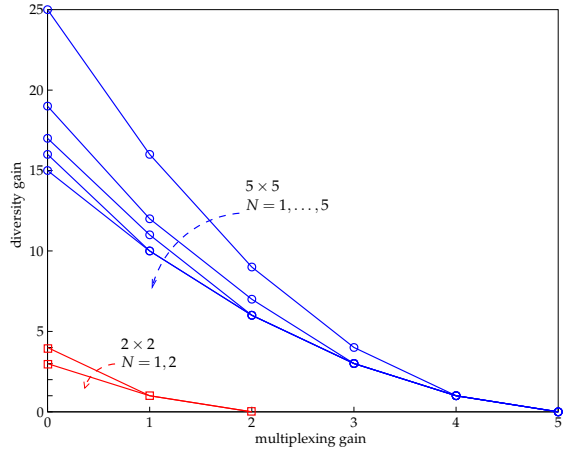}
\caption{Diversity-multiplexing tradeoff of $2\times2$ and $5\times5$ symmetric RP channels.}
\label{fig:RayleighProduct_n5}  
\end{center}
\end{figure}
\begin{itemize}
\item Interchanging layers does not influence the DMT.
\item The maximum diversity of the AF scheme is lower- and
  upper-bounded as
  \begin{equation}\label{eq:lbub}
    \frac{\ntilde_0 (\ntilde_1+1)}{2} 
    \leq 
    \dAF_{\max}
    \leq 
    \ntilde_0 \ntilde_1
  \end{equation}%
  which is obtained via the monotonicity from
  Corollary~\ref{coro:increase}. We have set $\ntilde_2\geq
  \ntilde_0+\ntilde_1-1$ for the upper bound and
  $\ntilde_{N}=\ntilde_{N-1}=\ldots=\ntilde_1$ for the lower bound.
  The upper bound shows that there exists a virtual
  $\ntilde_0\times\ntilde_1$ ``bottleneck'' channel that limits the AF
  scheme and that it is not necessarily one of the sub-channels. On
  the other hand, the lower bound is always strictly larger than half
  the upper bound and is independent of the number of hops $N$. In the
  symmetric case~(Corollary~\ref{coro:sym}), we observe that the DMT
  degrades with $N$ only when $N\leq n$ and that we have
  \begin{equation*}
    \dAF_{(n,\ldots,n)}(k) = \frac{(n-k)(n+1-k)}{2}
  \end{equation*}%
  for $N\geq n$. The observation can also be deduced from
  theorem~\ref{thm:reduction} applying which we infer that the order
  of any symmetric RP channel with $N > n$ is $N^*=n$. This
  non-trivial lower bound is somewhat anti-intuition, since it means
  that at this point introducing extra fading hops does not degrade
  the diversity any more. An example illustrating the DMT of the
  $2\times2$ and $5\times 5$ RP channels of different lengths is in
  \Fig{fig:RayleighProduct_n5}.
\item If one could increase the number of antennas at each relay layer
  without any constraint, then intuition tells us that the AF channel
  could be reduced to a $\nt\times \nr$ point-to-point Rayleigh MIMO
  channel and the diversity order is $\nt\mul\nr$. The relay layers
  ``disappear''. The intuition has been confirmed in
  \cite{Bolcskei_relay} in the single-layer case with the capacity
  results. Here, the result in Theorem~\ref{thm:reduction} indicates
  that this happens when there are exactly $\nt+\nr-1$ antennas at
  each relay layers from the diversity point of view. Further
  increasing the number of antennas is not necessary in the DMT sense.
  On the other hand, if the number of available antennas is fixed,
  then Corollary~\ref{coro:reduc} provides, through the minimal
  vertical form, the minimum numbers of antennas at each layer to
  achieve the diversity that could be achieved when all antennas were
  used. In both cases, our results yield simple guidelines to minimize
  the number of relay antennas~(also the number of relays in general)
  without loss of optimality of the DMT. In the same way, the number
  of transmit antennas at the source terminal can also be reduced to
  lower the coding complexity. A numerical example is given in
  Section~\ref{sec:nr}.
\end{itemize}

\subsubsection{Comparison to the Cut-Set Bound}
\label{sec:comparison}
A simple comparison between the DMT of the AF scheme and the cut-set
bound~\Eq{eq:dmt-ub} is carried out as follows. First, the AF scheme
is multiplexing optimal and achieves the maximum multiplexing gain
$\ntilde_0$ of the channel. Then, since $$(\ntilde_0-k)(\ntilde_1-k)
\leq \min_{i=1,\ldots,N} \{(n_{i-1}-k) (n_{i}-k)\},\quad\forall\,k,$$
the diversity upper bound is generally not achievable by the AF scheme
for integer multiplexing gain $k$. In particular, the best diversity
gain of the AF scheme is $\ntilde_0\mul\ntilde_1$, while the upper
bound is $\min_i\{n_i\mul n_{i+1}\}$. Finally, for any non-integer
multiplexing gain, say $r\in(k,k+1)$, $\dbar(r)$ is minimum of linear
functions and thus concave, while $\dAF(r)$ is linear. The comparison
shows that the \emph{bottleneck} of the channel is always one of the
hops~(inter-layer sub-channels), while the bottleneck of the AF scheme
is the virtual $\ntilde_0\times\ntilde_1$ channel that does not
correspond to any physical sub-channel in most cases. The following
remark states the necessary and sufficient condition for the AF scheme
to achieve the maximum diversity.
\begin{remark}\label{thm:cond_AF}
  The AF scheme achieves the diversity upper bound $\dmax$ if and only
  if it can be reduced to the bottleneck of the channel, \ie,
  \begin{equation}
    \label{eq:cond_AF}
    \min\{n_{i^*}, n_{i^*+1}\} = \ntilde_0,\quad \max\{n_{i^*}, n_{i^*+1}\} = \ntilde_1,\quad \text{and } \ntilde_2+1 \geq \ntilde_0+\ntilde_1,
  \end{equation}%
  where $i^*$ is such that $n_i\mul n_{i+1}$ is minimized.
\end{remark}
This condition is very stringent. It means that the two layers with
minimum numbers of antennas must stand one next to the other and that
the other layers must have a large number of antennas. Moreover, note
that the AF scheme achieving the maximum diversity does not
necessarily mean that it achieves $\dbar(r)$ for all $r$.

\subsubsection{Mismatch of Adjacent Sub-Channels}

\begin{figure*}
\begin{center}
  \subfigure[Canonical basis]{
\label{fig:ex_canon}
\epsfig{figure=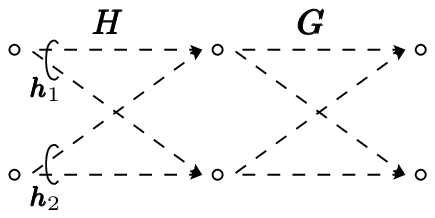,height=0.18\textwidth}}
\subfigure[Orthogonal basis~$\left\{\mh_1/\Norm{\mh_1},\mh_1^{\perp}/\|\mh_1^{\perp}\|\right\}$]{
\label{fig:ex_ortho}
\epsfig{figure=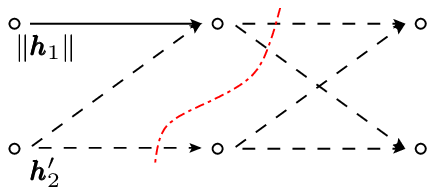,height=0.18\textwidth}}
\caption{The $(2,2,2)$ multihop channel in two different basis.}
\label{fig:ex}
\end{center}
\end{figure*}

In order to achieve the diversity upper bound, intuitively, one should
assure that the end-to-end channel is good if each sub-channel is
good.  However, this property does not hold for the AF scheme that
suffers from the \emph{mismatch} of adjacent sub-channels. A concrete
example is as follows.
\begin{example}
  In the symmetric two-hop channel with $n=2$~(Fig.~\ref{fig:ex}), the
  diversity order of the AF scheme is $3$ while the upper bound is
  $4$.
\end{example}
Note that the AF channel is in outage if the product channel $\mG\mH$
is bad, \ie, all the singular values of $\mG\mH$ are close to zero.
This probability can be decomposed as
\begin{align*}
  \Prob{\mG\mH \text{ is bad}} &= \Prob{\text{both } \mG\mH \text{
      and } \mH \text{ are bad}} 
   + \Prob{\mG\mH \text{ is bad, while } \mH \text{
      is not bad}}, 
\end{align*}
where we can verify that the first probability is essentially the
probability of the sub-channel $\mH$ being bad and that the second one
is essentially the probability of $\mG\mH$ being bad conditioned on
the event that $\mH$ is not bad. As we know, the first probability
decays with $\SNR$ as $\SNR^{-4}$. To find out the SNR exponent of the
second probability, we assume without loss of generality that the
vector $\mh_1$ is strong enough~(since $\mH$ is not bad), as shown in
\Fig{fig:ex_canon}. Now, we apply an orthogonal basis change from the
canonical basis to
$\left\{\mh_1/\Norm{\mh_1},\mh_1^{\perp}/\|\mh_1^{\perp}\|\right\}$
and get the equivalent channel in \Fig{fig:ex_ortho}. The basis change
being an unitary transformation that is independent of the remaining
parts of the channel, it does not affect the statistics of the rest of
the channel. As shown in \Fig{fig:ex_ortho}, the channel is bad if the
three independent edges crossed by the ``minimum cut'' are bad. The
probability for the latter to happen decays as $\SNR^{-3}$, from which
we conclude that the outage probability scales as
$\SNR^{-3}+\SNR^{-4}\asympteq\SNR^{-3}$.  Therefore, the mismatch
between $\mG$ and $\mH$ is the dominating outage event and the
end-to-end diversity of the $(2,2,2)$ channel with AF scheme is $3$,
as compared to $4$ given by the cut-set bound.

\section{Parallel Partition}
\label{sec:para-partition}

The naive AF scheme presented above can be seen as a space-only
processing. In the point-to-point MIMO channel, it has been shown that
space-only coding schemes~(\eg, the V-BLAST scheme~\cite{VBLAST}) are
suboptimal in diversity. Similarly, the AF scheme, as a space-only
relaying scheme, does not achieve the maximum diversity in the
multihop channel due to the mismatch between adjacent sub-channels.
The clue is, just like the space-time codes achieve the maximum
diversity in the point-to-point channel, space-time relay processing
should be utilized in order to exploit the maximum distributed
diversity in the multihop channel.

The first attempt was made in \cite{Jing} with a distributed
space-time coding scheme. In this scheme, each relay performs temporal
random unitary transformation on the received signal from the source
in an independent way. Then, they forward the transformed signal at
the same time as if they were jointly sending a space-time codeword.
The spatial correlation of the codewords is due to the fact that the
received signal at different relays is from the same source. The
temporal correlation, on the other hand, is brought in by the temporal
transformation. In their setting where a single layer of relays and
single-antenna terminals are assumed, the maximum diversity of the
channel is achieved. This scheme is then generalized to the
multi-antenna case~\cite{Jing2} with structured algebraic
transformations~\cite{Oggier_Hassibi} instead of random
transformations.  However, generalization of such schemes to the
multihop case is difficult and the DMT is hard, if not impossible, to
calculate.

In the following, we present a different approach to introduce the
temporal processing. This approach does not depend on the dimension of
the channel and thus suitable for multihop channels of arbitrary
number of hops.  The idea is to partition the relays in each layer.
Based on the partition, the relays coordinately amplify-and-forward
the received signal in a pre-assigned mode that changes periodically,
which creates a parallel channel in the time domain. Such partition is
thus called \emph{parallel partition}. We show that the mismatch is
removed in this way and the diversity upper bound is achieved.

In order to describe a parallel partition, some definitions and
notations are needed. A \emph{supernode} $\S$ is a set of indices
corresponding to a subset of antenna nodes in the same layer. The
cardinality of $\Scal$ is called the \emph{size} of the supernode. An
\emph{edge} is defined as the channel between two antenna nodes from
adjacent layers. An \emph{AF path} is defined as a sequence of
consecutive supernodes from the source to the destination, each
supernode performing the AF operation. A \emph{parallel partition}
$\P$ is defined as a set of AF paths. The number of AF paths in a
partition is called the \emph{partition size} and denoted by
$\Abs{\P}$. An \emph{independent parallel partition} is defined as a
parallel partition where any two different AF paths do not share
common edges. An independent partition of maximum size is called a
\emph{maximum partition}. An independent partition that achieves the
maximum diversity $\dmax$ is called a \emph{full diversity partition}.
\begin{lemma}\label{lemma:div}
  For any fading channel defined by $\mH$, we have
  \begin{equation}
    \label{eq:4}
    \Prob{\SNR\Frob{\mH}<1} \asympteq \SNR^{-d(0)},
  \end{equation}%
  where $d(r)$ is the DMT of the channel.
\end{lemma}
\begin{IEEEproof}
  See Appendix~\ref{sec:proof-lemma-div}.
\end{IEEEproof}

\begin{lemma} \label{lemma:div-para}
  Let us consider a set of $K$ independent parallel AF channels
  $$\my_k = \RP_k\mul\mx_k + \mz_k,\quad k=1,\ldots,K,$$
  where $\RP_k$'s
  are statistically independent. Then, the diversity order of the
  parallel channel is the sum of the diversity order of the individual
  AF sub-channels. Furthermore, if all the sub-channels have the same
  DMT $d_0(r)$, then the DMT of the parallel channel is $K\mul
  d_0(r)$.
\end{lemma}
\begin{IEEEproof}
  See Appendix~\ref{sec:proof-lemma-div-para}.
\end{IEEEproof}
  
\subsection{Independent Parallel Partition}

The independent parallel partition is accomplished in two steps~:
1)~partition each layer into supernodes, and 2)~find $K$ independent
AF paths connecting the supernodes. Each AF path defines a relaying
mode~: only the supernodes in this path are on and perform the AF
operation. Assume that a data frame of length $K\mul T$ is
transmitted. Then, the relays change the relaying mode every $T$
symbol times. We call it a \emph{parallel AF scheme}, since the
end-to-end channel is equivalent to a parallel AF channel in the time
domain. Note that the AF scheme is the trivial partition of size $1$
with a single ``wide'' AF path. As shown in remark~\ref{thm:cond_AF},
the trivial partition achieves the maximum diversity only when the
wide AF path satisfies the conditions in \Eq{eq:cond_AF}. This being
impossible in general, the parallel partition aims to find independent
``narrow'' paths each one of which satisfies the conditions in
\Eq{eq:cond_AF}. And if the number of independent paths is large
enough, then the maximum diversity order can be achieved according to
lemma~\ref{lemma:div-para}. Intuitively, the narrower the AF path is,
the easier the conditions \Eq{eq:cond_AF} are to be satisfied. In the
extreme case with the narrowest AF path $(1,1,\ldots,1)$, all
conditions in \Eq{eq:cond_AF} are met.
\begin{lemma} \label{lemma:maxi-par}
  In a $(n_0, n_1, \ldots, n_N)$ multihop channel, there are exactly
  $\dmax$ independent single-antenna AF paths.
\end{lemma} 
\begin{IEEEproof}
First, the converse is true, since otherwise, at least two AF paths
share the same edge in the bottleneck of the channel. Then, the
achievability is shown by construction~: we connect the multihop
channel in such a way that 1) there are $\dmax$ incoming and outgoing
edges for each intermediate layer, 2) the number of the incoming and
outgoing edges is the same for each antenna node~(say, in \layer{i})
and can be either $\floor{\dmax/n_i}$ or $\ceil{\dmax/n_i}$. This
partition contains $\dmax$ independent $(1,1,\ldots,1)$ AF paths each
one of which has diversity $1$.    
\end{IEEEproof}
The lemma implies that the maximum partition is of size $\dmax$. From
Lemma~\ref{lemma:div-para} and~\ref{lemma:maxi-par}, the following
proposition is immediate.
\begin{proposition}
  With the parallel AF scheme, the DMT
  \begin{equation}
    \label{eq:dmt_rateone}
    \dmax\mul (1-r)^+  
  \end{equation}%
  is always achievable in a multihop channel of arbitrary number of
  hops and antennas.
\end{proposition}
\begin{IEEEproof}
The DMT \Eq{eq:dmt_rateone} is simply achieved by applying the
parallel AF scheme with the maximum partition. In this case, $\dmax$
single-input-single-output~(SISO) parallel sub-channels are generated,
from which we have the DMT \Eq{eq:dmt_rateone}.   
\end{IEEEproof}
While the maximum diversity gain is achieved, this scheme only
exploits one out of $\ntilde_0$ degrees of freedom of the channel.
This is due to the SISO nature of the AF paths in the maximum
partition. In order to improve the achievable multiplexing gain, we
need parallel partitions with wider AF paths. Meanwhile, we still want
the maximum diversity, which requires that the AF paths should not be
too wide.  The following theorem states a necessary and sufficient
condition for an independent parallel partition to achieve the maximum
diversity.
\begin{theorem}\label{thm:cond_nece}
  Let the $n_{i^*}\times n_{i^*+1}$ channel be any bottleneck of the
  $(n_0,n_1,\ldots,n_N)$ multihop channel and $\Pcal$ be an
  independent partition of size $K$. Then, $\Pcal$ is a full diversity
  partition if and only if 1) $K=K_{i^*} K_{i^*+1}$ with $K_{i^*}\leq
  n_{i^*}$ and $K_{i^*+1}\leq n_{i^*+1}$, and 2) we have
  \begin{equation}
    \label{eq:cond_nece}
    \min_{i\notin\{i^*,i^*+1\}} n_{k,i} + 1 \geq n_{k,i^*} + n_{k,i^*+1},\quad \forall k,  
  \end{equation}%
  where $(n_{k,0},\ldots,n_{k,N})$ is the vector of numbers of
  antennas of the \th{k} AF path.
\end{theorem}
\begin{IEEEproof}
To prove the theorem, let us assume there are respectively $K_{i^*}$
and $K_{i^*+1}$ supernodes in the \layer{i^*} and \layer{i^*+1}, and
define $K'\defeq K_{i^*} K_{i^*+1}$. Then, we must have exactly
$K(\leq K')$ connections between the supernodes from these two layers.
The diversity of the partition ${\Pcal}$ is upper-bounded
\begin{align}
  d_{\Pcal} & \leq \sum_{k=1}^{K} n_{k,i^*} n_{k,i^*+1} \label{eq:tmp987}\\
  & \leq \sum_{k=1}^{K'} n_{k,i^*} n_{k,i^*+1} \label{eq:tmp876}\\
  & = n_{i^*} n_{i^*+1}.\nonumber
\end{align}%
Note that $\dmax = n_{i^*} n_{i^*+1}$ is achieved if and only if both
\Eq{eq:tmp987} and \Eq{eq:tmp876} have equality. Thus, we must have
both \Eq{eq:cond_nece} according to the conditions in \Eq{eq:cond_AF}
and $K=K_{i^*} K_{i^*+1}$ at the same time.
\end{IEEEproof}
Now, finding full diversity partitions with minimum size is an
optimization problem that minimizes the partition size $\Abs{\Pcal}$
subject to the constraint that $\Pcal$ must be an independent
partition and satisfy the conditions given by
theorem~\ref{thm:cond_nece}. Unfortunately, it remains an open problem
for a general multihop channel. The main difficulty lies in the lack
of knowledge on the mathematical structure of the independent
partitions for a general multihop channel. Nevertheless, the problem
is solved in the two-hop case.
\begin{proposition}\label{prop:2hop}
  For a $(n_0,n_1,n_2)$ channel, the minimum size of a full diversity
  partition is
  \begin{equation}
    \label{eq:2}
    K = \left\lceil \frac{n_1}{\Abs{n_0-n_2}+1} \right\rceil.
  \end{equation}%
\end{proposition}
\begin{IEEEproof}
  See Appendix~\ref{sec:proof-prop-2hop}.
\end{IEEEproof}
It is achieved by partitioning the relay layer into $K$ supernodes
of size $\left\lfloor\frac{n_1}{K}\right\rfloor$ or
$\left\lceil\frac{n_1}{K}\right\rceil$. For example, the minimum
partition size of the $(2,4,3)$ channel is $2$ as compared to the
maximum partition size $8$; and each AF path is a $(2,2,3)$ channel
instead of a $(1,1,1)$ channel.  Another example is the $(n,n,n)$
symmetric channel, where the minimum partition size is $n$ as compared
to the maximum partition size $n^2$; each AF path is a $(n,1,n)$
channel.

Some words regarding the related previous works before proceeding
further. In the relay channel with direct link and single layer of
relays, the $N$-relay non-orthogonal AF~(NAF)
scheme~\cite{ElGamal_coop} divides the data frame into $N$ sub-frames,
each one of which is relayed by one and only one relay. By creating a
parallel NAF channel, this scheme is optimal in diversity.  Similar
thought was shown in \cite{Elia_relay} in the same channel setting
with a different protocol called ND-RAF scheme. Removing the direct
link from the channel setting, the scheme in \cite{Elia_relay} becomes
the parallel AF scheme with the maximum partition in the
single-antenna single-layer case.

\subsection{Flip-and-Forward}
With the parallel AF scheme, the maximum multiplexing gain of the
channel is achieved only when every AF path in the partition achieves
the maximum multiplexing gain $\rmax=\ntilde_0$. In the following, we
propose a scheme that achieves both the maximum diversity gain and the
maximum multiplexing gain. Let us consider an example first.
\begin{figure*}
\begin{center}
\subfigure[parallel partition]{
\label{fig:ex_para}
\epsfig{figure=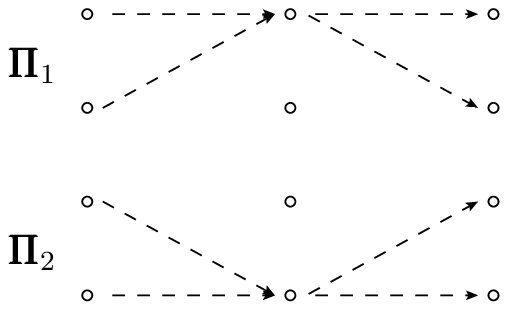,height=0.15\textwidth}}
\hspace{0.2\textwidth}
\subfigure[flip-and-forward]{
\label{fig:ex_ff}
\epsfig{figure=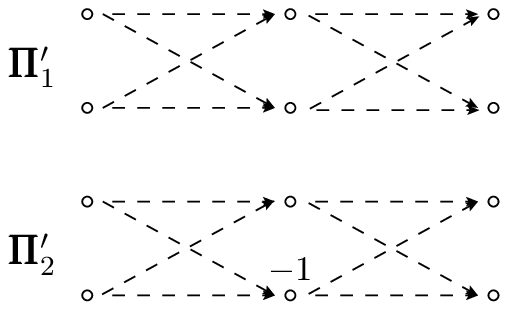,height=0.15\textwidth}}
\caption{Two sets of parallel channels from the $(2,2,2)$ multihop channel.}
\end{center}
\end{figure*}

\begin{example}\label{ex:FF222}
  The parallel channel $\{\H_1,\H_2\}$ in \Fig{fig:ex_para} has
  maximum diversity gain $4$ and multiplexing gain $1$, while the
  parallel channel $\{\H_1',\H_2'\}$ in \Fig{fig:ex_ff} has maximum
  diversity gain $4$ and multiplexing gain $2$.
\end{example}
In this example, $\{\H_1,\H_2\}$ corresponds to the parallel AF scheme
based on the full diversity partition proposed by
Proposition~\ref{prop:2hop}. However, it suffers from rate-deficiency,
since both sub-channels are of rank~$1$. An alternative is the channel
$\{\H_1',\H_2'\}$ shown in \Fig{fig:ex_ff}. Note that
\begin{align*}
  \H_1 &= \mH_2 \matrix{1&0\\0&0} \mH_1; &   \H_2 &= \mH_2 \matrix{0&0\\0&1} \mH_1; \\
  \H_1' &= \mH_2 \matrix{1&0\\0&1} \mH_1; &   \H_2' &= \mH_2 \matrix{1&0\\0&-1} \mH_1. 
\end{align*}
Hence, we have
\begin{equation*}
  \matrix{\H_1'&\H_2'} = \matrix{\H_1&\H_2} \matrix{\Id&\Id\\\Id&-\Id}
\end{equation*}%
from which $\Frob{\H_1'}+\Frob{\H_2'} = 2(\Frob{\H_1}+\Frob{\H_2}).$
Therefore, according to lemma~\ref{lemma:div}, they both achieve the
maximum diversity gain $4$ except that $\{\H_1',\H_2'\}$ has the
maximum multiplexing gain $2$ as well. This scheme is called the
\emph{Amplify-Flip-and-Forward}~(AFF)\footnote{The processing matrices
  $\mD_i$'s have been neglected for simplicity of demonstration.}
scheme, or simply the \emph{Flip-and-Forward}~(FF) scheme. The
intuition behind the FF scheme is as follows. It has been shown that
the mismatch between the two hops is the dominating outage event. Now,
suppose that $\H_1'$ is bad due to the bad ``angle'' between $\mH_1$
and $\mH_2$ both of which are not bad individually. Then, in the
second sub-channel, an independent ``rotation'' matrix $\diag\{1,-1\}$
is used to change the angle. With high probability, the new angle is
not bad and the mismatch is solved.

In the light of the example, we generalize the scheme to arbitrary
number of antennas and hops. Three steps are needed to describe the
construction.
\begin{enumerate}[\bfseries step 1]
\item Find a full diversity independent parallel partition $\Pcal$ of
  size $K$.  The partition defines the intermediate supernodes in each
  layer.
\item We denote the supernodes in \layer{i} by
  $\S_{i,1},\ldots,\S_{i,K_i}$ with $K_i$ being the number of
  supernodes in \layer{i}. And we define the flip matrices
  $\F{i}{k}$'s as $n_i\times n_i$ diagonal matrices with
  \begin{equation*}
    \F{i}{k}(j,j) =  
    \begin{cases}
      -1, & \text{if $j\in\S_{i,k}$ and $k\neq1$}, \\
      1, & \text{otherwise}.
    \end{cases}
  \end{equation*}%
\item The FF scheme is composed of $K'\defeq\prod_{i=1}^{N-1} K_i$
  parallel sub-channels $\{\H'_k\}_k$ with
  \begin{equation} \label{eq:tmp1111}
    \H'_k \defeq \mH_N \prod_{i=1}^{N-1} \left(\F{i}{f_i(k)} \mH_i \right),
  \end{equation}%
  where $f_1(k) \defeq (k-1)_{K_1}+1$ and
  \begin{equation*}
    f_i(k) \defeq \left(\left\lceil\frac{k-1}{\prod_{j=1}^{i-1} K_j}\right\rceil\right)_{K_i} + 1, \quad i=2,\ldots,N-1. 
  \end{equation*}%
\end{enumerate}
In other words, the set of relays works in $K'$ different flip modes,
each one being identified by a sequence of flip modes of individual
relay layers. And the mapping is effectuated by the functions
$f_1(k),f_2(k),\ldots,f_{N-1}(k)$. The exact DMT of the FF scheme
being difficult to obtain, we get a lower bound instead.

\begin{figure}[!t]
\begin{center}
  \includegraphics[angle=0,width=0.75\textwidth]{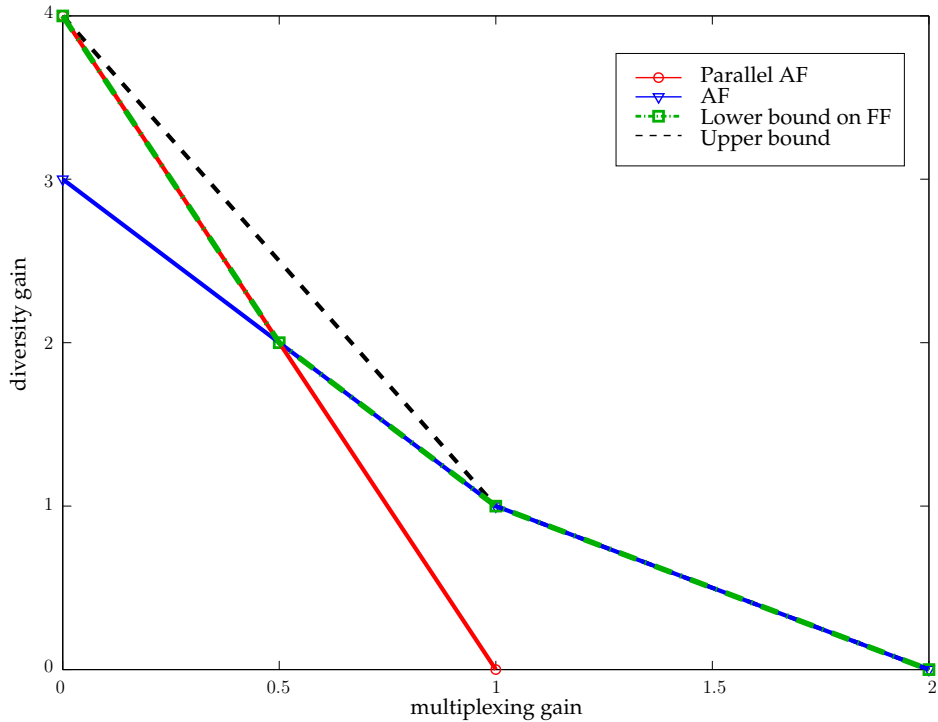}
\caption{Diversity-multiplexing tradeoff of $(2,2,2)$ channel with different schemes.}
\label{fig:dmt222}  
\end{center}
\end{figure}

\begin{theorem}\label{thm:DMT-FF}
  The FF scheme constructed above achieves the following DMT 
  \begin{equation}
    \label{eq:dmt-FF}
    \dFF(r) \geq \dAF(r) + \left(\dmax-\dAF(0)\right)(1-K'r)^+,\quad \forall\,r. 
  \end{equation}%
\end{theorem}
\begin{IEEEproof}
  See Appendix~\ref{sec:proof-theorem-dmt-ff}.
\end{IEEEproof}
We can verify that $\dFF(0)=\dmax$, that is, the maximum diversity of
the channel is achieved. Furthermore, the FF scheme is always better
than the AF scheme, especially at low multiplexing gain. This can be
explained by the intuition that the FF scheme solves the mismatch of
adjacent hops using all possible combinations of flip modes of
individual supernodes. The equivalent end-to-end channel of the FF
scheme can be bad only if at least one of the hops are bad. The
maximum diversity is thus achieved. \Fig{fig:dmt222} shows the DMT of
different schemes in the channel of Example~\ref{ex:FF222}. While the
AF and the parallel AF schemes achieve respectively the extreme of
maximum multiplexing gain $(2,0)$ and the extreme of maximum diversity
gain $(0,4)$, the FF scheme achieves both extremes.

\begin{remark}
  
The proposed FF scheme is constructed based on the flip matrices that
are diagonal with $\pm1$ entries. In fact, it can be shown that a
looser sufficient condition is for the matrices to be~1)~diagonal,
2)~linearly independent, and 3)~of unit absolute value~(power
constraints). Therefore, we can find infinitely many sets of ``flip''
matrices that satisfy the above conditions and they are all diversity
optimal. Intuitively, if the matrices are too ``close'', the FF scheme
tends to the AF scheme and the promised maximum diversity gain can be
achieved only when the SNR is very large. This is translated into a
poor power gain of the scheme. Hence, we should choose the matrices
such that they are ``far'' from each other. In this way, with high
enough probability, any mismatch can be solved by at least one
``rotation'' and the maximum diversity can be obtained in relatively
small SNR. However, what remains open is how to choose the distance
metric between the rotation matrices.

\end{remark}

\subsection{Non-Independent Partition}

With independent partition, the total diversity is the sum of the
diversity of each AF path. We also established some conditions that
independent partitions must satisfy to achieve the maximum diversity.
In the following, we investigate a particular case of non-independent
partition.

Let us consider a parallel channel defined by $\{\H_k\}_k$ with
\begin{equation}\label{eq:para_nonind}
  \H_k \defeq \mH_N\cdots\mH_{i+1} \mJ_k \mH_{i}\cdots\mH_1,
\end{equation}%
where the \emph{selection matrix} $\mJ_k$ is a $n_i\times n_i$
diagonal matrix whose entries are zero except that $\mJ_k(k,k)=1$. The
matrices $\H_k$'s are not independent, since they share the common
sub-channels $\G1\defeq \mH_{i-1}\cdots\mH_1$ and $\G2\defeq
\mH_N\cdots\mH_{i+2}$. However, the RP channels $\mH_{i+1} \mJ_k
\mH_{i}$'s are independent for different $k$'s. Despite the dependency
between the sub-channels, we can obtain the diversity order of the
parallel channel.
\begin{theorem}\label{thm:non-ind}
  The diversity order of the channel described above is 
  \begin{equation}
    \label{eq:dmax_nonind}
    \min\{\dAF_{(n_0,\ldots,n_i)}(0), \dAF_{(n_{i},\ldots,n_N)}(0)\}.
  \end{equation}%
\end{theorem}
\begin{IEEEproof}
  We use the DMT interpretation given in Section~\ref{sec:recur} to
  sketch the proof. One possibility for the parallel channel
  $\{\H_k\}_k$ to be in outage is that one of $\G_1$ and $\G2$ is bad.
  The diversity is either $\dAF_{(n_0,\ldots,n_{i-1})}(0)$ or
  $\dAF_{(n_{i+1},\ldots,n_{N})}(0)$. Another possibility is that both
  $\G_1$ and $\G_2$ are good and that $\{\H_k\}_k$ turns out to be
  bad.  Without loss of generality, we assume that the flow from the
  source to \layer{i-1} is $k_1$ and that from \layer{i+1} to the
  destination is $k_2$. And we call the outage event a
  \emph{type-$(k_1,k_2)$} event.  Then, it can be shown that
  $\{\H_k\}_k$ is equivalent to $\{\mH_{i+1}' \mJ_k \mH_{i}' \}_k$
  with $\mH_{i}'\in\CC^{n_i\times k_1}$ and $\mH_{i+1}'\in\CC^{k_2
    \times n_i}$ being Gaussian matrices with i.i.d. $\CN[1]$ entries.
  Now, we must disconnect all the sub-channels in $\{\mH_{i+1}' \mJ_k
  \mH_{i}' \}_k$, which costs $n_i \mul\min\{k_1,k_2\}$. Therefore,
  the total cost for the type-$(k_1,k_2)$ event is
  $$\dAF_{(n_0,\ldots,n_{i-1})}(k_1)+n_i\mul \min\{k_1,k_2\} +
  \dAF_{(n_{i+1},\ldots,n_{N})}(k_2).$$
  The typical outage event is
  the one that minimizes the above total cost. For $k_2\geq k_1$,
  using \Eq{eq:interp2}, we can show that the minimum total cost is
  $\dAF_{(n_0,\ldots,n_i)}(0)$. Similarly,
  $\dAF_{(n_{i+1},\ldots,n_N)}(0)$ is the minimum total cost for
  $k_1>k_2$. Since both costs are smaller than
  $\dAF_{(n_0,\ldots,n_{i-1})}(0)$ and
  $\dAF_{(n_{i+1},\ldots,n_{N})}(0)$ for the
  monotonicity~(Corollary~\ref{coro:increase}), we proved the theorem.
\end{IEEEproof}

Note that with this particular partition at \layer{i}, we achieve a
diversity order as if \layer{i} were clustered and the cooperative DF
scheme were used. This result implies that one might achieve the
maximum diversity with a partition of small size. For example, the
maximum diversity order of the $(3,2,2,2,3)$ channel is $4$ and all
the full diversity independent partitions are of size~$K=8$, \ie,
eight $(3,1,1,1,3)$ sub-channels. With the non-independent partition
described above, we get a couple of $(3,2,1,2,3)$ sub-channels, \ie,
size $2$. Since $\dAF_{(2,2,3)}(0)=4$, the maximum diversity $4$ is
achieved as well according to Theorem~\ref{thm:non-ind}.

We can apply the FF scheme to the case of non-independent partition.
Then, in this example, the parallel channel is $\{\H'_k\}_k$ with
$\H'_k \defeq \mH_N\cdots\mH_{i+1} \mF_k \mH_{i}\cdots\mH_1$ where the
flip matrix $\mF_k$ is a $n_i\times n_i$ diagonal matrix whose entries
are one except that $\mF_k(k,k)=-1$ if $k\neq1$. The channels
$\{\H'_k\}_k$ being a linear invertible transformation of
$\{\H_k\}_k$, the generalized FF scheme achieves the diversity given
by \Eq{eq:dmax_nonind}.

\subsection{Extensions}

With the nice parallel-channel structure, the FF scheme can be
extended to various cases. 

Let us first consider the extension to the MIMO relay channel with
direct link and a single layer of $N$ relays. By applying directly the
single-antenna NAF scheme~\cite{ElGamal_coop} to the multi-antenna
case, the source cooperates with one relay at a time. This is
equivalent to using the parallel AF scheme in the
source-relays-destination link. The DMT lower bound is obtained in
\cite{SY_JCB_coop} as
\begin{equation}
  \label{eq:tmp9888}
  d_{\mF}(r) + N\mul\dAF_{(\nt,n,\nr)}(2r),
\end{equation}%
where $d_{\mF}(r)$ is the DMT of the $\nt\times\nr$ source-destination
channel $\mF$ and each relay has $n$ antennas. In fact, this lower
bound can be improved to
\begin{equation}
  \label{eq:tmp9889}
  d_{\mF}(r) + \dFF_{(\nt,N\mul n,\nr)}(2r),
\end{equation}%
by replacing the parallel AF scheme in the source-relays-destination
link with the FF scheme. Comparing the second terms from
\Eq{eq:tmp9888} and \Eq{eq:tmp9889}, the gain in diversity of the new
scheme over the MIMO NAF is reflected by
\begin{align}
  N\dAF_{(\nt,n,\nr)}(0) &\leq N\mul n\mul\min\{\nt,\nr\} \label{eq:tmp8788}\\
  &= \dFF_{(\nt,N\mul n,\nr)}(0) \nonumber
\end{align}%
where the inequality \Eq{eq:tmp8788} becomes strict when $n$ is large.
The gain in multiplexing of the source-relays-destination link is
obvious when $n$ is small, \ie, $n<\min\{\nt,\nr\}$. In this case, the
FF scheme pools the relay antennas together to provide more degrees of
freedom.

Another extension is to the multiuser case. Let us take the multiple
access channel as an example. For simplicity, we assume that $M$ users
try to communicate with the common destination through the same layers
of relays. Then, we the FF scheme, we have an equivalent parallel
multiple access channel with 
\begin{equation}
  \label{eq:para-MAC}
  \my_k = \sum_{i=1}^M \H_{k,i}\mul \mx_i + \mz_k,\quad k=1,\ldots,K',
\end{equation}%
where $\left\{\H_{k,i}\right\}_k$ is similarly defined as in
\Eq{eq:tmp1111} with 
\begin{equation}
  \label{eq:tmp7768}
    \H_{i,k} \defeq \mH_N \F{N-1}{f_{N-1}(k)} \mH_{N-1}\cdots \mH_{2}\F{1}{f_{1}(k)} \mH_{1,k}.
\end{equation}%
Note that only the first hop is distinct for different users. Using
the techniques of \cite{Tse_DMT_MAC} and the our results for the
single-user FF scheme, it is possible to analyze the DMT of the FF
scheme in the multiple access channel. It is trivial to show that
similar extension also holds for the broadcast channels with minor
modifications.

\section{The Clustered Case Revisited}
\label{sec:clustered}

In Section~\ref{sec:CCnDF}, it has been shown that the cooperative DF
scheme achieves the DMT cut-set bound in the clustered case. In this
section, we would like to study some alternative schemes, since it
might be impossible or unnecessary for all the clusters to decode the
source message in some cases.

\subsection{Serial Partition}
\label{sec:int-decoding}
The AF and the cooperative DF schemes can in fact be seen as two
extremes of what we call the \emph{serial partition} of multihop
channels, defined as follows.
\begin{definition}
  A \emph{serial partition} is defined by a set of layer indices
  $\Dcal\defeq\{\Dcal_1,\Dcal_2,\ldots,\Dcal_{\Abs{\Dcal}}\}$ with
  $0<\Dcal_1<\Dcal_2<\ldots\Dcal_{\Abs{\Dcal}-1}<\Dcal_{\Abs{\Dcal}}\defeq
  N$, each layer performing cooperative decoding-and-forward
  operation.
\end{definition}%
With a serial partition, the multihop channel becomes a serial
concatenation of $\Abs{\Dcal}$ AF channels. As in \Eq{eq:dmt-ub}, the
DMT of the multihop channel with any partition $\Dcal$ is easily
derived as
\begin{equation}
  \label{eq:dmt-mixed}
  d_\Dcal(r) = \min_{i=1,\ldots,\Abs{\Dcal}} \dAF_{(n_{\Dcal_{i-1}},\ldots,n_{\Dcal_i})}(r),
\end{equation}%
where we defined $\Dcal_0\defeq 0$.  To get the maximum diversity
gain, the question of \emph{when to decode} has been answered
earlier~: when the conditions in \Eq{eq:cond_AF} are not met. Another
question is \emph{where to decode}, \ie, how to find the partition of
minimum size that achieves a given diversity order.
\begin{proposition}\label{prop:where_to_decode}
  Let us take $\Dcal_0 = 0$ and we succesively decide $\Dcal_i$ as the
  maximum integer in $(\Dcal_{i-1},N]$ such that
  \begin{equation}
    \label{eq:tmp098}
    \dAF_{(n_{\Dcal_{i-1}},\ldots, n_{\Dcal_{i}})}(0) \geq d. 
  \end{equation}%
  Then, the decoding set $\{\Dcal_i\}$ defines the partition of
  minimum size that achieves a given diversity $d$~($\leq d_{\max}$).
\end{proposition}
\begin{IEEEproof}
  From \Eq{eq:dmt-mixed}, it is easy to show that the proposed
  partition achieves diversity $d$. Now, we would like to show that
  the size of the proposed partition is minimized. To this end, it is
  enough to show that for any set $\Dcal'$ of decoding points that
  achieves diversity $d$, we have $\Dcal'_i\leq\Dcal_i$, $\forall\,i$.
  This is obviously true for $\Dcal'_1$, since the diversity of the AF
  channel degrades with the number of hops. By induction on $i$, it is
  shown that $\Dcal'_{i+1}\leq\Dcal_{i+1}$ because otherwise
  $(n_{\Dcal_i},\ldots,n_{\Dcal_{i+1}}) \subseteq
  (n_{\Dcal'_i},\ldots,n_{\Dcal'_{i+1}})$ and the corresponding
  diversity of the AF scheme cannot be larger than $d$ according to
  the monotonicity of the DMT~(Corollary~\ref{coro:increase}).
\end{IEEEproof}
The proposition matches the intuition that we should only decode when
we have to, in the diversity sense. In other words, we allow for the
degradation of diversity introduced by the AF operation, as long as
the resulting diversity is larger than the target $d$.

\subsection{CSI Aided Linear Processing}

Another option is to linear process the received signal at each
cluster without decoding it. Unlike the AF scheme in the non-clustered
case, where trivial antenna-wise normalization is performed, we can
run inter-antenna processing based on the available CSI at the
cluster. With receiver CSI at the relays, let us consider the
following project-and-forward~(PF) scheme. At \layer{i}, the received
signal is first projected to the signal subspace spanned by the
columns of the channel matrix $\umH_i$. The dimension of the subspace
is $\r_i$, the rank of $\umH_i$.  After the component-wise
normalization, the projected signal is transmitted using $\r_i$~(out
of $n_i$) antennas.  It is now clear that
$\umH_{i+1}\in\CC^{n_{i+1}\times \r_i}$ is actually composed of the
$\r_i$ columns of the previously defined $\mH_{i+1}$, with $r_0\defeq
n_0$. More precisely, the $\umQ_i\in\CC^{n_i\times \r_i}$ is an
orthogonal basis with $\transc{\umQ}_i \umQ_i=\Id$. We can rewrite
$$\umH_i = \umQ_i \umG_i$$
with $\umG_i\in\CC^{\r_i\times r_{i-1}}$.
For simplicity, we let $\umQ_i$ be obtained by the QR
decomposition~\cite{Horn} of $\umH_i$ if $n_i>\r_{i-1}$ and be
identity matrix if $n_i\leq \r_i$.  The spirit of the PF scheme is not
to use more antennas than necessary to forward the signal. Since the
useful signal lies only in the $\r_i$-dimensional signal subspace, the
projection of the received signal provides sufficient statistics and
reduces the noise power by a factor $\frac{n_i}{\r_i}$. In this case,
only $\r_i$ antennas are needed to forward the projected signal. Let
us define $\mP_i \defeq \umD_i\transc{\umQ}_i$.  Then, as in the AF
case, the PF multihop channel is equivalent to the channel defined by
$$\RPPF = \umH_N\mP_{N-1}\cdots\umH_2\mP_1\umH_1.$$
The following
proposition states that receiver CSI and inter-antenna processing do
not improve the DMT of the AF scheme.
\begin{proposition}\label{thm:PF} 
  The PF scheme is equivalent to the AF scheme.
\end{proposition}
\begin{IEEEproof}
  See Appendix~\ref{sec:proof-thm-PF}.
\end{IEEEproof}
While the PF and AF have the same DMT, the PF outperforms the AF in
power gain for two reasons. One reason is, as stated before, that the
projection reduces the average noise power.  The other reason is that
the accumulated noise in the AF case is more substantial than that in
the PF case. This is because in the PF case, less relay antennas are
used than in the AF case. Since the power of independent noises from
different transmit antennas add up at the receiver side, the
accumulated noise in the AF case ``enjoys'' a larger ``transmit
diversity order'' than in the PF case.

On the other hand, if we could have receiver \emph{and} transmitter
CSI at the clusters, the DMT could be improved as shown by the
following example.
\begin{example}\label{ex:MF}
  For a $(n,n,\ldots,n)$ clustered multihop channel, the DMT cut-set
  bound can be achieved by linear processing within clusters if both
  transmitter and receiver CSI are available at each cluster.
\end{example}
The optimum linear relaying scheme is defined by the processing
matrices $\mT_i$'s with $\mT_i\defeq \transc{\mV}_{i+1} \mU_i$ where
we assume that $\mH_i=\transc{\mU}_i\mSigma_i\mV_i$ is the singular
value decomposition of $\mH_i$. The diagonal elements in the singular
value matrix $\mSigma_i$ are in increasing order. This scheme matches
the adjacent hops by aligning the singular values in the same order.
It is then equivalent to the channel defined by $\prod_i \mSigma_i$,
whose DMT can be shown\footnote{The proof, that is essentially as the
  proof in \cite{Zheng_Tse}, is omitted here.} to be as the $n\times
n$ Rayleigh channel.

\section{Codes Construction}
\label{sec:stc}

Now, we need codes that actually attain the DMT promised by the
studied relaying strategies. To this end, the construction of Perfect
STBCs~\cite{Oggier_perfect,Elia} for MIMO channels is extended to the
multihop relay channels. The constructed codes are approximately
universal~\cite{Tavildar}.

\subsection{The Clustered Case}
The relay clusters that perform the cooperative DF operation partition
the multihop channel into a series of $\Abs{\Dcal}$ MIMO channels,
say, $\sr{\mH}_1,\sr{\mH}_2,\ldots,\sr{\mH}_{\Abs{\Dcal}}$ with
$\sr{\mH}_i\in\CC^{n_{\Dcal_i}\times n_{\Dcal_{i-1}}}$.  An obvious
coding scheme that achieves the DMT is described as follows. Let $r$
be the target multiplexing gain. First, the source terminal encodes
the message of $T\mul r\log\SNR$ bits with a $n_0\times T$ Perfect
STBC $\Xcal_0(r)$. Then, in a successive manner, \layer{\Dcal_i} tries
to decode the message. When a success decoding is assumed, the $T\mul
r\log\SNR$ bits are encoded with a $n_{\Dcal_i}\times T$ Perfect STBC
$\Xcal_i(r)$ and forwarded. We can show that as long as $T\geq \Tmin$
with
$$\Tmin\defeq\max_{i=1,\ldots,{\Abs{\Dcal}}} n_{\Dcal_{i-1}},$$
the
series of Perfect STBCs $\{\Xcal_i\}_i$ can be found~\cite{Elia}. With
the union bound, the end-to-end error probability is upper-bounded
\begin{equation}
  \label{eq:170}
  \Pe(r,\SNR) \leq \sum_{i=1}^{\Abs{\Dcal}} \Pe^{(i)}(r,\SNR),  
\end{equation}%
where $\Pe^{(i)}$ is the error probability of $\Xcal_i(r)$ in the MIMO
sub-channel $\sr{\mH}_i$. Since $\Xcal_i(r)$ is DMT-achieving for any
fading statistics, we have
\begin{equation}
  \label{eq:176}
  \Pe^{(i)}(r,\SNR) \asympteq
\SNR^{-\dAF_{(n_{\Dcal_{i-1}},\ldots,n_{\Dcal_{i}})}(r)}.
\end{equation}%
From \Eq{eq:170} and \Eq{eq:176}, the DMT \Eq{eq:dmt-mixed} is
achieved with coding delay $\Tmin$. Since the Perfect STBCs are
approximately universal~\cite{Tavildar}, so is this coding scheme.
Note that this scheme can be used for the AF and PF schemes with
$\Abs{\Dcal}=1$.

\subsection{The Non-Clustered Case}
In the non-clustered case, the parallel AF and the FF schemes are
used. Note that both schemes share the common parallel MIMO channel
structure
\begin{equation}
  \label{eq:parallel}
  \my_k = \prl{\mH}_k\mul \mx_k + \mz_k, \quad k=1,\ldots,K,  
\end{equation}%
where $\prl{\mH}_k\in\CC^{\nrk\times \ntk }$ and $K$ is the
number of the parallel sub-channels. Let $\Xcal$ be a code for the
parallel channel. A codeword is defined by a set of matrices
$\{\mX_k\}_{k=1}^K$ with $\mX_k\in\CC^{\ntk\times T}$. We define
a parallel STBC with non-vanishing determinant~(NVD) as follows.
\begin{definition}
  Let $\Bcal$ be an alphabet that is scalably dense, \ie, for $0\leq
  a \leq 1$,
\begin{eqnarray*}
\Abs{\Bcal(\SNR)} \asympteq \SNR^{a}, \quad \textrm{and} \\
s \in \Bcal(\SNR) \Rightarrow \Abssqr{s} \asymptleq \SNR^{a}. 
\end{eqnarray*}%
Then, a parallel STBC $\Xcal$ is called a \emph{parallel NVD code} if
it
\begin{enumerate}
\item is $\Bcal$-linear\footnote{$\Xcal$ is $\Bcal$-linear means that
    each entry of any codeword in $\Xcal$ is a linear combination of
    symbols from $\Bcal$.};
\item has full symbol rate, \ie, it transmits on average $\sum_k \ntk$
  symbols per channel use from the signal constellation $\Bcal$;
\item has the NVD property, \ie, for any pair of different codewords
  $\{\mX_k\}_k,\{\hat{\mX}_k\}_k\in\Xcal$,
  \begin{equation}
    \label{eq:NVD}
    \prod_k 
    \det\left(
      (\mX_k-\hat{\mX}_k)\transc{(\mX_k-\hat{\mX}_k)} \right)
    \geq\kappa>0,
  \end{equation}
  with $\kappa$ a constant independent of the SNR.
\end{enumerate}
\end{definition}
We have the following result. 
\begin{theorem} \label{thm:dmt-para}
  The parallel NVD codes are approximately universal over the parallel
  channel defined by \Eq{eq:parallel}.
\end{theorem}%
\begin{IEEEproof}
  See Appendix~\ref{sec:proof-thm-dmt-para}.
\end{IEEEproof}
Thus, to achieve the DMT of the parallel AF and the FF schemes, it is
enough to construct a parallel NVD codes. Several remarks are made
before proceeding to the code construction.
\begin{remark}
  The actual data rate of the NVD codes is controlled by the size of
  the alphabet $\Bcal$ and the symbol rate. Efficient decoding
  schemes~(\eg, sphere decoding) may not be implementable when the
  channel is under-determined or, alternatively speaking,
  rank-deficient in the sense that $\sum_k
  \text{rank}(\prl{\mH}_k)< \sum_k \ntk$. Practical schemes
  include reducing the symbol rate while increasing the size of the
  alphabet $\Bcal$. This, however, does not guarantee the
  DMT-achievability.
\end{remark}
\begin{remark}
  Explicit parallel NVD codes for asymmetric parallel channel~(\ie,
  $\nti\neq \ntj$ for some $i\neq j$) being hard to construct
  algebraically, we focus on the symmetric case. Note that in the FF
  scheme, the equivalent parallel channel is always symmetric. In the
  parallel AF scheme, the numbers of transmit antennas of different
  sub-channels may be different. However, the problem can be overcome
  by using the same number of antennas~(\ie, $\max_k \ntk$). The
  resulting parallel channel has at least the same DMT as the original
  channel. Nevertheless, an alternative code construction that is
  suitable for both symmetric and asymmetric parallel channels is
  provided in Appendix~\ref{sec:altercod} for completeness.
\end{remark}
\begin{remark}
  From a given parallel partition with size $S$, the number of the
  parallel sub-channels $K$ is $S$ in the parallel AF scheme, generally
  larger than $S$ in the FF scheme. Since the minimum coding delay is
  $K\mul\max_k \ntk$ that grows linearly with $K$, it grows at
  least linearly with $S$. Moreover, the complexity of decoding can
  grow up to exponentially with $K$ if ML decoding is used. That is
  why it is important to find partitions of small size $S$.
\end{remark}

\subsection{Algebraic Construction of Parallel NVD Codes}
\label{sec:paraNVD}
A systematic way to construct NVD codes is the construction from
cyclic division algebra~(CDA). For more details on the concept, the
readers can refer to \cite{Rajan-03}. In the following, we aim to
construct the Perfect symmetric parallel NVD codes with quadrature
amplitude modulation~(QAM)
constellations.\footnote{The construction was first reported in
  \cite{SY_JCB_ISIT_parallel} and is included for sake of
  completeness.} The generalization to hexagonal constellations is
straightforward.

\subsubsection{$K=1$}
We start by the construction of NVD codes for MIMO channels~($K=1$).
Let $\LL\defeq \QQ(i,\theta)$ be a cyclic extension of degree $\nt$ on
the base field $\QQ(i)$. We denote $\sigma$ the generator of the
Galois group $\Gal(\LL/\QQ(i))$. Let $\gamma\in \QQ(i)$ be such that
$\gamma,\gamma^2,\ldots,\gamma^{\nt-1}$ are non-norm elements in
$\LL$. Then, we can construct a CDA $\Acal = (\LL/\QQ(i), \sigma,
\gamma)$ of degree $\nt$. Each element in $\Acal$ has the following
matrix representation
\begin{equation}
  \label{eq:Xi}
 \mXi=\left(\begin{array}{cccc}
    x_{0} & x_{1} & \ldots & x_{\nt-1}\\
    \gamma\sigma\left(x_{\nt-1}\right) & \sigma\left(x_{0}\right) & \ldots & \sigma\left(x_{\nt-2}\right)\\
    \vdots & \vdots & \ddots & \vdots\\
    \gamma\sigma^{\nt-1}\left(x_{1}\right) &
    \gamma\sigma^{\nt-1}\left(x_{2}\right) & \ldots &
    \sigma^{\nt-1}\left(x_{0}\right)\end{array}\right) ,
\end{equation}%
where $x_i\in\Ocal_{\LL}$, $\forall\,i$. Since $\Acal$ is a CDA, we
can show that
$\det\mXi \in \ZZ[i]$
and that the determinant is zero only when
$\mXi$ is a zero matrix. Thus, the NVD property is proved by
considering that the difference matrix of each pair of codewords is
in the form of $\mXi$.

It is usually desirable to get a STBC with good shaping. To this end,
we can impose the additional constraint that the vectorized codeword
is a rotated version of a $\textrm{QAM}^{N\mul \nt^{2}}$
constellation, as known as the cubic constellation. Rotated
constellations
constructions from algebraic number fields are well-known now (see,
\eg, \cite{Oggier-2} for a comprehensive tutorial on this topic).
This can be made possible if~1)~$x_i$'s in the matrix $\mXi$ belong to
some properly chosen ideal $\Ical\subseteq
\Ocal_{\LL}$~\cite{Bayer}, and~2)~$\Abs{\gamma}=1$~(see \cite{Elia}
for a
general method). The thus-constructed NVD codes are
well-known as the Perfect STBCs.

\subsubsection{$K>1$}

\begin{figure}
\begin{center}\includegraphics[%
  width=0.3\textwidth]{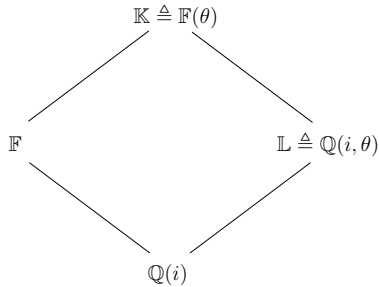}\end{center}
\caption{\label{fig:fields-extension}Field extension tower.}
\end{figure}
The construction of parallel NVD codes is similar to the construction
presented above. First, we construct a CDA in the same manner as the
previous case by simply 1)~replacing the base field $\QQ(i)$ by a new
field $\FF$, a Galois extension of degree $K$ over $\QQ(i)$;
2)~replacing the field $\LL$ by $\KK\defeq \FF(\theta)$, a cyclic
extension of degree $\nt$ over $\FF$ (same $\theta$ as the previous
case); and 3)~choosing $\gamma$ such that
$\gamma,\gamma^2,\ldots,\gamma^{\nt-1}$ are non-norm elements in
$\KK$. We impose that $\FF\cap\LL=\QQ(i)$. Note that the extension
$\KK/\FF$ remains cyclic
with the same Galois group as
$\Gal(\LL/\QQ(i))$~(\Fig{fig:fields-extension}). Thus, the
constructed CDA is $\Acal(\KK/\FF, \sigma, \gamma)$. Now, let
$\{\tau_1,\tau_2,\ldots,\tau_K\}$ be the Galois group of the extension
$\FF/\QQ(i)$ and define
$$\mXi_k \defeq \tau_k(\mXi),\quad k=1,\ldots,K,$$
where $\mXi$ is the
matrix representation of some element in $\Acal$ and is in the form
\Eq{eq:Xi}. Now, we have
\begin{align*}
  \prod_k \det\mXi_k &=
  \prod_k \tau_k\left(\det\mXi\right) \\
  &= N_{\FF/\QQ(i)} \left(\det\mXi\right)
\end{align*} 
that is in $\ZZ[i]$. Finally, we construct codewords $\{\mX_k\}_k$ in
the form of $\{\mXi_k\}_k$ with QAM symbols and we can show that the
difference matrix of a pair of different codewords is also in the form of 
$\{\mXi_k\}_k$ with symbols in $\ZZ[i]$. The
NVD condition~\Eq{eq:NVD} is thus met. Similarly, the
cubic shaping can be obtained with the same kind of conditions
mentioned before. An explicit code construction is provided in the
following example.
\begin{example}[Two transmit antennas and $K=2^m$ sub-channels]\label{ex:code_construction}
  Let us define $\zeta_{2^{m+1}}\defeq e^{-i\pi 2^{m}}$. Then, we
  consider the base field
  $\mathbb{F}=\mathbb{Q}\left(\zeta_{2^{m+2}}\right)$, an extension of
  $\QQ(i)$ of degree $2^m$ and take
  $\mathbb{K}=\FF(\sqrt{5})=\mathbb{Q}\left(\zeta_{2^{m+2}},\sqrt{5}\right)$.
  We can verify that $\gamma\defeq\zeta_{2^{m+2}}$ is a non-norm
  element in $\KK$~(see Appendix~\ref{zeta2m-not-norm}). Let
  $\theta=\frac{1+\sqrt{5}}{2}$ and
  $\sigma:\theta\mapsto\bar{\theta}=\frac{1-\sqrt{5}}{2}$. The ring of
  integers of $\mathbb{K}$ is $\mathcal{O}_{\mathbb{K}}=\left\{
    a+b\theta\mid a,b\in\mathbb{Z}\left[\zeta_{2^{m+2}}\right]\right\}
  $. And the chosen ideal is principle, \ie,
  $\Ical=(\alpha)\Ocal_{\KK}$ with $\alpha=1+i-i\theta$.  The matrix
  $\mXi$ is given by
  \begin{equation}
    \mXi=\left[\begin{array}{cc}
        \alpha\cdot\left(a+b\theta\right) & \alpha\cdot\left(c+d\theta\right)\\
        \gamma\bar{\alpha}\cdot\left(c+d\bar{\theta}\right) & \bar{\alpha}\cdot\left(a+b\bar{\theta}\right)\end{array}\right],\label{eq:Golden-2}
  \end{equation}%
  where $a,b,c,d \in \ZZ[\zeta_{2^{m+2}}]$. We can show that the
  shaping property is satisfied and finally, this code is a perfect
  STBC for the parallel channel.
\end{example}

\section{Numerical Examples}
\label{sec:nr}

In this section, we present the numerical results on the proposed
schemes. The performance measures are either the outage probability or
the symbol error rate probability versus the average received SNR per
bit. The results are obtained with Monte-Carlo simulations.

\begin{figure*}[!t]
\begin{center}
\subfigure[Outage probability]{
\label{fig:vertical_reduction}  
\epsfig{figure=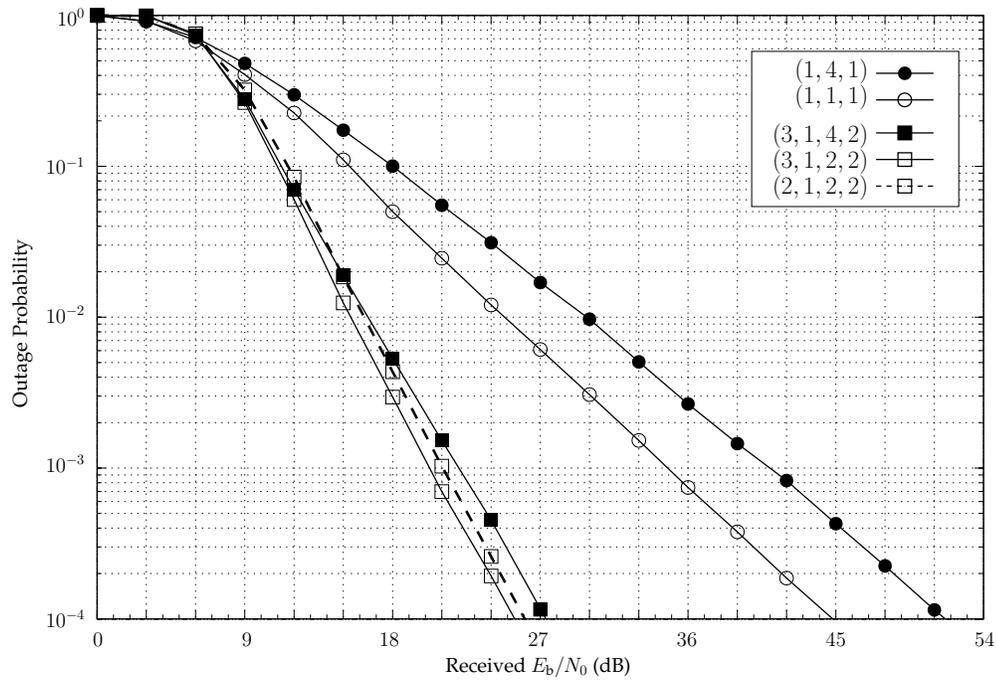,angle=0,width=0.85\textwidth}}\\
\subfigure[Symbol error rate]{
\label{fig:coded_performance}  
\epsfig{figure=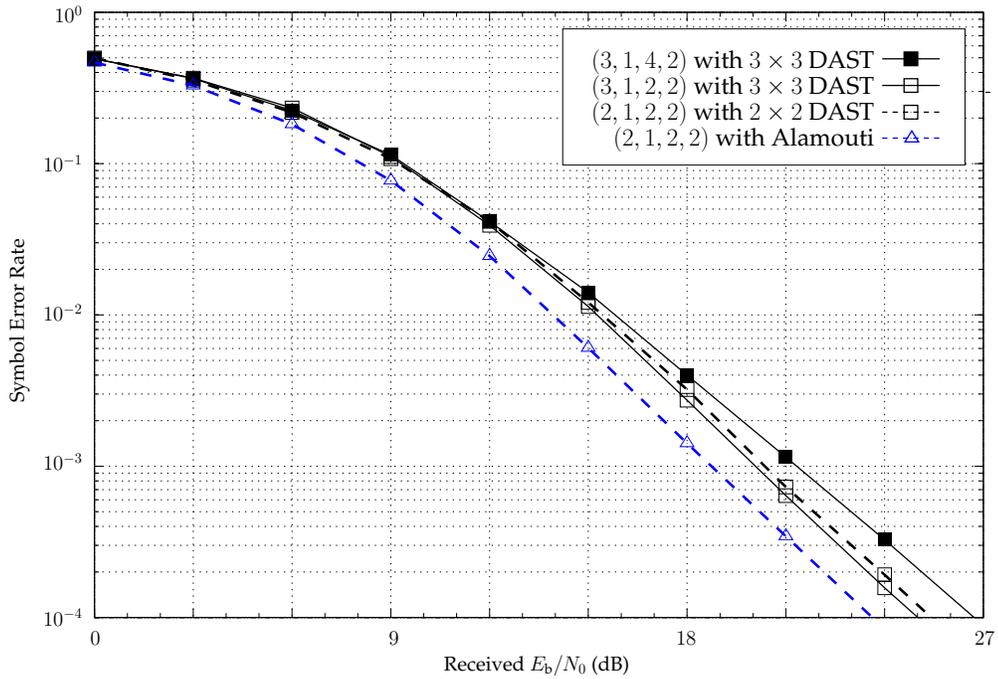,angle=0,width=0.85\textwidth}}
\caption{Vertical reduction~: target data rate $2$~bits per channel use in the outage performances or $4$-QAM
  constellation in the coded cases.}  
\end{center} 
\end{figure*}

The first example is to illustrate the impact of vertical reduction of
multihop channels, as shown in \Fig{fig:vertical_reduction}. In a
$(1,4,1)$ channel, the necessary antenna number $\bar{n}$ from
\Eq{eq:minimum_number} is $1$ and the minimal vertical form is thus
$(1,1,1)$. We observe that, with the same diversity order $1$, an
asymptotic power gain of $7$~dB is obtained by using only one relay
antennas out of four, if the AF scheme is used. The gain is due to the
fact that using more relaying antennas hardens of relayed noise. In
the $(3,1,4,2)$ channel, the necessary number of antennas $\bar{n}$
from \Eq{eq:minimum_number} is $2$. As shown in
\Fig{fig:vertical_reduction}, by restricting the number of relay
antennas to $2$, we have a $(3,1,2,2)$ channel and an asymptotic power
gain of $2$ dB is observed. We can further reduce the number of
transmit antennas to $2$ to get a $(2,1,2,2)$ channel. Unlike the
reduction of relay antennas, the reduction of transmit antennas does
not provide any gain because it does not impact the relayed noise. In
contrast, it degrades the performance since the first hop $(2,1)$ is
faded more seriously than the original first hop $(3,1)$.
Nevertheless, the $(2,1,2,2)$ channel is still better than the
$(3,1,4,2)$ channel and is only $0.7$ dB from the $(3,1,2,2)$ channel.
The coded performance of the $(3,1,4,2)$ channel is then
studied~\Fig{fig:coded_performance}. The diagonal algebraic
space-time~(DAST) code\footnote{Note that the DAST code is the
  diagonal version of the rate-one Perfect code proposed in
  \cite{Oggier_perfect}.}~\cite{Damen_DAST} can be used. As shown in
\Fig{fig:coded_performance}, with the DAST code, the symbol error rate
performances of in the $(3,1,4,2)$, $(3,1,2,2)$ and $(2,1,2,2)$
channels have exactly the same behavior as the outage performances of
the channels do~\Fig{fig:vertical_reduction}. Moreover, the reduction
in the number of transmit antenna allows us to use the Alamouti
code~\cite{Alamouti}~(the $(2,1,2,2)$ channel). As we can see in the
figure, the Alamouti code, besides the advantage of lower decoding
complexity, outperforms all the DAST codes. The potential benefits
from the vertical reduction are thus highlighted.

\begin{figure*}[!t]
\begin{center}
\subfigure[Outage probability]{
\label{fig:outage_aff_af}
\epsfig{figure=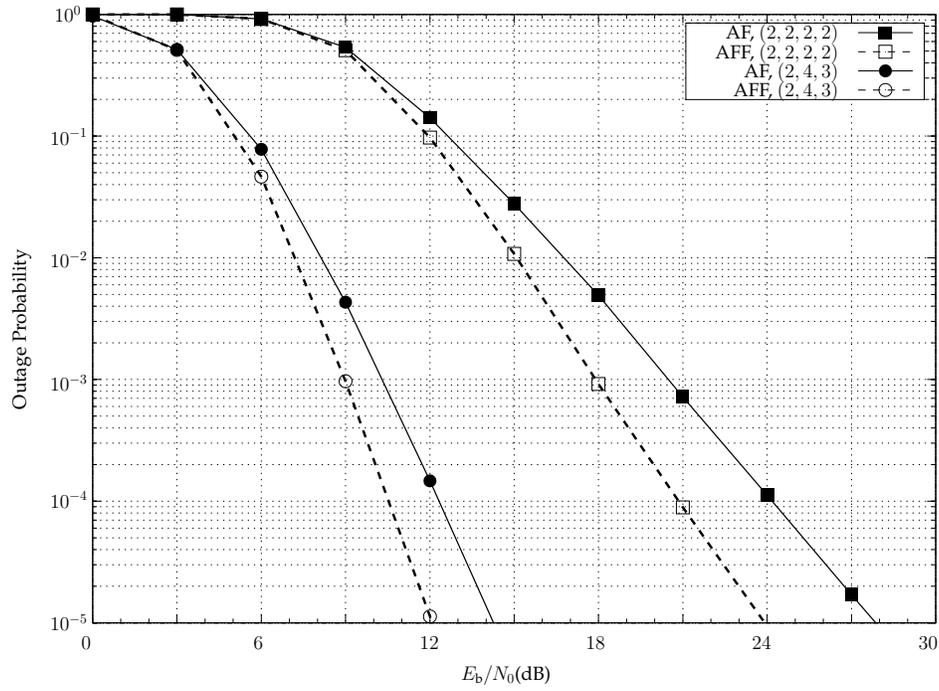,angle=0,width=0.75\textwidth}}\\
\subfigure[Symbol error rate]{
\label{fig:coded_aff_af}
\epsfig{figure=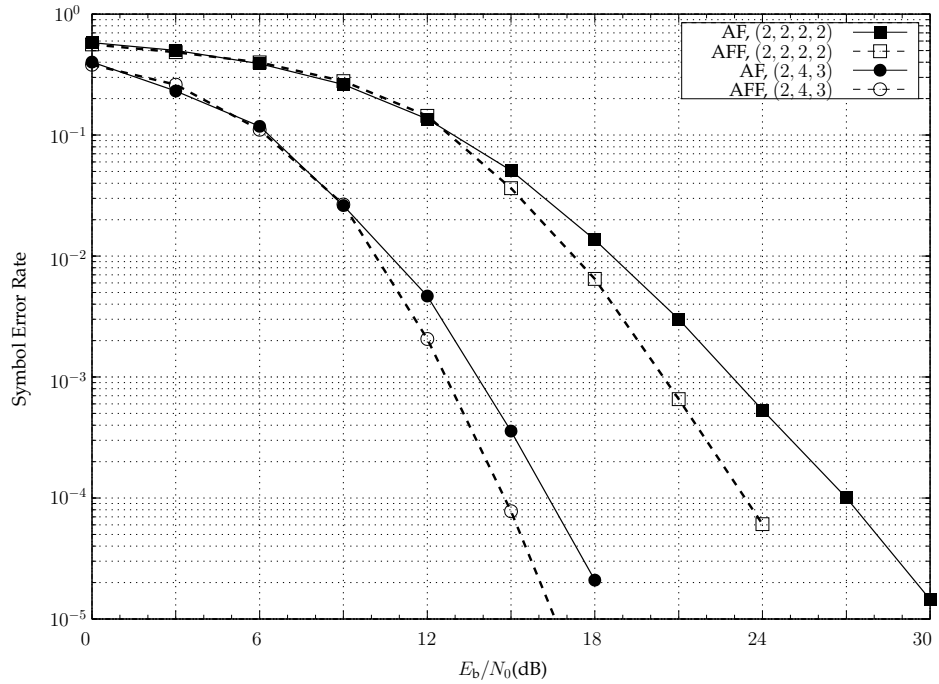,angle=0,width=0.75\textwidth}}
\caption{AF vs. AFF~: target data rate $4$~bits per channel use in the outage performances or $4$-QAM
  constellation in the coded cases.}  \label{fig:outage_multihop3}
\end{center} 
\end{figure*}

Then, we consider the parallel partition of two multihop channels~:
the $(2,2,2,2)$ and $(2,4,3)$ channels. The resulting AFF scheme is
compared to the AF scheme in terms of both the outage probability and
the symbol error rate. With the AFF scheme, we create respectively
four and two parallel sub-channels with two transmit antennas for the
$(2,2,2,2)$ and $(2,4,3)$ channels. Specifically, the AFF scheme for
the $(2,2,2,2)$ channel is based on a partition of four $(2,1,1,2)$
sub-channels and for the $(2,4,3)$ channel is a partition of two
$(2,2,3)$ sub-channels. As shown in \Fig{fig:outage_aff_af}, the
diversity order of the AFF scheme for the $(2,2,2,2)$~(respectively,
$(2,4,3)$ channel) is $4$~(respectively, $8$), as compared to that of
the AF scheme~($3$ and $6$, respectively). The coded performance is
also studied. We apply the construction provided by
Example~\ref{ex:code_construction} to get Perfect parallel STBCs for
two and four sub-channels. As we can observe in
\Fig{fig:coded_aff_af}, with the use of Perfect codes, the symbol
error rate performance has similar behaviors as the outage
performance.

\begin{figure*}[!t]
\begin{center}
  \includegraphics[angle=0,width=0.75\textwidth]{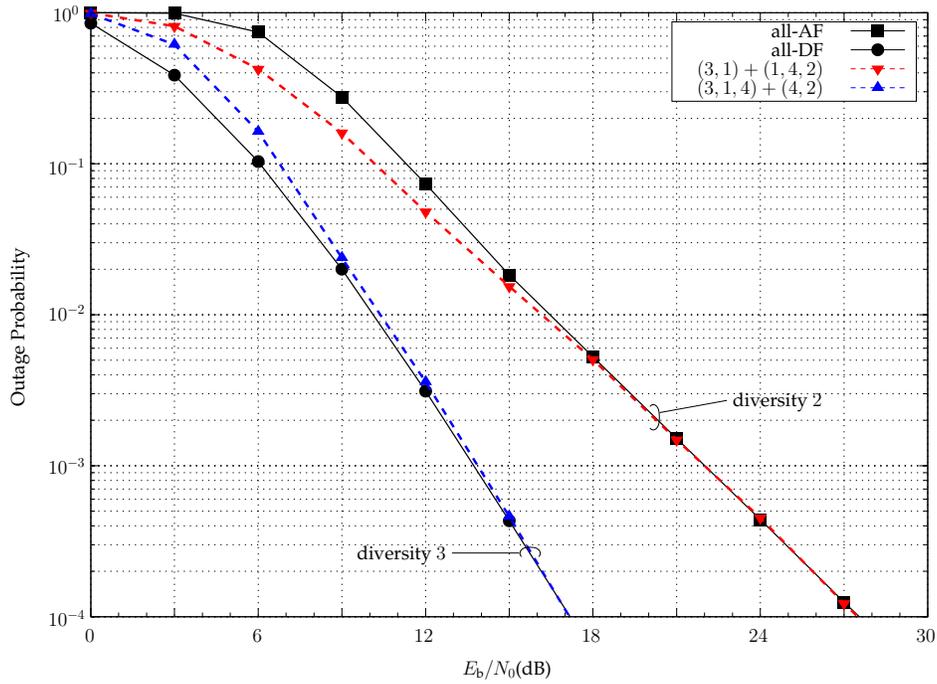}
  \caption{The $(3,1,4,2)$ multihop channel~: outage probability of
    the serial partition with various numbers of decoding clusters,
    target data rate $2$~bits per channel use.}
  \label{fig:outage_3142}
\end{center} \end{figure*}

The last example is a $(3,1,4,2)$ channel in the clustered case.
Through this example, we would like to address the impact of ``where
to decode'' on the end-to-end performance. The all-AF and all-DF
schemes correspond respectively to the case with no decoding relay
cluster and that with two decoding relay clusters. With one decoding
cluster, the choice is made between decoding at the first cluster and
decoding at the second one. As shown in \Fig{fig:outage_3142}, the
all-AF scheme has diversity order two and the all-DF scheme has
diversity order $3$ as analytically expected. With only one decoding
cluster, the diversity order is also predictable~: diversity two in
the single-antenna cluster and diversity 3 in the four-antenna
cluster. What is impressive in this example is that the two curves
with different choices of decoding cluster joins the all-AF and all-DF
curves respectively at high SNR. Therefore, only one decoding cluster
is enough to achieve good performance in this case. And the decoding
cluster should not be the single-antenna node.

\section{Conclusion}
\label{sec:conclusion}

The diversity of MIMO multihop relay channels has been investigated in
both the clustered and non-clustered cases. Our results showed that,
in both cases, the maximum diversity gain and the maximum multiplexing
gain of the multihop channel can be achieved. In the clustered case,
the optimal scheme is cooperative decode-and-forward that achieves the
upper bound on the diversity-multiplexing tradeoff of the channel. In
the non-clustered case, the key to achieve the maximum diversity is
space-time relay processing. Our approach is to introduce temporal
processing to the amplify-and-forward scheme by creating a parallel
channel in the time domain. We proposed a flip-and-forward that
achieves both the maximum diversity and multiplexing gain of an
arbitrary multihop channel in a distributed manner. We also showed
that the FF scheme can be easily extended to the multiuser case. With
its low relaying and signaling complexity, the FF scheme is suitable
for wireless \emph{ad hoc} networks. Approximately universal coding
schemes have been proposed for all the relaying strategies studied in
this work.

\newpage
\appendices

\section{Preliminaries}

The followings are some preliminary results that are essential to the
proofs.

\begin{lemma}[Calculation of diversity-multiplexing tradeoff]
  \label{lemma:cal-dmt}
  Consider a linear fading Gaussian channel defined by $\mH$ for which
  $\det\left(\Id+\SNR\mul\mH\transc{\mH}\right))$ is a function of
  $\mlambda$, a vector of positive random variables. Then, the DMT
  $d_\mH(r)$ of this channel can be calculated as
  \begin{equation*}
    d_\mH(r) = \inf_{\Ocal(\malpha, r)} \E(\malpha)
  \end{equation*}%
  where $\alpha_i\defeq-\log v_i/\log\SNR$ is the exponent of $v_i$,
  $\Ocal(\malpha, r)$ is the outage event set in terms of $\malpha$
  and $r$ in the high SNR regime, and $E(\malpha)$ is the exponential
  order of the pdf $p(\malpha)$, \ie,
  \begin{equation*}
    p(\malpha) \asympteq \SNR^{-E(\malpha)}.
  \end{equation*}%
\end{lemma}%
\begin{IEEEproof}
  This lemma can be justified by \Eq{eq:dmt-compact} using Laplace's
  method, as shown in \cite{Zheng_Tse}.
\end{IEEEproof}

\begin{definition}[Wishart Matrix]
  The $m\times m$ random matrix $\mW = \mH\transc{\mH}$ is a (central)
  complex Wishart matrix with $n$ degrees of freedom and covariance
  matrix $\mR$~(denoted as $\mW\sim\Wcal_m(n,\mR)$), if the
  columns of the $m\times n$ matrix $\mH$ are zero-mean independent
  complex Gaussian vectors with covariance matrix $\mR$.
\end{definition}

\begin{lemma}\label{lemma:eq-wishart}
  The joint pdf of the eigenvalues of $\mW \defeq \mH\transc{\mH}
  \sim\Wcal_m(n,\mR_{m\times m})$ is identical to that of any
  $\mW' \sim\Wcal_{m'}(n,\diag(\mu_1,\ldots,\mu_{m'}))$ if
  $\mu_1\geq\ldots\geq\mu_{m'}>\mu_{m'+1}=\ldots=\mu_m = 0$ are the
  eigenvalues of $\mR_{m\times m}$.
\end{lemma}
\begin{IEEEproof}
  Apply the eigenvalue decomposition on $\mR$ and the result is
  immediate using the unitary invariance property~\cite{Edelman} of
  Wishart matrices.
\end{IEEEproof}

\begin{lemma}[\!\!\cite{James,Gao_Smith,Simon,Tulino_Verdu}]\label{lemma:Wishart}
  Let $\mW$ be a central complex Wishart matrix
  $\mW\sim\Wcal_m(n,\mR)$, where the eigenvalues of $\mR$ are
  distinct\footnote{In the particular case where some eigenvalues of
    $\mR$ are identical, we apply the l'Hospital rule to the pdf
    obtained, as shown in \cite{Simon}.} and their ordered values are
  $\mu_1>\ldots>\mu_m>0$. Let $\lambda_1>\ldots>\lambda_q>0$ be the
  ordered positive eigenvalues of $\mW$ with $q\defeq\min\{m,n\}$. The
  joint pdf of $\mlambda$ conditioned on $\mmu$ is
\begin{subnumcases}{p(\mlambda|\mmu)=} 
  K_{m,n} {\Det(\mOmega_1)} \prod_{i=1}^m
  \mu_i^{m-n-1} \lambda_i^{n-m} \prod_{i<j}^m
  \frac{\lambda_i-\lambda_j}{\mu_i-\mu_j}, & \text{if $n\geq m$,}
  \label{eq:Wishart:n>m}
  \\
  G_{m,n} {\Det(\mOmega_2)} \prod_{i<j}^m \frac{1}{(\mu_i-\mu_j)}
  \prod_{i<j}^n (\lambda_i-\lambda_j), &\text{if $n<m$,}
  \label{eq:Wishart:n<m}
\end{subnumcases}
where $K_{m,n}$ and $G_{m,n}$ are normalization factors; $\Det(\cdot)$
denotes the absolute value of the determinant $\det(\cdot)$;
$\mOmega_1 \defeq \left[e^{-\lambda_j/\mu_i}\right]_{i,j=1}^m$ and
  \begin{equation}
    \label{eq:def-Xi2}
    \mOmega_2 \defeq \matrix{1 & \mu_1 & \cdots & \mu_1^{m-n-1} & \mu_1^{m-n-1}e^{-\frac{\lambda_1}{\mu_1}} & \cdots & \mu_1^{m-n-1}e^{-\frac{\lambda_n}{\mu_1}} \\
      \vdots & \vdots & \ddots &\vdots& \vdots& \ddots & \vdots \\
        1 & \mu_m & \cdots & \mu_m^{m-n-1} & \mu_m^{m-n-1}e^{-\frac{\lambda_1}{\mu_m}} & \cdots & \mu_m^{m-n-1}e^{-\frac{\lambda_n}{\mu_m}}
}.
  \end{equation}%
  In the non-correlated case with $\mR=\Id$, the joint pdf is
  \begin{equation}
    \label{eq:Wishart:Id}
    P_{m,n}  e^{-\sum_i \lambda_i} \prod_{i=1}^q \lambda_i^{\Abs{m-n}} \prod_{i<j}^q (\lambda_i-\lambda_j)^2.    
  \end{equation}
\end{lemma}
Now, let us define the \emph{eigen-exponents} $$\alpha_i\defeq-\log
\lambda_i/\log\SNR,\ i=1,\ldots,q,\ \text{and}\ 
\beta_i\defeq-\log\mu_i/\log\SNR,\ i=1,\ldots,m.$$

\begin{lemma}\label{lemma:Det}
  \eqncasesasymptlabel{\Det(\mOmega_1)}{\SNR^{-\E_{\mOmega_1}(\malpha,\mbeta)}}{for
    $(\malpha,\mbeta)\in\Rcal^{(1)}$}{\SNR^{-\infty}}{otherwise,}{eq:detexp}
where 
\begin{equation}
  \label{eq:EmOmega1}
  \E_{\mOmega_1}(\malpha,\mbeta) \defeq  \sum_{j=1}^m \sum_{i<j} (\alpha_i-\beta_j)^+,
\end{equation}
and
\begin{equation}
  \label{eq:R1}
  \Rcal^{(1)} \defeq \left\{\alpha_1\leq\ldots\leq\alpha_m,\ \beta_1\leq\ldots\leq\beta_m,\ \text{and}\ \beta_i\leq\alpha_i,\ \text{for}\ i=1,\ldots,m \right\}.
\end{equation}%
\end{lemma}

\begin{IEEEproof}
Please refer to \cite{SY_JCB_coop} for details.
\end{IEEEproof}

\begin{lemma}\label{lemma:DetXi}
  \eqncasesasymptlabel{\Det\left(\mOmega_2\right)}{\SNR^{-\E_{\mOmega_2}(\malpha,\mbeta)}}{for
    $(\malpha,\mbeta)\in
    \Rcal^{(2)}$}{\SNR^{-\infty}}{otherwise,}{eq:lemma2} where
  \begin{equation}
    \label{eq:EmOmega2}
    \E_{\mOmega_2}(\malpha,\mbeta) \defeq \sum_{i=1}^n
        (m-n-1)\beta_i + \sum_{i=n+1}^m (m-i)\beta_i +
        \sum_{j=1}^n\sum_{i<j}\pstv{(\alpha_i-\beta_j)} +
        \sum_{j=n+1}^m\sum_{i=1}^n\pstv{(\alpha_i-\beta_j)}
  \end{equation}
and
  \begin{equation}
    \label{eq:R2}
    \Rcal^{(2)} \defeq \left\{\alpha_1\leq\ldots\leq\alpha_n,\ \beta_1\leq\ldots\leq\beta_m,\ \text{and}\ \beta_i\leq\alpha_i,\ \text{for}\ i=1,\ldots,n \right\}.
  \end{equation}%
\end{lemma}

\begin{IEEEproof}
First, we have 
\begin{equation}\label{eq:lemmas:tmp1}
  \Det{(\mOmega_2)} = \prod_{i=1}^m \mu_i^{m-n-1} 
  \Det \matrix{
    \mu_1^{-(m-n-1)} &\cdots& 1 & e^{-\lambda_1/\mu_1}&\cdots&e^{-\lambda_n/\mu_1}\\
    \vdots & \ddots & \vdots& \vdots& \ddots & \vdots \\
    \mu_m^{-(m-n-1)} &\cdots& 1 & e^{-\lambda_1/\mu_m}&\cdots&e^{-\lambda_n/\mu_m}\\
  }.
\end{equation}
Then, let us denote the determinant in the RHS of \Eq{eq:lemmas:tmp1}
as $D$ and we rewrite it as
\begin{align}
  D                                 %
  &= \Det \matrix{
    d_{1,m}^{(m-n-1)} &\cdots& 0 & e^{-\lambda_1/\mu_1}-e^{-\lambda_1/\mu_m}&\cdots&e^{-\lambda_n/\mu_1}-e^{-\lambda_n/\mu_m}\\
    \vdots & \ddots & \vdots& \vdots& \ddots & \vdots \\
    d_{m-1,m}^{(m-n-1)} &\cdots& 0 & e^{-\lambda_1/\mu_{m-1}}-e^{-\lambda_1/\mu_{m}}&\cdots&e^{-\lambda_n/\mu_{m-1}}-e^{-\lambda_n/\mu_{m}}\\
    \mu_m^{-(m-n-1)} &\cdots& 1 & e^{-\lambda_1/\mu_m}&\cdots&e^{-\lambda_n/\mu_m}\\
  } \label{eq:lemmas:tmp3} \\
  &\asympteq \Det \matrix{
    d_{1,m}^{(m-n-1)} &\cdots& d_{1,m}^{(1)} & e^{-\lambda_1/\mu_1}&\cdots&e^{-\lambda_n/\mu_1}\\
    \vdots & \ddots & \vdots& \vdots& \ddots & \vdots \\
    d_{m-1,m}^{(m-n-1)} &\cdots& d_{m-1,m}^{(1)} & e^{-\lambda_1/\mu_{m-1}}&\cdots&e^{-\lambda_n/\mu_{m-1}}\\
  } \prod_{i=1}^n \left(1-e^{-\lambda_i/\mu_m}\right)
  \label{eq:lemmas:tmp2}
\end{align}%
where $d_{i,j}^{(k)}\defeq \mu_i^{-k} - \mu_j^{-k}$ and the product
term in \Eq{eq:lemmas:tmp2} is obtained since
$1-e^{-(\lambda_i/\mu_m-\lambda_i/\mu_j)}\asympteq
1-e^{-\lambda_i/\mu_m}$ for all $j<m$. Let us denote the determinant
in \Eq{eq:lemmas:tmp2} as $D_m$. Then, by multiplying the first column
in $D_m$ with $\mu_m^{m-n-1}$ and noting that $\mu_m^{m-n-1}
d_{i,m}^{(m-n-1)}=1-\left({\mu_m}/{\mu_i}\right)^{m-n-1}\approx
1$, the first column of $D_m$ becomes all $1$. Now, by eliminating the
first $m-2$ ``$1$''s of the first column by subtracting all rows by
the last row as in \Eq{eq:lemmas:tmp3} and \Eq{eq:lemmas:tmp2}, we
have $\mu_m^{m-n-1} D_m\asympteq \prod_{i=1}^n
\left(1-e^{-\lambda_i/\mu_m}\right) D_{m-1}$. By continuing reducing
the dimension, we get
\begin{equation*}
  \begin{split}
    \Det(\mOmega_2) &\asympteq  \Det\left[e^{-\lambda_j/\mu_i}\right]_{i,j=1}^n \prod_{i=1}^{n+1}\mu_i^{m-n-1}\prod_{i=n+2}^m \mu_i^{m-i}\\
    &\quad \cdot\prod_{i=1}^n\prod_{j=n+1}^m
    \left(1-e^{-\lambda_i/\mu_j}\right)
  \end{split}
\end{equation*}
from which we prove the lemma, by applying \Eq{eq:detexp}.    
\end{IEEEproof}
With the preceding lemmas, we have the following lemma that provides
the asymptotical pdf of $\malpha$ conditioned on $\mbeta$ in the high
SNR regime.
\begin{lemma}\label{lemma:pdfRayleighCond}
  \eqncasesasympt{p(\malpha|\mbeta)}{\SNR^{-\E(\malpha|\mbeta)}}{for
    $(\malpha,\mbeta)\in\Rcal_{\malpha|\mbeta}$,}{\SNR^{-\infty}}{otherwise,}
where 
\begin{equation}
  \label{eq:expcond}
  \E(\malpha|\mbeta) \defeq \sum_{i=1}^q (n+1-i) \alpha_i + \sum_{i=1}^q (i-n-1)\beta_i
                 + \sum_{j=1}^q\sum_{i<j} (\alpha_i-\beta_j)^+ 
                 + \sum_{j=q+1}^m\sum_{i=1}^q (\alpha_i-\beta_j)^+, 
\end{equation}
and
\begin{equation}
  \label{eq:feasibility}
  \Rcal_{\malpha|\mbeta} \defeq \left\{\alpha_1\leq\ldots\leq\alpha_q,\ \beta_1\leq\ldots\leq\beta_m,\ \text{and}\ \beta_i\leq\alpha_i,\ \text{for}\ i=1,\ldots,q \right\}.
\end{equation}%
\end{lemma}
\begin{IEEEproof}
  Let us replace $\Det(\mOmega_1)$ and $\Det(\mOmega_2)$ in
  \Eq{eq:Wishart:n>m} and \Eq{eq:Wishart:n<m} using the results of
  Lemma~\ref{lemma:Det} and Lemma~\ref{lemma:DetXi}.  Then, by
  applying variable changes as done in \cite{Zheng_Tse},
  \Eq{eq:expcond} can be obtained after some elementary manipulations.
\end{IEEEproof}
When $\mR=\Id$, \ie, $\mu_1=\ldots=\mu_m=1$, the joint pdf of
$\malpha$ is found in \cite{Zheng_Tse} as shown in the following
lemma.
\begin{lemma}\label{lemma:pdfRayleigh}
  \eqncasesasymptlabel{p(\malpha)}{\SNR^{-\sum_{i=1}^q
      (m+n+1-2i)\alpha_i}}{for
    $\malpha\in\Rcal_{\malpha}$,}{\SNR^{-\infty}}{otherwise,}{eq:pdfRayleigh}
with $\Rcal_{\malpha}\defeq\left\{0\leq\alpha_1\leq\ldots\leq\alpha_q\right\}$.
\end{lemma}
This lemma can be justified either by using \Eq{eq:Wishart:Id} or by
setting $\beta_i=0,\ \forall\,i$ in \Eq{eq:expcond}.

\begin{lemma}[\cite{SY_JCB_ds}]\label{lemma:invariance-asymp}
  Let $\mM$ be any $m\times n$ random matrix and $\mT$ be any $m\times
  m$ non-singular matrix whose singular values satisfy
  $\sigma_{\min}(\mT)\asympteq\sigma_{\max}(\mT)\asympteq\SNR^0$.
  Define $q\defeq\min\{m,n\}$ and $\mbs{\tilde{M}} \defeq \mT\mM$. Let
  $\sigma_1(\mM)\geq\ldots\geq\sigma_q(\mM)$ and
  $\sigma_1(\mbs{\tilde{M}})\geq\ldots\geq\sigma_q(\mbs{\tilde{M}})$
  be the ordered singular values of $\mM$ and $\mbs{\tilde{M}}$,
  Then, we have
  \begin{equation*}
    \sigma_i(\mbs{\tilde{M}}) \asympteq \sigma_i(\mM),\quad\forall i.
  \end{equation*}%
\end{lemma}

\section{Proof of Theorem~\ref{thm:dmt-rp}}\label{sec:proof-thm-dmt-rp}

\begin{proposition}\label{prop:asympt-pdf}
  Let us denote the non-zero ordered eigenvalues of $\RP\transc{\RP}$
  by $\lambda_1\geq\cdots\geq\lambda_{\nmin}>0$ with
  ${\nmin}\defeq\D\min_{i=0,\ldots,N} n_i$.  Then, the joint pdf of
  the eigen-exponents $\malpha$ satisfies
  \eqncasesasymptlabel{p(\malpha)}{\SNR^{-\E(\malpha)}}{for
    $0\leq\alpha_1\leq\ldots\leq\alpha_{\nmin}$,}{\SNR^{-\infty}}{otherwise,}{eq:p-malpha}
  where
  \begin{equation}
    \label{eq:Ea}
    \E(\malpha) \defeq \sum_{i=1}^{\nmin}c_i\alpha_i
  \end{equation}%
  with $c_i$'s defined by \Eq{eq:ci}. 
\end{proposition}
From Lemma~\ref{lemma:cal-dmt}, we can derive the DMT with the following
optimization problem
$$
d(r) = \min_{\malpha \in \Ocal_0(r)} \sum_i c_i\mul \alpha_i$$
with $\Ocal_0(r) \defeq \{\sum_i (1-\alpha_i)^+ \leq r\}$ being the
outage region. Note that $c_i$ is decreasing and $\alpha_i$ is
increasing with respect to $i$. Then, the proof of
Theorem~\ref{thm:dmt-rp} is immediate.

Now, what remains is the proof of Proposition~\ref{prop:asympt-pdf}.
The following lemma will be needed in the proof.
\begin{lemma}~\label{lemma:ci}
  Let $\Ical_k\defeq[\,p_k,p_{k-1}]$, $k=1,\ldots,N$, be $N$
  consecutively joint intervals with $p_N\defeq -\infty$, $p_0 \defeq
  \ntilde_0$, and
  \begin{equation}\label{eq:pk}
    p_k \defeq
      \sum_{l=0}^k \tilde{n}_l - k \tilde{n}_{k+1}\quad k=1,\ldots,N-1.
  \end{equation}%
  Then, we have
  \begin{equation}\label{eq:ci3}
    c_i = 1-i + \left\lfloor\frac{\sum_{l=0}^{k}\ntilde_l - i}{k} \right\rfloor, \quad \text{for}\ i\in\Ical_{k}.
  \end{equation}%
\end{lemma}
\begin{IEEEproof}
  $c_i$ defined by \Eq{eq:ci} is the minimum of $N$ sequences
  corresponding to the $N$ values of $k$. It is enough to show that
  each of the $N$ sequences dominates in a consecutive manner. We omit
  the details here.
\end{IEEEproof}

\subsection{Sketch of the Proof of Proposition~\ref{prop:asympt-pdf}}
\label{sec:sketch-proof}
The proof will be by induction on $N$. From
lemma~\ref{lemma:pdfRayleigh}, the proposition is trivial for $N=1$.
Suppose the proposition holds for some $N$ and
$\RP\defeq\mH_1\cdots\mH_N$, we would like to show that it is also
true for $N+1$ and $\RP' \defeq \mH_{1}\cdots\mH_{N+1}$. For
simplicity, the ``primed'' notations~(\eg, $\malpha'$, $\mn'$,
$\mnt'$, $\mc'$, $\nminp$, etc.) will be used for the respective
parameters of $\RP'$.  Note that
$\RP'\transc{(\RP')}\sim\Wcal_{n_0}(n_{N+1},\RP\transc{\RP})$ for a
given $\RP$, since $\RP' = \RP \mH_{N+1}$.  According to
lemma~\ref{lemma:eq-wishart}, the pdf of the eigenvalues $\mlambda'$
of $\RP'\transc{(\RP')}$ is identical to that of
$\Wcal_{\nmin}(n_{N+1},\diag(\mlambda))$.  Hence, the pdf of
$\malpha'$ can be obtained as the marginal pdf of $(\malpha',\malpha)$
\begin{align}
  p(\malpha') &= \int_{\RR^{{\nmin}}} p(\malpha',\malpha) \d \malpha \nonumber\\
  &= \int_{\RR^{{\nmin}}} p(\malpha'|\malpha) p(\malpha) \d \malpha \nonumber\\
  &\asympteq \int_{\Rcal} \SNR^{-\E(\malpha'|\malpha)}
  \SNR^{-\E(\malpha)}
  \d \malpha \label{eq:tmp31}\\
  &\asympteq \SNR^{-\Ec(\malpha')}, \label{eq:tmp32}
\end{align}%
where \Eq{eq:tmp31} comes from lemma~\ref{lemma:pdfRayleighCond} and
our assumption that \Eq{eq:p-malpha} holds for $N$, with
\begin{align}
  \Rcal
  &\defeq \Rcal_{\malpha'|\malpha}\cap\Rcal_{\malpha} \nonumber\\
  &=\left\{0\leq\alpha'_1\leq\ldots\leq\alpha'_{\nminp},\ 
    0\leq\alpha_1\leq\ldots\leq\alpha_{\nmin},\ \text{and}\ 
    \alpha_i\leq\alpha'_i,\ \text{for}\ i=1,\ldots,\nminp \right\} \label{eq:feasibleRegion}
\end{align}%
being the feasible region; the exponent $\Ec(\malpha')$ in
\Eq{eq:tmp32} is
\begin{align}
  \Ec(\malpha') &= \min_{\malpha\in\Rcal} \E(\malpha',\malpha)
  \label{eq:minprob}
\end{align}%
with $\E(\malpha',\malpha) \defeq \E(\malpha'|\malpha) +
\E(\malpha)$. From \Eq{eq:expcond} and \Eq{eq:Ea},
\begin{multline}
  \E(\malpha',\malpha) = \sum_{i=1}^{\nminp} (n_{N+1}-i+1) \alpha_i'
  + \sum_{j=1}^{\nminp} \left( (j-1-n_{N+1}+c_j)\alpha_j + \sum_{i<j}\pstv{(\alpha_i'-\alpha_j)} \right) \\
  + \sum_{j=\nminp+1}^{{\nmin}} \left( c_j \alpha_j +
    \sum_{i=1}^{\nminp} \pstv{(\alpha_i'-\alpha_j)} \right).
  \label{eq:jointexp}
\end{multline}%
It remains to show that $\Ec(\malpha')=\E'(\malpha')\defeq \sum_i c_i
\alpha'_i$ with
\begin{equation}\label{eq:ci2}
  c'_i \defeq 1-i + \min_{k=1,\ldots,N+1} \left\lfloor\frac{\sum_{l=0}^{k}\ntilde'_l - i}{k} \right\rfloor,\quad i=1,\ldots,\nminp
\end{equation}%
by solving the optimization problem \Eq{eq:minprob}, which is
accomplished in the rest of the section.


\subsection{Solving the Optimization Problem}
We need to distinguish three cases, according to how the value of
$n_{N+1}$ affects the ordered dimension $\mnt'$.
\begin{figure*}
  \subfigure[Case~1]{
  \epsfig{figure=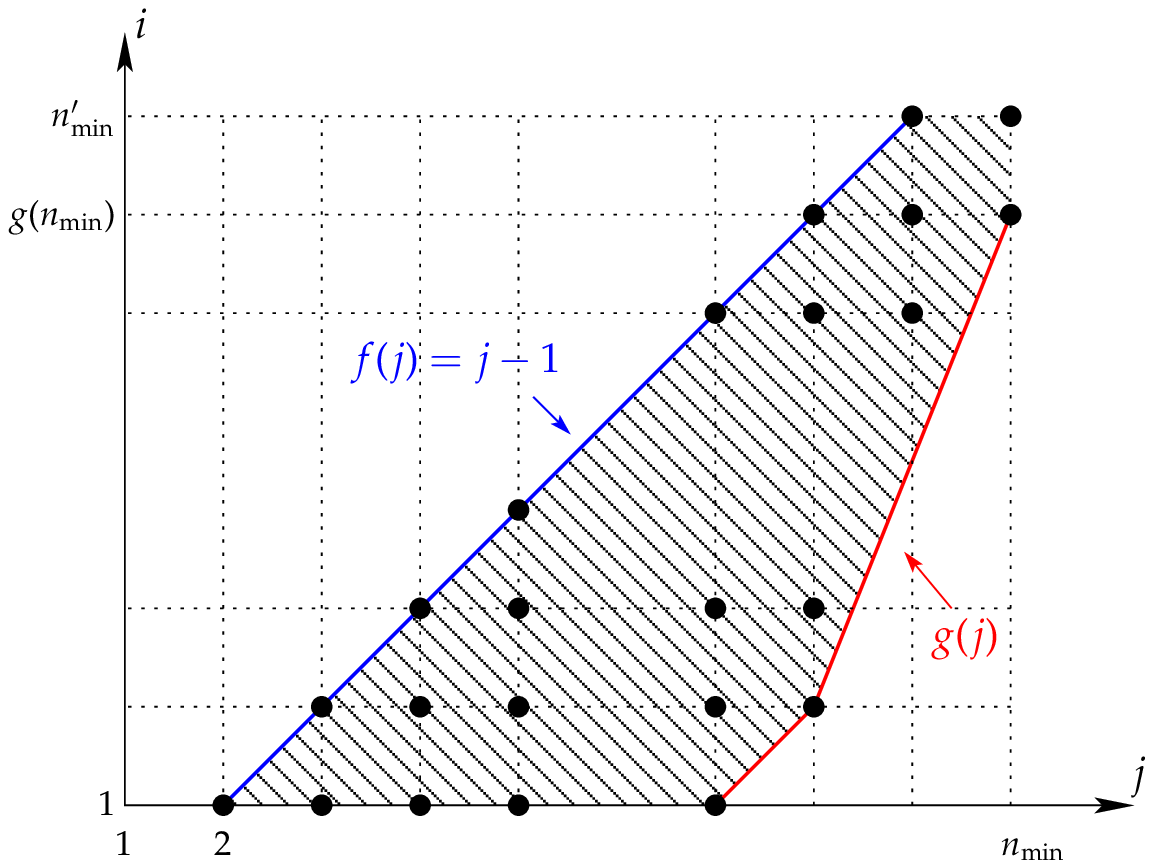,height=0.25\textwidth}
  \label{fig:find-ci1}    }
  \subfigure[Case~2]{
  \epsfig{figure=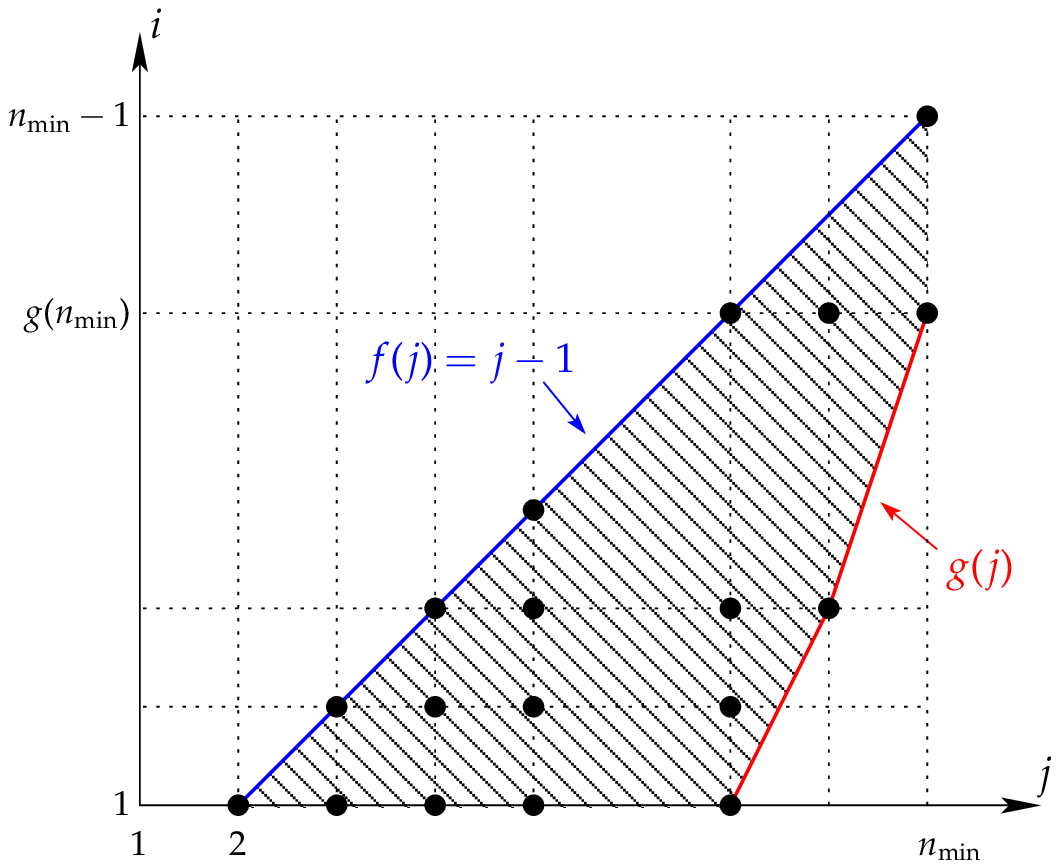,height=0.25\textwidth}
  \label{fig:find-ci2}  }
  \subfigure[Case~3]{
  \epsfig{figure=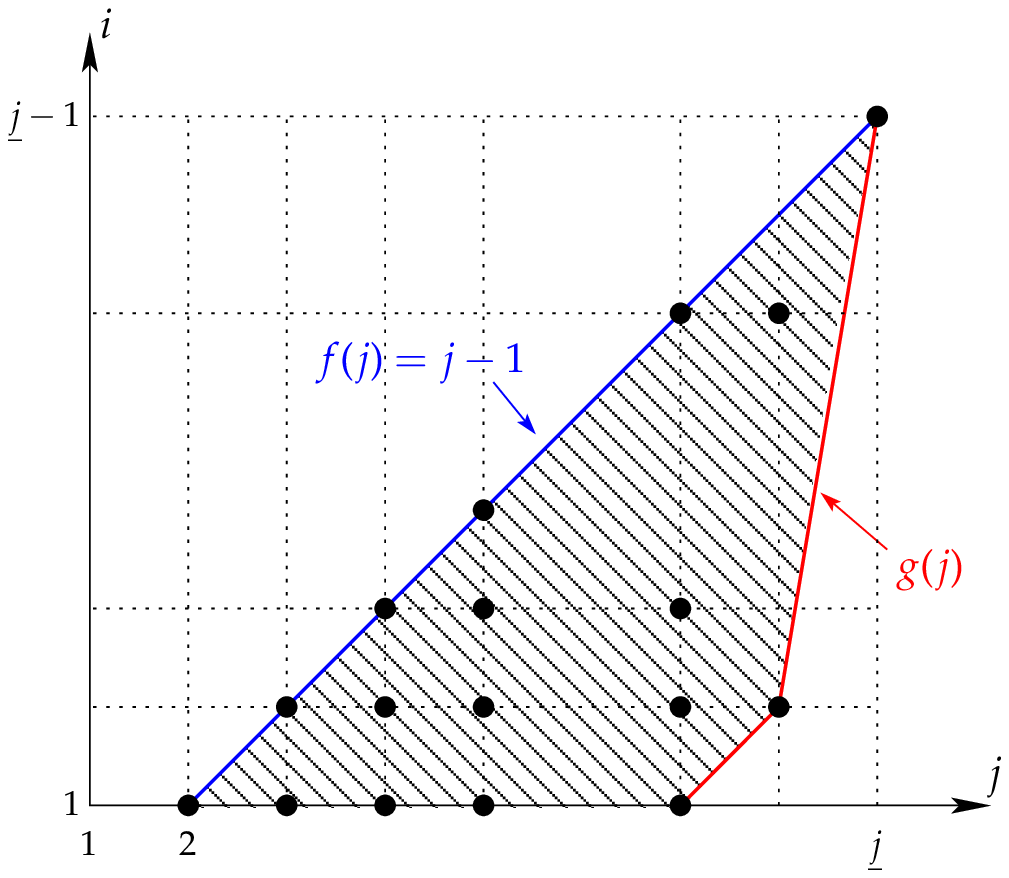,height=0.25\textwidth}
  \label{fig:find-ci3}  }
\caption{For each $j$, the black dots represent the $\alpha'$'s that are freed 
  by $\alpha_j$. Therefore, we can get the total number of freed
  $\alpha'_i$ by counting the black dots in row $i$. More precisely,
  there are $\left\lfloor{g^{-1}(i)}\right\rfloor - \left\lceil{
      f^{-1}(i) }\right\rceil + 1=\left\lfloor{g^{-1}(i)}\right\rfloor
  - i$ black dots for $i\leq g({\nmin})$, and ${\nmin} - \left\lceil{ f^{-1}(i)
    }\right\rceil + 1={\nmin} - i$ black dots for $i > g({\nmin})$.}
\end{figure*}%

\subsubsection{Case 1~[$n_{N+1}<\ntilde_0$]}
\label{sec:n_n+1in1-tilden_0}

In this case, we have $\nminp=\ntilde'_0=n_{N+1}$.  Minimization of
$\E(\malpha,\malpha')$ of \Eq{eq:jointexp} with respect to $\malpha$
can be decomposed into ${\nmin}$ minimizations with respect
to~$\alpha_1,\ldots,\alpha_{{\nmin}}$ successively, \ie,
$\min_{\malpha} = \min_{\alpha_{\nmin}}\cdots\min_{\alpha_1}$.  We
start with $\alpha_1$. From \Eq{eq:feasibility}, the feasible region
of $\alpha_1$ is $0\leq\alpha_1\leq\alpha'_1$.  Since the only
$\alpha_1$-related term in \Eq{eq:jointexp} is $(c_1-n_{N+1})\alpha_1$
and $c_1-n_{N+1}>0$ for $n_{N+1}<\ntilde_0$, we have $\alpha_1^* = 0$.
Now, suppose that the minimization with respect
to~$\alpha_1,\ldots,\alpha_{j-1}$ is done and that we would like to
minimize with respect to~$\alpha_j$. For $\alpha_j$, $j\leq \nminp$,
we set the initial region as
\begin{equation*}
  0\leq\alpha'_1\leq\cdots\leq\alpha'_{j-1}\leq\alpha_j\leq\alpha'_j
\end{equation*}%
in which we have $\sum_{i<j}\pstv{(\alpha'_i-\alpha_j)}=0$. The
feasibility conditions in \Eq{eq:feasibleRegion} require that
$\alpha_j$ must not go right across $\alpha'_j$. The only choice is
therefore to go to the left. Each time $\alpha_j$ goes across an
$\alpha'_i$ from the right to the left, $(\alpha'_i-\alpha_j)^+$
increases by $\alpha'_i-\alpha_j$, which increases the coefficient of
$\alpha'_i$ by $1$ and decreases the coefficient of $\alpha_j$ by $1$.
It can be shown that, to minimize the value of $\E(\malpha,\malpha')$
with respect to~$\alpha_j$, $\alpha_j$ is allowed to cross $\alpha'_i$ only when
the current coefficient of $\alpha_j$ in \Eq{eq:jointexp} is
positive.\footnote{When the coefficient of $\alpha_i$ in
  \Eq{eq:jointexp} is positive, decreasing $\alpha_i$ decreases
  $\E(\malpha,\malpha')$.} So, $\alpha_j$ stops moving only in the
following two cases~: 1) it hits the left extreme, $0$; and 2) its
coefficient achieves $0$ when it is in the interval
$[\alpha'_k,\alpha'_{k+1}]$ for some $k<j$. Either case,
$\alpha_j$-related terms are gone and what remain are the
$\alpha'_i$'s ``freed'' by $\alpha_j$ from $\sum_{i<j}
\pstv{(\alpha'_i-\alpha_j)}$. Same reasoning applies to $\alpha_j$ for
$j> \nminp$, except that the initial region is set to
$0\leq\alpha'_1\leq\cdots\leq\alpha'_{\nminp}\leq\alpha_j$.

Therefore, the optimization problem can be solved by counting the
total number of freed $\alpha'_i$'s. As shown in \Fig{fig:find-ci1},
when $j$ is small, the initial coefficient of $\alpha_j$ is large and
thus $\alpha_j$ can free out $\alpha'_{j-1},\ldots, \alpha'_{1}$. We
have $\alpha_j^*=0$, which corresponds to the first stopping
condition. For large $j$, the initial coefficient of $\alpha_j$ is not
large enough and only $\alpha'_{j-1},\ldots,\alpha'_{g(j)}$ is freed,
which corresponds to the second stopping condition. With the above
reasoning, we can get $g(j)$
\eqncaseslabel{g(j)}{j-1-(j-1-n_{N+1}+c_j)+1}{$j\leq
  \nminp$,}{n_{N+1}-c_j+1}{$j>\nminp$.}{eq:tmp48} From \Eq{eq:tmp48}
and \Eq{eq:ci}, we get
\begin{equation}
  \label{eq:gi}
  g(j) = n_{N+1} - \min_{k=1,\ldots,N} \left\lfloor\frac{\sum_{l=0}^{k}\ntilde_l - (k+1)j}{k} \right\rfloor,
\end{equation}
and 
\begin{align}
  \left\lfloor g^{-1}(i)\right\rfloor &= \min_{k=1,\ldots,N}
  \left\lfloor\frac{\sum_{l=0}^{k}\ntilde_l - k(n_{N+1}-i)}{k+1}
  \right\rfloor \label{eq:gi_inv1}.
\end{align}%
Now, $\Ec(\malpha')$ can be obtained\footnote{In the above
  minimization procedure, we ignored the feasibility condition
  $\alpha_j\geq\alpha_k,\ \forall\,j>k$. A more careful analysis
  reveals that it is always satisfied with the described procedure.}
from \Fig{fig:find-ci1}
\begin{align}
  \Ec(\malpha') &= \sum_{i=1}^{\nminp} (n_{N+1}-i+1) \alpha'_i +
  \sum_{i=1}^{g({\nmin})} (\left\lfloor
    g^{-1}(i)\right\rfloor-i)\alpha'_i +
  \sum_{i=g({\nmin})+1}^{\nminp}
  (\nmin-i)\alpha'_i \nonumber\\
  &= \sum_{i=1}^{g({\nmin})} \left( 1-2i+ n_{N+1} + \left\lfloor
      g^{-1}(i)\right\rfloor \right) \alpha'_i +
  \sum_{i=g({\nmin})+1}^{\nminp} \left( 1-2i+
    n_{N+1} + \nmin \right) \alpha'_i \nonumber\\
  &= \sum_{i=1}^{g({\nmin})} \left( 1 - i + \min_{k=2,\ldots,N+1}
    \left\lfloor\frac{\sum_{l=0}^{k}\ntilde'_l - i}{k}\right\rfloor
  \right) \alpha'_i +
  \sum_{i=g({\nmin})+1}^{\nminp} \left(1-2i+ n_{N+1} + \nmin \right) \alpha'_i\label{eq:tmp89}\\
  &= \sum_{i=1}^{\nminp} \left( 1-i + \min_{k=1,\ldots,N+1}
    \left\lfloor\frac{\sum_{l=0}^{k}\ntilde'_l - i}{k}\right\rfloor \right) \alpha'_i \label{eq:tmp54}\\
  &= \E'(\malpha'),
\end{align}%
where \Eq{eq:tmp89} is from \Eq{eq:gi_inv1} and the fact that
$\ntilde'_0=n_{N+1}$, $\ntilde'_l=\ntilde_{l-1}$, $l=1,\ldots,N+1$; \Eq{eq:tmp54}
can be derived from lemma~\ref{lemma:ci}, since $p'_1 =
n_{N+1}+\ntilde_0-\ntilde_1 = g({\nmin})$ and therefore the term
$\min_k$ in \Eq{eq:tmp54} is dominated by $k\geq2$ for $i\leq g({\nmin})$
and by $k=1$ for $i>g({\nmin})$, corresponding to the two terms in
\Eq{eq:tmp89}, respectively.

\subsubsection{Case 2~[$n_{N+1}\in[\ntilde_0,\ntilde_1)$]}
\label{sec:n_n+1in1-tilden_1}
In this case, we have $\nminp={\nmin}$ and $\ntilde'_1 = n_{N+1}$. From
\Eq{eq:jointexp},
\begin{multline}
  \E(\malpha',\malpha) = \sum_{i=1}^{\nminp} (n_{N+1}-i+1) \alpha_i' +
  \sum_{j=1}^{\nminp} \left( (j-1-n_{N+1}+c_j)\alpha_j + \sum_{i<j}
  \pstv{(\alpha_i'-\alpha_j)}\right). \label{eq:jointexp2}
\end{multline}
Since $j-1-n_{N+1}+c_j>0$, $\forall\,j\leq {\nminp}$, the minimization
of $\E(\malpha',\malpha)$ with respect to~$\malpha$ is in exactly the
same manner as in the previous case. Therefore, $\Ec(\malpha')$ can be
obtained from \Fig{fig:find-ci2} with $g(j)$ in the same form as
\Eq{eq:gi}
\begin{align}
  \Ec(\malpha') &= \sum_{i=1}^{\nminp} (n_{N+1}-i+1) \alpha'_i +
  \sum_{i=1}^{g({\nmin})} (\left\lfloor
    g^{-1}(i)\right\rfloor-i)\alpha'_i +
  \sum_{i=g({\nmin})+1}^{\nminp}
  (\nmin-i)\alpha'_i \nonumber\\
  &= \E'(\malpha').
\end{align}%

\subsubsection{Case 3~[$n_{N+1}\in[\ntilde_1,\infty)$]}
\label{sec:n_n+1in1-tilden_3}
As in the previous case, we have $\nminp={\nmin}$ and the same
$\E(\malpha',\malpha)$ as defined in \Eq{eq:jointexp2}. Without loss
of generality, we assume that
$n_{N+1}\in[\ntilde_{k^*},\ntilde_{{k^*}+1})$ for some
${k^*}\in[1,N]$~(we set $\ntilde_{N+1} \defeq \infty$). Then, we have
\begin{equation}
  \label{eq:tmp98}
  \ntilde'_l = \ntilde_l,\quad \text{for}\ l=1,\ldots,{k^*},
\end{equation}%
and
\begin{equation}
  \label{eq:tmp99}
  p_{k^*} < p'_{k^*}\leq p_{{k^*}-1}=p'_{{k^*}-1}\leq\cdots\leq p_1=p'_1.
\end{equation}%
Unlike the previous case, $j-1-n_{N+1}+c_j$ is not always positive.
Let $\jlb$ be the smallest integer such that the coefficient
$j-1-n_{N+1}+c_j$ of $\alpha_j$ in \Eq{eq:jointexp2} is zero. It is
obvious that for $j\geq\jlb$, $\alpha^*_j=\alpha'_j$. Hence, we have
\begin{align}
  \Ec(\malpha') &= \sum_{i=1}^{\nminp} (n_{N+1}-i+1) \alpha'_i +
  \sum_{i=1}^{\jlb-1} (\left\lfloor g^{-1}(i)\right\rfloor-i)\alpha'_i
  + \sum_{j=\jlb}^{\nminp}
  (j-1-n_{N+1}+c_j)\alpha'_j, \nonumber
\end{align}%
where the second term is from \Fig{fig:find-ci3}. Furthermore, we can
show that $\jlb \leq p'_{k^*}$, since $p'_{k^*}-1-n_{N+1}+c_{p'_{k^*}}=0$.
Therefore, we get
\begin{align}
  \Ec(\malpha') &= \sum_{i=1}^{\jlb-1} \left( 1-2i+ n_{N+1} +
    \left\lfloor g^{-1}(i)\right\rfloor \right) \alpha'_i +
  \sum_{i=\jlb}^{p'_{k^*}-1} (n_{N+1}-i+1) \alpha'_i +
  \sum_{i=p'_{k^*}}^{\nminp} c_i \alpha'_i. \label{eq:tmp56}
\end{align}%
Now, we would like to show that the coefficient of $\alpha'_i$ in
\Eq{eq:tmp56} coincides with $c'_i$.  First, for $i\leq\jlb-1$, $i\in
\,\Ical'_{{k^*}+1}\cup\cdots\cup\Ical'_N$ and
lemma~\ref{lemma:ci} implies that
\begin{align*}
  1 - 2i + n_{N+1} + \left\lfloor g^{-1}(i)\right\rfloor &= 1 - i +
  \min_{k=2,\ldots,N+1} \left\lfloor\frac{\sum_{l=0}^{k}\ntilde'_l -
      i}{k}\right\rfloor \\
  &= 1 - i + \min_{k=1,\ldots,N+1}
  \left\lfloor\frac{\sum_{l=0}^{k}\ntilde'_l -
      i}{k}\right\rfloor \\
  &=c'_i.
\end{align*}
Then, for $i\geq p'_{k^*}$, we have 
\begin{equation*}
  i\,\in\,\left(\Ical'_{{k^*}}\cup\cdots\cup\Ical'_1\right)\cap\left(\Ical_{{k^*}}\cup\cdots\cup\Ical_1\right).
\end{equation*}%
Hence, 
\begin{align}
  c'_i &= 1 - i + \min_{k=1,\ldots,k^*}
  \left\lfloor\frac{\sum_{l=0}^{k}\ntilde'_l -
      i}{k}\right\rfloor \nonumber\\
       &= 1 - i + \min_{k=1,\ldots,k^*}
  \left\lfloor\frac{\sum_{l=0}^{k}\ntilde_l -
      i}{k}\right\rfloor \label{eq:tmp112}\\
       &= c_i, \nonumber
\end{align}
where \Eq{eq:tmp112} is from \Eq{eq:tmp98} and \Eq{eq:tmp99}. Finally,
for $i\in[\jlb,p'_{k^*})$, let us rewrite $i=p'_{k^*}-\Delta_i$. Since
$i-1-n_{N+1}+c_i=0$, $\forall\,i\in[\jlb,p'_{k^*})$, we have
\begin{align*}
  \left\lfloor \frac{\sum_{l=0}^{k^*} \ntilde_l - i - {k^*} n_{N+1}}{{k^*}}
  \right\rfloor &= \left\lfloor \frac{\sum_{l=0}^{k^*} \ntilde_l - p'_{k^*}
      + \Delta_i - {k^*} n_{N+1}}{{k^*}} \right\rfloor \nonumber\\
  &=  \left\lfloor \frac{\Delta_i}{{k^*}} \right\rfloor \nonumber\\
  &= 0,
\end{align*}%
from which we have $\Delta_i\in[0,{k^*}-1]$ and
\begin{align*}
  c'_i &= \left\lfloor \frac{\sum_{l=0}^{k^*} \ntilde_l + n_{N+1}
      -i}{{k^*}+1}\right\rfloor + 1 - i \nonumber\\
  &= \left\lfloor \frac{\sum_{l=0}^{k^*} \ntilde_l + n_{N+1}
      - p'_{k^*} + \Delta_i}{{k^*}+1}\right\rfloor + 1 - i \nonumber\\
  &= 1 + n_{N+1} - i.
\end{align*}%
The proof is complete.

\subsection{Proof of Proposition~\ref{prop:pdf-general-RP}}\label{sec:proof-pdf-general-RP}
Let $\malpha(\mM)$ denote the vector of the eigen-exponents of a
matrix $\mM$ as previously defined.  To prove the first case, we use
induction on $N$.  Suppose that it is true for $N$, which means that
the joint pdf of $\malpha(\RP_g \transc{\RP}_g)$ is the same as that
of $\malpha(\RP \transc{\RP})$.  Furthermore, we know by
lemma~\ref{lemma:invariance-asymp} that $\malpha(\RP_g\mT_{N,N+1}
\transc{\mT_{N,N+1}}\transc{\RP}_g) = \malpha(\RP_g \transc{\RP}_g)$.
Same steps as \Eq{eq:tmp31}\Eq{eq:tmp32} complete the proof.  To prove
the second statement, we perform a singular value decomposition on the
matrices $\mT_{i,i+1}$'s and then apply the first statement.

\section{Proof of Theorem~\ref{thm:reduction} and Theorem~\ref{thm:minimal}}
\subsection{Proof of Theorem~\ref{thm:reduction}}
\label{sec:proof-thm-reduction}
Let
\begin{equation}
  \label{eq:010}
  c_i^{(m)} \defeq 1-i + \min_{k=1,\ldots,m}
\left\lfloor\frac{\sum_{l=0}^{k}\ntilde_l - i}{k} \right\rfloor,\quad
i=1,\ldots,{\nmin}.  
\end{equation}
What we should prove is that $c_i^{(N)}=c_i^{(k)},\quad\text{for}\ 
i=1,\ldots,\nmin$ if and only if \Eq{eq:reduction-cond} is true. To
this end, it is enough to show that
\begin{equation}
  \label{eq:tmp111}
  c_i^{(N)} = c_i^{(N-1)} \quad\text{for}\ i=1,\ldots,\nmin
\end{equation}%
if and only if $p_{N-1}\leq N-1$, that is,
$(N-1)\left(\ntilde_N+1\right)\geq \sum_{l=0}^{N-1} \ntilde_l$, and
then apply the result successively to show the theorem. Note that we
need Lemma~\ref{lemma:ci} to eliminate the minimization in
\Eq{eq:010}. The detailed proof is omitted here.

\subsection{Proof of Theorem~\ref{thm:minimal}}
\label{sec:proof-thm-minimal}

The direct part of the theorem is trivial. To show the converse, let
$\mntilde\defeq (\ntilde_0,\ntilde_1,\ldots,\ntilde_N)$ and
$\mntilde'\defeq (\ntilde'_0,\ntilde'_1,\ldots,\ntilde'_{N'})$ be the
two concerned minimal forms. In addition, we assume, without loss of
generality, that
\begin{align*}
  \ntilde_1&=\cdots=\ntilde_{i_1},\ldots,\ntilde_{i_{M-1}+1}=\cdots=\ntilde_{i_M}\\
  \ntilde'_1&=\cdots=\ntilde'_{i'_1},\ldots,\ntilde'_{i'_{M'-1}+1}=\cdots=\ntilde'_{i'_{M'}}
\end{align*}
with $i_M\leq N$ and $i'_{M'}\leq N'$. Now, let us define
$c_{0i}\defeq c_i - (1-i)$ with $c_i$ defined in \Eq{eq:ci3}. It can
be shown that $M$ intervals are non-trivial with
$\Abs{\Ical_{i_k}}\neq0$, $k=1,\ldots,M$. The values of $c_{0i}$'s are
in the following form
\begin{equation*}
  \overbrace{\ldots, \underbrace{\ntilde_{i_M},\ldots,\ntilde_{i_M}}_{i_M}}^{\Abs{\Ical_{i_M}}},\ \overbrace{\underbrace{\ntilde_{i_M}-1,\ldots,\ntilde_{i_M}-1}_{i_{M-1}}, \ldots, \underbrace{\ntilde_{i_{M-1}},\ldots,\ntilde_{i_{M-1}}}_{i_{M-1}}}^{\Abs{\Ical_{i_{M-1}}}},\ldots,\overbrace{\ntilde_2-1,\ldots,\ntilde_1+1,\ntilde_1}^{\Abs{\Ical_1}}.
\end{equation*}
Same arguments also apply to $\mntilde$ with $M'$ and $i'$, etc. It
is then not difficult to see that to have exactly the same
$c_{0i}$'s~(thus, same $c_{i}$'s), we must have $N=N'$ and
\begin{equation*}
  \ntilde_i = \ntilde'_i,\ \forall i=0,\ldots,N,
\end{equation*}%
that is, the same minimal form.

\section{Proof of Theorem~\ref{thm:recursive}}\label{sec:proof-thm-recursive}
\subsection{Sketch of the Proof}
To prove the theorem, we will first show the following equivalence
relations~:
\begin{align*} 
   (\R_1^{(N)}(k),\R_3^{(N)}(i,k)) &\stackrel{(a)}{\llra}   (\R_1^{(N)}(k),\R_2^{(N)}(i)),\quad \forall i,k;\\
   \R_3^{(N)}(i,k) &\stackrel{(b)}{\llra}   \R_3^{(N)}(N-1,k),\quad \forall i,k;\\
   (\R_1^{(N)}(k),\R_2^{(N)}(N-1)) &\stackrel{(c)}{\llra}   (\R_1^{(N)}(k),\R_2^{(N)}(i)\ \text{with ordered $\mn$});\\
   (\R_1^{(N)}(k),\R_2^{(N)}(i)\ \text{with ordered $\mn$}) &\stackrel{(d)}{\llra} (\R_1^{(N)}(k),\R_2^{(N)}(N-1)\ \text{with ordered and minimal $\mn$}).
\end{align*} 
\subsubsection{Equivalences $(a)$ and $(b)$} 
The direct parts of $(a)$, $(b)$, and $(d)$ are immediate since the
RHS are particular cases of the left hand side~(LHS).  To show the
reverse part of (a), we rewrite
\begin{align}
  \dRP_{(n_0,\ldots,n_N)}(k) &= \dRP_{(n_0-k,\ldots,n_N-k)}(0) \label{eq:tmp01}\\
  &= \min_{j\geq 0} \left\{ \dRP_{(n_0-k,\ldots,n_i-k)}(j) + \dRP_{(j,n_{i+1}-k,\ldots,n_N-k)}(0) \right\}\label{eq:tmp02}\\
  &= \min_{j'\geq k} \left\{ \dRP_{(n_0,\ldots,n_i)}(j') +
  \dRP_{(j',n_{i+1},\ldots,n_N)}(k) \right\}, \label{eq:tmp03}
\end{align}
where $\R_1$ is used twice in \Eq{eq:tmp01} and \Eq{eq:tmp03}; $\R_2$
is used in \Eq{eq:tmp02}. As for (b), if $\R_3^{(N)}(N-1,k)$ holds, then
\begin{align}
  \dRP_{(n_0,\ldots,n_N)}(k) &= \min_{j\geq k} \left\{ \dRP_{(n_0,\ldots,n_{N-1})}(j) + \dRP_{(j,n_N)}(k) \right\} \label{eq:tmp04}\\
  &= \min_{j'\geq j\geq k} \left\{ \dRP_{(n_0,\ldots,n_{N-2})}(j')  + \dRP_{(j',n_{N-1})}(j) + \dRP_{(j,n_N)}(k) \right\} \label{eq:tmp05}\\ 
  &= \min_{j'\geq k} \left\{ \dRP_{(n_0,\ldots,n_{N-2})}(j')  + \dRP_{(j',n_{N-1},n_N)}(k) \right\} \label{eq:tmp06}
\end{align}%
which proves $\R_3^{(N)}(N-2,k)$. By continuing the process, we can show
that $\R_3^{(N)}(i,k)$ is true for all $i$, provided $\R_3^{(N)}(N-1,k)$
holds. 
\subsubsection{Equivalences $(c)$ and $(d)$}
Through $(a)$ and $(b)$, one can verify that the LHS of $(c)$ is
equivalent to the RHS of $(a)$ of which the RHS of $(c)$ is a
particular case. Hence, the direct part of $(c)$ is shown. The
reverse part of (c) can be proved by induction on $N$. For $N=2$,
$\R_2^{(N)}(N-1)$ can be shown explicitly using the direct
characterization \Eq{eq:dk}. Now, assuming that $\R_2^{(N)}(N-1)$ for
non-ordered $\mn$, we would like to show that $\R_2^{N+1}(N)$ holds.
Let us write
\begin{align}
  \min_{j\geq 0} \left\{\dRP_{(n_0,\ldots,n_{N})}(j) +
    \dRP_{(j,n_{N+1})}(0)\right\}
  &= \min_{j\geq 0} \left\{\dRP_{(\ntilde_0,\ldots,\ntilde_{i-1},\ntilde_{i+1},\ldots,\ntilde_{N+1})}(j)  + \dRP_{(j,\ntilde_i)}(0) \right\}\label{eq:tmp001}\\
  &= \min_{k\geq j\geq 0} \left\{\dRP_{(\ntilde_0,\ldots,\ntilde_{i-1},\ntilde_{i+1},\ldots,\ntilde_{N})}(k)  + \dRP_{(k,\ntilde_{N+1})}(j) + \dRP_{(j,\ntilde_i)}(0) \right\} \label{eq:tmp002}\\
  &= \min_{k\geq j'\geq 0} \left\{\dRP_{(\ntilde_0,\ldots,\ntilde_{i-1},\ntilde_{i+1},\ldots,\ntilde_{N})}(k)  + \dRP_{(k,\ntilde_{i})}(j') + \dRP_{(j',\ntilde_{N+1})}(0) \right\}\label{eq:tmp003}\\
  &= \min_{j'\geq 0} \left\{\dRP_{(\ntilde_0,\ldots,\ntilde_{N})}(j')  + \dRP_{(j',\ntilde_{N+1})}(0) \right\} \label{eq:tmp003} \nonumber\\
  &= \dRP_{(n_0,\ldots,n_{N+1})}(0), \nonumber
\end{align}%
where the permutation invariance property is used in \Eq{eq:tmp001};
$\R_3^{(N)}(N-1,k)$ is used in \Eq{eq:tmp002} since we assume that
$\R_2^{(N)}(N-1)$ is trues; $\ntilde_i$ and $\ntilde_{N+1}$ can be
permuted according to $\R_2^{(2)}(1)$. Finally, we should prove the
reverse part of (d), \ie, 
\begin{equation}
  \label{eq:tmp004}
  \dRP_{(\ntilde_0,\ldots,\ntilde_{N})}(0) = \min_{j\geq 0} \left\{\dRP_{(\ntilde_0,\ldots,\ntilde_{N-1})}(j)  + j\ntilde_N \right\} 
\end{equation}%
provided that $\R_2^{(N)}(N-1)$ holds for minimal $\mn$.

If $\mn$ is not minimal, then showing (c) is equivalent to showing 
\begin{equation}
  \label{eq:tmp005}
    \dRP_{(\ntilde_0,\ldots,\ntilde_{N^*})}(0) = \min_{j\geq 0} \left\{ \dRP_{(\ntilde_0,\ldots,\ntilde_{N^*})}(j)  + j\ntilde_N \right\},
\end{equation}%
where $N^*$ is the order of $\mn$ with $\ntilde_{N^*+1}\leq
\ntilde_{N}$. Therefore, we should show that the minimum is achieved
with $j=0$. According the direct characterization~\Eq{eq:dk}, this is
true only when $\ntilde_N\geq c_1$. Let us rewrite $c_1$ as
\begin{align*}
  c_1 &= \left\lfloor\frac{\sum_{l=0}^{N^*}\ntilde_l - 1}{N^*} \right\rfloor \\
&= \left\lfloor\frac{N^*\ntilde_{N^*+1} + p_{N^*} - 1}{N^*} \right\rfloor.
\end{align*}%
Since $p_{N^*}\geq N^*$ is always true according to the reduction
theorem, we have $c_1\leq \ntilde_{N^*+1}\leq \ntilde_N$. The rest
of this section is devoted to proving that \Eq{eq:tmp004} holds for
minimal $\mn$.

\subsection{Minimal $\mn$}
Now, we restrict ourselves in the case of minimal and ordered $\mn$,
\ie, we would like to prove
\begin{equation}
  \label{eq:tmp006}
    \dRP_{(\ntilde_0,\ldots,\ntilde_{N^*})}(0) = \min_{j\geq 0} \left\{\dRP_{(\ntilde_0,\ldots,\ntilde_{N^*-1})}(j)  + j\ntilde_N\right\}.
\end{equation}%
Since $c_{p_{N^*-1}} \leq \ntilde_{N^*}$, the optimal $j$ is in the
interval $\Ical_{N^*}\defeq [1,p_{N^*-1}]$. Now, showing
\Eq{eq:tmp006} is equivalent to showing
\begin{equation*}
\sum_{i=1}^{p_{N^*-1}} 1-i+ \left\lfloor\frac{\sum_{l=0}^{N^*}\ntilde_l - i}{N^*} \right\rfloor = \min_{p_{N^*-1}\geq j \geq 0} \sum_{i=j+1}^{p_{N^*-1}} 1-i+ \left\lfloor\frac{\sum_{l=0}^{N^*-1}\ntilde_l - i}{N^*-1} + j\ntilde_{N^*}\right\rfloor 
\end{equation*}%
which, after some simple manipulations, is reduced to
\begin{equation}\label{eq:tmp007}
\sum_{i=1}^{p_{M}} \left( i - p_M + \left\lfloor\frac{i-1}{M+1} \right\rfloor \right) = \min_{k} \sum_{i=1}^{k} \left( i - p_M + \left\lfloor\frac{i-1}{M} \right\rfloor \right),
\end{equation}%
where we set $M\defeq N^*-1$ for simplicity of notation. Obviously,
the minimum of the RHS of \Eq{eq:tmp007} is achieved with such $k^*$
that 
\begin{align}
  k^* - p_M + \left\lfloor\frac{k^*-1}{M} \right\rfloor &\leq 0, \label{eq:tmp008-1}\\
\text{and}\  (k^*+1) - p_M + \left\lfloor\frac{k^*}{M} \right\rfloor &> 0. \label{eq:tmp008-2}
\end{align}%
Let us decompose $k^*$ as $k^*=aM+b$ with $b\in[1,M]$. Then, \Eq{eq:tmp008-1} becomes
\begin{equation}
  \label{eq:tmp009-1}
  aM+b-p_M+a \leq 0
\end{equation}%
which also implies that $aN+1-p_M+a\leq 0$ from which 
$a = \left\lfloor \frac{p_M-1}{M+1} \right\rfloor.$
The form of $a$ suggests that $p_M$ can be decomposed as
\begin{equation}
  p_M = a(M+1) + \bar{b}. \label{eq:tmp009-2}
\end{equation}%
From \Eq{eq:tmp009-1} and \Eq{eq:tmp009-2}, we have $b\leq \bar{b}$ and
thus $b=\min\left\{M,\bar{b}\right\}$. With the form of optimal $k$
and some basic manipulations, we have finally
\begin{equation*}
  \sum_{i=1}^{p_{M}} \left( i - p_M + \left\lfloor\frac{i-1}{M+1} \right\rfloor \right) -  \sum_{i=1}^{k^*} \left( i - p_M + \left\lfloor\frac{i-1}{M} \right\rfloor \right)
 = 0
\end{equation*}%
which ends the proof.

\section{Proof of Lemmas~\ref{lemma:div} and~\ref{lemma:div-para}}


\subsection{Proof of Lemma~\ref{lemma:div}}
\label{sec:proof-lemma-div}
First, we have $$ \SNR\lambda_{\max}(\transc{\mH}\mH) \leq
\SNR\Frob{\mH} \leq \det(\Id + \SNR\transc{\mH}\mH),$$ from which
\begin{equation}\label{eq:tmp666}
  \Prob{\SNR\lambda_{\max}(\transc{\mH}\mH) < 1+\epsilon} \geq \Prob{\det(\Id + \SNR\transc{\mH}\mH) < 1+\epsilon}
\end{equation}%
with $\epsilon$ being some strictly positive constant. Then, we also
have
\begin{equation}\label{eq:tmp667}
\Prob{\SNR\lambda_{\max}(\transc{\mH}\mH) <
  1+\epsilon} \leq \Prob{\det(\Id + \SNR\transc{\mH}\mH) <
  (2+\epsilon)^{\text{rank}(\mH)}},
\end{equation}%
since $\det(\Id+\SNR \transc{\mH}\mH) = \prod_i
(\Id+\SNR\lambda_i(\transc{\mH}\mH))$. From \Eq{eq:tmp666} and
\Eq{eq:tmp667}, we have
\begin{align*}
  \Prob{\SNR\lambda_{\max}(\transc{\mH}\mH) < 1+\epsilon} 
  &\asympteq \Prob{\det(\Id + \SNR\transc{\mH}\mH) < 1+\epsilon'} \\
  &\asympteq \SNR^{-d(0)},
\end{align*}%
where $\epsilon'$ is another strictly positive constant. Hence, $
\Prob{\SNR\Frob{\mH}<1+\epsilon} \asympteq \SNR^{-d(0)}. $ The lemma
is proved since $\Prob{\SNR\Frob{\mH}<1+\epsilon} \asympteq
\Prob{\SNR\Frob{\mH}<1}$.

\subsection{Proof of Lemma~\ref{lemma:div-para}}
\label{sec:proof-lemma-div-para}

Let us consider a parallel channel $\{\mH_k\}_{k=1}^K$, each
sub-channel of rank $M_k$ and with eigen-exponents
$\{\alpha_{1,k},\alpha_{2,k},\ldots,\alpha_{M_k,k}\}$. Since each
sub-channel is an AF path, the joint pdf of the eigen-exponents in the
high SNR regime is $p_k(\malpha_k) \asympteq \SNR^{-\sum_i
  c_{i,k}\mul\alpha_{i,k}}.$ From Lemma~\ref{lemma:cal-dmt}, the DMT is
$$d_{\Pcal}(r)\defeq \min_{\{\malpha_k\}_k \in \Ocal(r)} \sum_k\sum_i
c_{i,k}\mul\alpha_{i,k}$$
with $\Ocal(r) \defeq \left\{\sum_k\sum_i
  (1-\alpha_{i,k})^+ \leq K r \right\} $ being the outage region.
First, we can deduce that
\begin{align*}
  d_{\Pcal}(0) &=\sum_k\sum_i c_{i,k} \\
  & =\sum_k d_k(0).
\end{align*}
Then, if all AF paths have the same DMT, they have the same set
$\{c_{i,k}\}_i$, \ie, $c_{i,k}= c_i,\ \forall\,k$. We can verify that
setting $\alpha_{i,k}=\alpha_i$, $\forall\,k$ is without loss of
optimality, since 1)~the objective function is linear and symmetric on
different $k$, and 2)~the constraints are convex and symmetric on
different $k$. Finally, the optimization problem becomes
$$
\min_{\malpha \in \Ocal_0(r)} K\mul\sum_i c_i\mul \alpha_i$$
with
$\Ocal_0(r) \defeq \{\sum_i (1-\alpha_i)^+ \leq r\}$ is the outage
region of each single AF path. The lemma can be proved immediately
from here.

\section{Other Proofs}

\subsection{Proof of Proposition~\ref{prop:2hop}} 
\label{sec:proof-prop-2hop}
Without loss of generality, we assume that $n_0\geq n_2$. Then, the
bottleneck of the channel is the $n_1\times n_2$ channel. Since the
partition achieves the maximum diversity, by
theorem~\ref{thm:cond_nece}, the partition size is $K=K_{1} K_{2}$
with the $n_1$~(respectively, $n_2$) antennas being partitioned into
$K_{1}$~(respectively, $K_{2}$) supernodes. Moreover, for any AF path
$k$ in the partition, we have $n_{k,0}+1 \geq n_{k,1}+n_{k,2}.$ Adding
all the $K$ inequalities up gives
  \begin{equation}
    \label{eq:tmp432}
    \sum_{k=1}^{K} n_{k,0} + K_{1} K_{2} \geq K_{2} n_1 + K_{1} n_2.     
  \end{equation}%
  The sum in the LHS of \Eq{eq:tmp432} can be upper-bounded by $K_{1}
  n_0$, since each supernode in the transmitter cannot be connected to
  more than $K_{1}$ nodes. Hence, we have the following inequality
  after some simple manipulations
  \begin{equation*}
    K_{1} \geq \left\lceil \frac{K_{2} n_1}{K_{2}+n_0-n_2} \right\rceil,
  \end{equation*}%
  from which we have the lower bound on the partition size 
  \begin{equation*}
    K_{1}K_{2} \geq K_{2} \left\lceil \frac{K_{2} n_1}{K_{2}+n_0-n_2} \right\rceil,
  \end{equation*}%
  which is obviously increasing with $K_{2}$. Therefore, the minimum
  lower bound is obtained by setting $K_{2}=1$ and it coincides with
  \Eq{eq:2}. It can be shown that this lower bound is achieved by
  partitioning the intermediate layer into $K$ supernodes with $K$
  defined by \Eq{eq:2} without partitioning either of the source and
  the destination antennas.

\subsection{Proof of Theorem~\ref{thm:DMT-FF}}
\label{sec:proof-theorem-dmt-ff}

Let us define the selection matrices $\J{i}{k}$'s as $n_i\times n_i$
diagonal matrices with
  \begin{equation*}
    \J{i}{k}(j,j) =  
    \begin{cases}
      1 & \text{if $j\in\S_{i,k}$}, \\
      0 & \text{otherwise}.
    \end{cases}
  \end{equation*}%
  First, we would like to prove that the maximum diversity gain is
  achieved. This can be done in two steps. The first step is to prove
  that the parallel channel $\{\H''_k\}_k$ with $\H''_k \defeq \mH_N
  \prod_{i=1}^{N-1} \left(\J{i}{f_i(k)} \mH_i \right)$ achieves the
  maximum diversity. To this end, note that by partitioning the
  rows~(respectively, columns) of $\mH_N$~(respectively, $\mH_1$)
  according to the indices in
  $\S_{N,1},\ldots,\S_{N,K_N}$~(respectively,
  $\S_{0,1},\ldots,\S_{0,K_1}$), the matrix $\H''_k$ can be
  partitioned into $K_0 K_N$ blocks, each one being an AF path from
  the source to the destination. Therefore, $\{\H''_k\}_k$ comprises
  $K_0 K_1 \cdots K_N$ AF paths, \ie, all possible paths. Obviously,
  these paths include the $K$ independent paths $\{\H_k\}_k$ in the
  independent partition. Therefore, the maximum diversity is achieved
  since $\sum_{k=1}^{K'} \Frob{\H''_k} \geq \sum_{k=1}^{K}
  \Frob{\H_k}. $

  The key of the second step is to show that the set of matrices
  $\{\H'_k\}_k$ defined in \Eq{eq:tmp1111} is actually an invertible
  constant linear transformation of $\{\H''_k\}_k$, \ie,
  \begin{equation*}
    \matrix{\H'_1&\cdots&\H'_{K'}} = \matrix{\H''_1&\cdots&\H''_{K'}} \mT.
  \end{equation*}%
  In this case, we have 
  \begin{align*}
    \sum_{k=1}^{K'} \Frob{\H'_k} &\geq \lambda_{\min}(\mT\transc{\mT})
    \sum_{k=1}^{K'} \Frob{\H''_k} \\
    &\asympteq \sum_{k=1}^{K'} \Frob{\H''_k}
  \end{align*}%
  and the diversity is lower-bounded by the maximum diversity,
  according to Lemma~\ref{lemma:div-para}. Hence, the FF scheme also
  achieves the maximum diversity. The key point is shown in the
  following. First, let us divide the set of indices $\{1,\ldots,K'\}$
  into $K'/{K_1}$ groups, each one comprising exactly $K_1$ integers
  $i_1,\ldots,i_{K_1}$ such that $ f_j(i_1)=\ldots=f_j(i_{K_1}),\quad
  \forall\,j=2,\ldots,N-1, $ and $f_1(i_j)$ varies from $1$ to $K_1$.
  Then, we partition the set $\{\H'_k\}_k$ according to the partition
  of the indices described above. Hence, the matrices in the same
  group can be rewritten as $\left\{\mG \F{1}{0}\mH_1, \ldots, \mG
    \F{1}{K_1}\mH_1\right\}$ with $\mG$ being some matrix. We have
  \begin{equation*}
    \matrix{\mG \F{1}{1}\mH_1 & \cdots & \mG \F{1}{K_1}\mH_1} = 
    \matrix{\mG \J{1}{1}\mH_1 & \cdots & \mG \J{1}{K_1}\mH_1} \mT_1,
  \end{equation*}%
  where $\mT_1$ is composed of $K_1\times K_1$ blocks of matrices with
  the $(i,j)$-th block being $-\Id$ if $i=j\geq 2$ and $\Id$
  otherwise. We can verify that $\mT_1$ is invertible and with the
  transformation, the matrices $\F{1}{k}$'s are replaced by
  $\J{1}{k}$'s with the same indices. In the same manner, we can
  successively replace the matrices $\F{2}{k},\ldots,\F{N-1}{k}$ with
  $\J{2}{k},\ldots,\J{N-1}{k}$ by similar invertible transformations
  $\mT_2,\ldots,\mT_{N-1}$ as $\mT_1$. Finally, we obtain $\{\H''_k\}_k$
  and the total transformation is invertible, constant and linear.
  
  Note that the parallel channel of the FF scheme is in outage for a
  target rate $K'r\log\SNR$ implies that at least one of the
  sub-channels is in outage for a target rate $r\log\SNR$. Therefore,
  one can show that $\SNR^{-\dFF(r)} \asymptleq \SNR^{-\dAF(r)}$, from
  which $\dFF(r)\geq\dAF(r)$. Finally, by showing that $\dFF(r)$ is
  piece-wise linear with $K'\mul \ntilde_0$ sections, we prove the
  theorem.

\subsection{Proof of Theorem~\ref{thm:PF}}\label{sec:proof-thm-PF}
Let $\mlambda(\mM)$ and $\malpha(\mM)$ denote the vector of the
ordered eigenvalues and the corresponding eigen-exponents of a matrix
$\mM$. The theorem can be proved by showing a stronger result~: the
asymptotical pdf of $\malpha({\RPPFtran} {\RPPF})$ in the high SNR
regime is identical to that of $\malpha(\transc{\RP}\mul {\RP})$. We
show it by induction on $N$. For $N=1$, since $\umH_1 = \mH_1$, the
result is direct. Suppose that the theorem holds for $N$. Let us show
that it also holds for $N+1$. Note that $\RPpPF=\umH_{N+1} \mP_N \RPPF
= \umH_{N+1} \umD_N \transc{\umQ}_N \RPPF,$ from which we have
  \begin{align*}
    {\RPpPFtran}\mul {\RPpPF} &\sim \Wcal_{n_0}(n_{N+1},\transc{(\umD_N
      \transc{\umQ}_N
      \RPPF)}(\umD_N \transc{\umQ}_N \RPPF)) \\
    &\sim \Wcal_{\underline{\nmin}}(n_{N+1},\mlambda(\transc{(\umD_N
      \transc{\umQ}_N \RPPF)}(\umD_N \transc{\umQ}_N \RPPF)))
  \end{align*}
  for a given $\RP$. Similarly, $\transc{\RP'}\mul {\RP'} \sim
  \Wcal_{{\nmin}}(n_{N+1},\mlambda(\transc{\RP}\mul \RP))$. At high SNR,
  we can show that
  \begin{align*}
    \malpha(\transc{(\umD_N \transc{\umQ}_N \RPPF)}(\umD_N
    \transc{\umQ}_N \RPPF)) & = \malpha(\transc{(\transc{\umQ}_N
      \RPPF)}(\transc{\umQ}_N \RPPF)) \\
    &= \malpha({\RPPFtran}\mul\RPPF),
  \end{align*}%
  where the first equality comes from
  lemma~\ref{lemma:invariance-asymp} and the second one holds because
  $\transc{(\transc{\umQ}_N \RPPF)}(\transc{\umQ}_N \RPPF) =
  {\RPPFtran}\mul\RPPF.$ Finally, since we suppose that the joint pdf of
  $\malpha({(\RPPFtran)}\mul\RPPF)$ is the same as that of
  $\malpha(\transc{\RP}\mul\RP)$, we can draw the same conclusion for
  $\malpha({\RPpPFtran}\mul\RPpPF)$ and
  $\malpha(\transc{(\RP')}\mul\RP')$.

\subsection{Proof of Theorem~\ref{thm:dmt-para}}
\label{sec:proof-thm-dmt-para}
Let us consider an equivalent block-diagonal channel of the parallel
channel~\Eq{eq:parallel} in the following form 
\begin{equation}\label{eq:BD}
  \my_e = \diag(\prl{\mH}_k)\mul \mx_e + \mz_e,
\end{equation}
where $\mx_e\defeq
\trans{[\trans{\mx_1}\ \trans{\mx_2}\ \ldots\ \trans{\mx_K}]}$, and
$\my_e, \mz_e$ are defined in the same manner. Now, from the parallel
NVD code $\Xcal$, we can build a block-diagonal code
$\Xcal_{\text{BD}}$ with codewords defined by $\mX_{\text{BD}}\defeq
\diag\{\mX_k\}$. We can verify that $\Xcal_{\text{BD}}$ is actually
a rate-$\nav$ NVD code defined in \cite{SY_JCB_coop} with
$\nav\defeq \sum_k \ntk/K$. From \cite[Th.~3]{SY_JCB_coop},
we have
\begin{align*}
  d_{\Xcal_{\text{BD}}}(r) &\geq d\left(\frac{\sum_k \ntk}{\nav}\mul r\right) \\
  &= d(K\mul r) ,
\end{align*}%
where $d(r)$ is the DMT of the parallel channel (and thus the
block-diagonal channel). Finally, it is obvious that $d_{\Xcal}(K\mul
r)=d_{\Xcal_{\text{BD}}}(r),$ since using $\Xcal$ will have the same
error performance\footnote{This is due to the block-diagonal nature of
  the equivalent channel.} as using $\Xcal_{\text{BD}}$ except that the
transmission rate is $K$ times higher. We have thus $d_{\Xcal}(r)\geq
d(r)$. It is shown in \cite{SY_JCB_coop} that the achievability holds
for any fading statistics. Thus, the code is approximately universal.

\subsection{An Alternative Code Construction}
\label{sec:altercod}

A simple alternative construction that is approximately universal is
described as follows. Let $\XcalF$ be a $\nsum\times T$ full rate NVD
code with $\nsum \defeq \sum_k \ntk$ and $T\geq \nsum$. Then, $\XcalF$
achieves the DMT $d(r)$ of the channel \Eq{eq:BD}. It means that by
partitioning every codeword matrix $\XF\in\XcalF$ into $K\times1$
blocks in such a way that the \th{k} block is of size $\ntk\times T$
and sending the \th{k} block in the \th{k} sub-channel, the DMT of the
original parallel channel is achieved. Although this construction is
simple and suitable for both symmetric and asymmetric channels, the
main drawback is that the coding delay is roughly $K$ times larger
than the parallel NVD code constructed in Section~\ref{sec:paraNVD}.
Decoding complexity of such codes is sometimes prohibitive.

\subsection{$\zeta_{2^{m}}$ is not a norm in $\mathbb{K}$\label{zeta2m-not-norm}}

Assume that $\zeta_{2^{m}}$ is a norm in $\KK$, which
means\begin{equation} \exists
  x\in\mathbb{K},N_{\mathbb{K}/\mathbb{Q}\left(\zeta_{2^{m}}\right)}(x)=\zeta_{2^{m}}.\label{eq:x-existence}\end{equation}
Consider now the extensions described in \Fig{fig:fields-extension}
with the proper fields.  From \Eq{eq:x-existence} and the left
extension of \Fig{fig:fields-extension}, we deduce that $
N_{\KK/\QQ(i)}(x)=N_{\QQ\left(\zeta_{2^{m}}\right)/\QQ(i)}\left(N_{\KK/\QQ\left(\zeta_{2^{m}}\right)}(x)\right)=-i$,
since the minimal polynomial of $\zeta_{2^{m}}$ is $X^{2^{m-2}}-i$.
Meanwhile, from the right extension of \Fig{fig:fields-extension}, we
have
$N_{\KK/\QQ(i)}(x)=N_{\QQ\left(i,\sqrt{5}\right)/\QQ(i)}\left(N_{\KK/\QQ\left(i,\sqrt{5}\right)}(x)\right)=-i.$
Denote
$y=N_{\KK/\QQ\left(i,\sqrt{5}\right)}(x)\in\QQ\left(i,\sqrt{5}\right)$.
Then the number $z=\frac{1+\sqrt{5}}{2}\mul y$ has an algebraic norm
equal to $i$, and belongs to $\QQ\left(i,\sqrt{5}\right)$ which is in
contradiction with the result obtained in \cite{Belfiore_golden}. So,
$\zeta_{2^{m}}$ is a non-norm element.


\newpage

\newpage

\newpage

\begin{thebibliography}{10}
\providecommand{\url}[1]{#1}
\csname url@samestyle\endcsname
\providecommand{\newblock}{\relax}
\providecommand{\bibinfo}[2]{#2}
\providecommand{\BIBentrySTDinterwordspacing}{\spaceskip=0pt\relax}
\providecommand{\BIBentryALTinterwordstretchfactor}{4}
\providecommand{\BIBentryALTinterwordspacing}{\spaceskip=\fontdimen2\font plus
\BIBentryALTinterwordstretchfactor\fontdimen3\font minus
  \fontdimen4\font\relax}
\providecommand{\BIBforeignlanguage}[2]{{%
\expandafter\ifx\csname l@#1\endcsname\relax
\typeout{** WARNING: IEEEtran.bst: No hyphenation pattern has been}%
\typeout{** loaded for the language `#1'. Using the pattern for}%
\typeout{** the default language instead.}%
\else
\language=\csname l@#1\endcsname
\fi
#2}}
\providecommand{\BIBdecl}{\relax}
\BIBdecl

\bibitem{vanderMeulen}
E.~C. {van der Meulen}, ``Three-terminal communication channels,'' \emph{Adv.
  Appl. Prob.}, vol.~3, no.~1, pp. 120--154, 1971.

\bibitem{Cover_relay}
T.~M. Cover and A.~{El Gamal}, ``Capacity theorems for the relay channel,''
  \emph{{IEEE} Trans. Inf. Theory}, vol.~25, no.~5, pp. 572--584, Sep. 1979.

\bibitem{Kramer_relay}
G.~Kramer, M.~Gastpar, and P.~Gupta, ``Cooperative strategies and capacity
  theorems for relay networks,'' \emph{{IEEE} Trans. Inf. Theory}, vol.~51,
  no.~9, pp. 3037--3063, Sep. 2005.

\bibitem{Wang_relay}
B.~Wang, J.~Zhang, and A.~{H{\o}st-Madsen}, ``On the capacity of {MIMO} relay
  channels,'' \emph{{IEEE} Trans. Inf. Theory}, vol.~1, no.~1, pp. 29--43, Jan.
  2005.

\bibitem{Gupta_Kumar}
P.~Gupta and P.~R. Kumar, ``The capacity of wireless networks,'' \emph{{IEEE}
  Trans. Inf. Theory}, vol.~46, no.~2, pp. 388--404, Mar. 2000.

\bibitem{Xie_Kumar}
L.~Xie and P.~R. Kumar, ``A network information theory for wireless
  communication~: {S}caling laws and optimal operation,'' \emph{{IEEE} Trans.
  Inf. Theory}, vol.~50, no.~5, pp. 748--767, May 2004.

\bibitem{Gastpar_Vetterli}
M.~Gastpar and M.~Vetterli, ``On the capacity of large {Gaussian} relay
  networks,'' \emph{{IEEE} Trans. Inf. Theory}, vol.~51, no.~3, pp. 765--779,
  Mar. 2005.

\bibitem{Morgen_Bolcskei}
V.~I. Morgenshtern and H.~B{\"o}lcskei, ``Crystallization in wireless
  networks,'' \emph{{IEEE} Trans. Inf. Theory}, vol.~53, no.~10, Oct. 2007, to
  appear.

\bibitem{Ayfer_Adhoc}
A.~{\"O}zg{\"u}r, O.~{L\'ev\^eque}, and D.~N.~C. Tse, ``Hierarchical
  cooperation achieves optimal capacity scaling in ad hoc networks,''
  \emph{{IEEE} Trans. Inf. Theory}, Sep. 2006, submitted.

\bibitem{Sendonaris1}
A.~Sendonaris, E.~Erkip, and B.~Aazhang, ``User cooperation diversity---{P}art
  {I}: {S}ystem description,'' \emph{{IEEE} Trans. Commun.}, vol.~51, no.~11,
  pp. 1927--1938, Nov. 2003.

\bibitem{Sendonaris2}
------, ``User cooperation diversity---{P}art {II}: {I}mplementation aspects
  and performance analysis,'' \emph{{IEEE} Trans. Commun.}, vol.~51, no.~11,
  pp. 1939--1948, Nov. 2003.

\bibitem{LTW1}
J.~N. Laneman and G.~W. Wornell, ``Distributed space-time-coded protocols for
  exploiting cooperative diversity in wireless networks,'' \emph{{IEEE} Trans.
  Inf. Theory}, vol.~49, no.~10, pp. 2415--2425, Oct. 2003.

\bibitem{LTW2}
J.~N. Laneman, D.~N.~C. Tse, and G.~W. Wornell, ``Cooperative diversity in
  wireless networks: Efficient protocols and outage behavior,'' \emph{{IEEE}
  Trans. Inf. Theory}, vol.~50, no.~12, pp. 3062--3080, Dec. 2004.

\bibitem{Hunter2}
T.~Hunter, S.~Sanayei, and A.~Nosratinia, ``Outage analysis of coded
  cooperation,'' \emph{{IEEE} Trans. Inf. Theory}, vol.~52, no.~2, pp.
  375--391, Feb. 2006.

\bibitem{Nabar}
R.~U. Nabar, H.~{B\"olcskei}, and F.~W. {Kneub\"uhler}, ``Fading relay
  channels: {P}erformance limits and space-time signal design,'' \emph{{IEEE}
  J. Sel. Areas Commun.}, vol.~22, no.~6, pp. 1099--1109, Aug. 2004.

\bibitem{ElGamal_coop}
K.~Azarian, H.~{El Gamal}, and P.~Schniter, ``On the achievable
  diversity-multiplexing tradeoff in half-duplex cooperative channels,''
  \emph{{IEEE} Trans. Inf. Theory}, vol.~51, no.~12, pp. 4152--4172, Dec. 2005.

\bibitem{SY_JCB_SAF}
\BIBentryALTinterwordspacing
S.~Yang and J.-C. Belfiore, ``Towards the optimal amplify-and-forward
  cooperative diversity scheme,'' \emph{{IEEE} Trans. Inf. Theory}, vol.~53,
  no.~9, Sep. 2007, to appear. [Online]. Available:
  \url{http://arxiv.org/pdf/cs.IT/0603123}
\BIBentrySTDinterwordspacing

\bibitem{Elza_coop}
M.~Yuksel and E.~Erkip, ``Multi-antenna cooperative wireless systems: {A}
  diversity multiplexing tradeoff perspective,'' \emph{{IEEE} Trans. Inf.
  Theory}, Oct. 2007, special Issue on Models, Theory, and Codes for Relaying
  and Cooperation in Communication Networks, to appear.

\bibitem{Bolcskei_relay}
H.~B{\"o}lcskei, R.~U. Nabar, {\"O}.~Oyman, and A.~J. Paulraj, ``Capacity
  scaling laws in {MIMO} relay networks,'' \emph{{IEEE} Trans. Wireless
  Commun.}, vol.~5, no.~6, pp. 1433--1444, Jun. 2006.

\bibitem{Jing}
Y.~Jing and B.~Hassibi, ``Distributed space-time coding in wireless relay
  networks,'' \emph{{IEEE} Trans. Wireless Commun.}, vol.~5, no.~12, pp.
  3524--3536, Dec. 2006.

\bibitem{Jing2}
------, ``Cooperative diversity in wireless relay networks with
  multiple-antenna nodes,'' \emph{{IEEE} Trans. Signal Process.}, 2006,
  submitted.

\bibitem{Yeh_Leveque}
S.~Yeh and O.~L\'ev\^eque, ``Asymptotic capacity of multi-level
  amplify-and-forward relay networks,'' in \emph{Proc. IEEE International
  Symposium on Information Theory}, Nice, France, Jun. 2007.

\bibitem{Zheng_Tse}
L.~Zheng and D.~N.~C. Tse, ``Diversity and multiplexing: {A} fundamental
  tradeoff in multiple-antenna channels,'' \emph{{IEEE} Trans. Inf. Theory},
  vol.~49, no.~5, pp. 1073--1096, May 2003.

\bibitem{Tse_DMT_MAC}
D.~N.~C. Tse and P.~Viswanath, ``Diversity-multiplexing tradeoff in multiple
  access channels,'' \emph{{IEEE} Trans. Inf. Theory}, vol.~50, no.~9, pp.
  1859--1874, Sep. 2004.

\bibitem{Cover3}
T.~M. Cover and J.Thomas, \emph{Elements of Information Theory}.\hskip 1em plus
  0.5em minus 0.4em\relax New York: Wiley, 1991.

\bibitem{Oggier_perfect}
F.~Oggier, G.~Rekaya, J.-C. Belfiore, and E.~Viterbo, ``Perfect space-time
  block codes,'' \emph{{IEEE} Trans. Inf. Theory}, vol.~52, no.~9, pp.
  3885--3902, Dec. 2006.

\bibitem{Elia}
P.~Elia, K.~R. Kumar, S.~A. Pawar, P.~V. Kumar, and H.~Lu, ``Explicit,
  minimum-delay space-time codes achieving the diversity-multiplexing gain
  tradeoff,'' \emph{{IEEE} Trans. Inf. Theory}, vol.~52, no.~9, pp. 3869--3884,
  Sep. 2006.

\bibitem{Tavildar}
S.~Tavildar and P.~Viswanath, ``Approximately universal codes over slow fading
  channels,'' \emph{{IEEE} Trans. Inf. Theory}, vol.~52, no.~7, pp. 3233--3258,
  Jul. 2006.

\bibitem{Borade_Zheng_Gallager}
S.~Borade, L.~Zheng, and R.~Gallager, ``Amplify and forward in wireless relay
  networks: Rate, diversity and network size,'' \emph{{IEEE} Trans. Inf.
  Theory}, Oct. 2007, special Issue on Relaying and Cooperation in
  Communication Networks, to appear.

\bibitem{Muller_productmatrix}
R.~R. M{\"u}ller, ``On the asymptotic eigenvalue distribution of concatenated
  vector-valued fading channels,'' \emph{{IEEE} Trans. Inf. Theory}, vol.~48,
  no.~7, pp. 2086--2091, Jul. 2002.

\bibitem{VBLAST}
P.~Wolniansky, G.~Foschini, G.~Golden, and R.~Valenzuela, ``{V-BLAST}: {An}
  architecture for realizing very high data rates over the rich-scattering
  wireless channel,'' in \emph{Proc. of the URSI International Symposium on
  Signal, Systems, and Electronics Conference}, New York, 1998, pp. 295--300.

\bibitem{Oggier_Hassibi}
F.~Oggier and B.~Hassibi, ``An algebraic coding scheme for wireless relay
  networks with multiple-antenna nodes,'' \emph{{IEEE} Trans. Signal Process.},
  Mar. 2006, submitted.

\bibitem{Elia_relay}
\BIBentryALTinterwordspacing
P.~Elia and P.~V. Kumar, ``Approximately universal optimality over several
  dynamic and non-dynamic cooperative diversity schemes for wireless
  networks.'' [Online]. Available: \url{http://fr.arxiv.org/pdf/cs.IT/0512028}
\BIBentrySTDinterwordspacing

\bibitem{SY_JCB_coop}
S.~Yang and J.-C. Belfiore, ``Optimal space-time codes for the {MIMO}
  amplify-and-forward cooperative channel,'' \emph{{IEEE} Trans. Inf. Theory},
  vol.~53, no.~2, pp. 647--663, Feb. 2007.

\bibitem{Horn}
R.~A. Horn and C.~R. Johnson, \emph{Matrix Analysis}.\hskip 1em plus 0.5em
  minus 0.4em\relax New York: Cambridge, 1985.

\bibitem{Rajan-03}
B.~A. Sethuraman, B.~S. Rajan, and V.~Shashidhar, ``Full-diversity, high-rate
  space-time block codes from division algebras,'' \emph{{IEEE} Trans. Inf.
  Theory}, vol.~49, no.~10, pp. 2596--2616, Oct. 2003.

\bibitem{SY_JCB_ISIT_parallel}
S.~Yang, J.-C. Belfiore, and G.~{Rekaya-Ben Othman}, ``Perfect space-time block
  codes for parallel {MIMO} channels,'' in \emph{Proc. IEEE International
  Symposium on Information Theory}, Seattle, WA, Jul. 2006.

\bibitem{Oggier-2}
F.~Oggier and E.~Viterbo, ``Algebraic number theory and code design for
  {R}ayleigh fading channels,'' in \emph{Foundations and Trends in
  Communications and Information Theory}, 2004, vol.~1, no.~3, pp. 333--415.

\bibitem{Bayer}
E.~{Bayer-Fluckiger}, F.~Oggier, and E.~Viterbo, ``New algebraic constructions
  of rotated $\mathbb{Z}^n$-lattice constellations for the rayleigh fading
  channel,'' \emph{{IEEE} Trans. Inf. Theory}, vol.~50, no.~4, pp. 702--714,
  Apr. 2004.

\bibitem{Damen_DAST}
M.~O. Damen, K.~{Abed-Meraim}, and J.-C. Belfiore, ``Diagonal algebraic space
  time block codes,'' \emph{{IEEE} Trans. Inf. Theory}, vol.~48, no.~3, pp.
  628--636, March 2002.

\bibitem{Alamouti}
S.~Alamouti, ``Space-time block coding: A simple transmitter diversity
  technique for wireless communications,'' \emph{{IEEE} J. Sel. Areas Commun.},
  vol.~16, pp. 1451--1458, Oct. 1998.

\bibitem{Edelman}
A.~Edelman, ``Eigenvalues and condition numbers of random matrices,'' Ph.D.
  Dissertation, MIT, 1989.

\bibitem{James}
A.~T. James, ``Distributions of matrix variates and latent roots derived from
  normal samples,'' \emph{Annals of Math. Statistics}, vol.~35, pp. 475--501,
  1964.

\bibitem{Gao_Smith}
H.~Gao and P.~J. Smith, ``A determinant representation for the distribution of
  quadratic forms in complex normal vectors,'' \emph{J. Multivariate Analysis},
  vol.~73, pp. 155--165, May 2000.

\bibitem{Simon}
S.~H. Simon, A.~L. Moustakas, and L.~Marinelli, ``Capacity and character
  expansions: {Moment} generating function and other exact results for {MIMO}
  correlated channels,'' \emph{{IEEE} Trans. Inf. Theory}, vol.~52, no.~12, pp.
  5336--5351, Dec. 2006.

\bibitem{Tulino_Verdu}
A.~M. Tulino and S.~Verdu, ``Random matrix theory and wireless
  communications,'' in \emph{Foundations and Trends in Communications and
  Information Theory}, 2004, vol.~1, no.~1, pp. 1--182.

\bibitem{SY_JCB_ds}
\BIBentryALTinterwordspacing
S.~Yang and J.-C. Belfiore, ``Diversity-multiplexing tradeoff of double
  scattering {MIMO} channels,'' \emph{{IEEE} Trans. Inf. Theory}, Mar. 2006,
  submitted for publication. [Online]. Available:
  \url{http://arxiv.org/pdf/cs.IT/0603124}
\BIBentrySTDinterwordspacing

\bibitem{Belfiore_golden}
J.-C. Belfiore, G.~Rekaya, and E.~Viterbo, ``The {G}olden code: {A} $2\times2$
  full-rate space-time code with non-vanishing determinants,'' \emph{{IEEE}
  Trans. Inf. Theory}, vol.~51, no.~4, pp. 1432--1436, Apr. 2005.

\end{thebibliography}
\end{document}